\documentclass[iop]{emulateapj-rtx4}
\shortauthors{Sekanina \& Kracht}
\shorttitle{Comet C/1995 O1:\ Orbit, Jovian Encounter, and Activity}
\slugcomment{Version \today }

\newcommand{\lapeq}{$\;$\raisebox{0.3ex}{$<$}\hspace{-0.28cm}\raisebox{-0.75ex}{$\sim$}$\;$}
\newcommand{\gapeq}{$\;$\raisebox{0.3ex}{$>$}\hspace{-0.28cm}\raisebox{-0.75ex}{$\sim$}$\;$}

\begin{document}
%
%
\title{ORBITAL EVOLUTION, ACTIVITY, AND MASS LOSS OF COMET C/1995 O1
(HALE-BOPP).\\I.\ CLOSE ENCOUNTER WITH JUPITER IN THIRD MILLENNIUM BCE
AND EFFECTS\\OF OUTGASSING ON THE COMET'S MOTION AND PHYSICAL PROPERTIES}
\author{Zdenek Sekanina$^1$ \& Rainer Kracht$^2$}
\affil{$^1$Jet Propulsion Laboratory, California Institute of Technology,
  4800 Oak Grove Drive, Pasadena, CA 91109, U.S.A.\\
$^2$Ostlandring 53, D-25335 Elmshorn, Schleswig-Holstein, Germany}
\email{Zdenek.Sekanina@jpl.nasa.gov\\
{\hspace*{2.59cm}}R.Kracht@t-online.de{\vspace{-0.1cm}}}

\begin{abstract}
This comprehensive study of comet C/1995 O1 focuses first on investigating
its~orbital~\mbox{motion}~over a~period of 17.6~yr (1993--2010).  The comet
is suggested to have approached Jupiter to 0.005~AU~on $-$2251 November 7,
in general conformity with Marsden's (1999) proposal of a Jovian encounter
nearly 4300~yr ago.  The variations of sizable nongravitational effects with
heliocentric distance correlate~with the evolution of outgassing, asymmetric
relative to perihelion.  The future orbital~period~will~shorten~to
$\sim$1000~yr because of orbital-cascade resonance effects.  We find that
the sublimation curves of parent molecules are fitted with the type of a law
used for the nongravitational acceleration, determine~their \mbox{orbit-integrated} mass loss, and conclude that the share of water ice was at most 57\%,
and possibly less than 50\%, of the total outgassed mass.  Even though
organic parent mole\-cules (many still unidentified) had very low abundances
relative to water individually, their high molar mass and sheer number made
them, summarily, important potential mass contributors to the total production
of gas.  The~mass~loss of dust per orbit exceeded that of water ice by a
factor of $\sim$12, a dust loading high enough to imply a major role for
heavy organic molecules of low volatility in accelerating the minuscule dust
particles~in the expanding halos to terminal velocities as high as 0.7 km
s$^{-1}\!$.  In Part II, the comet's nucleus will~be modeled as a compact
cluster of massive fragments to conform to the integrated nongravitational
effect.
\end{abstract}


\section{Introduction:\ Making Case for A Close Encounter with Jupiter}
The celebrated comet C/1995 O1 (Hale-Bopp) was exceptional in a number of
ways, one of which was its enormous intrinsic brightness.  This trait rendered
it possible to discover and observe the comet at very large heliocentric
distance and to collect astrometric observations over a long period of
time.  Discovered visually on 1995 July 23, more than 600 days before
perihelion, as an object of mag 10.5--10.8 (Hale \& Bopp 1995), the comet
was subsequently recognized at mag 18 on a plate taken on 1993 April 27
with the 124-cm UK Schmidt telescope of the Siding Spring Observatory
near Coonabarabran, N.S.W., Australia (McNaught 1995), 3.9~yr before
perihelion and 13.0~AU from the Sun.  With the latest astrometrically
measured images from 2010 December~4 (when 30.7~AU from the Sun!), taken
with the 220-cm f/8.0 Ritchey-Chr\`etien reflector of the European Southern
Observatory's Station at La Silla, Chile (Sarneczky et al.\ 2011), the
observed orbital arc covers a period of 17.6~yr, with 13.7~yr of
post-perihelion astrometry --- an unprecedented feat for a single-apparition
comet.

No less astonishing is the orbital history of the comet (Marsden 1999). Soon
after its discovery, the orbit could substantially be improved thanks to the
Siding Spring pre-discovery observation.  Although Marsden eventually
determined that the near-perihelion osculating orbital period equalled about
2530~yr, he noticed that this was due in part to the perturbations by
Jupiter~\mbox{during} the comet's approach to within 0.8~AU of the planet
about one year before the 1997 perihelion and that the orbit had originally
been more elongated, with a period of slightly longer than 4200~yr.
Obviously, the object did not arrive from the Oort Cloud; its previous
perihelion passage had occurred in the course of the 23rd century BCE.

As described by Marsden, this was only a minor part of the story.  Because
the comet's ascending node places it practically on Jupiter's orbit, extremely
close encounters with the planet are possible, and Marsden (1999) argued that
this was precisely what happened 42~centuries ago on the comet's approach to
perihelion.  He even ventured to propose that the comet had been captured by
Jupiter's gravity from its original Oort-Cloud-type, highly elongated orbit.

In his investigation, Marsden (1999) was handicapped by a relatively short
interval of time that the comet's observations then spanned.  The last
observation he employed was dated 1998 February 8, so that the observed
orbital arc was only 4.8~yr long, not much more than a quarter of the
interval of time available nowadays.
%
%

The fascinating problem of the long-term orbital evolution of C/1995~O1 and the
related issues warrant a new investigation, given in particular that the much
greater span of astrometric observations currently available allows us to
conduct a considerably more profound examination of the orbital scenario and
its potential implications than it was possible a dozen or more years ago [for
some later efforts similar to Marsden's (1999), see, e.g., Szutowicz et al.\
(2002) or Kr\'olikowska (2004)].  We further propose to confront our approach
and results with evidence from the comet's physical observations made in the
course of the 1997 apparition and to draw conclusions from this data interaction.
\section{Data, Tools, and Method of Approach}
We began by collecting 3581 astrometric observations of comet C/1995~O1
from {\vspace{-0.05cm}}the database maintained by the {\it Minor Planet
Center\/}.\footnote{See {\tt http://www.minorplanetcenter.net/db\_search/}.} 
This is a complete list of the comet's currently available astrometric data in
equinox J2000.  Besides the single pre-discovery observation in 1993 (McNaught
1995), the data set contains 830 observations from the year 1995, 1458 from
1996, 653 from 1997, 295 from 1998, 198 from 1999, 50 from 2000, 51 from 2001,
9 from 2002, 14 from 2003, 14 from 2005 (of which 6 from January--February and
8 from July--August), 5 observations from 2007 October, and 3 from 2010 December
4.\footnote{On a website {\tt https://groups.yahoo.com/neo/groups/comets-
ml/conversations/messages/19755/} D.\,Herald, Murrumbateman, Australia, reported
a detection of C/1995~O1 on CCD images of a total exposure time of 144~minutes
taken on 2012 August 7 with a 40-cm Schmidt-Cassegrain telescope.  However, in
response to our more recent inquiry, Mr.\,Herald stated in an e-mail message
dated 2014 August 30 that he had been unable to independently confirm the
detection.  For additional information, see the Appendix A.}

The primary objective of the first phase of our orbit-determination effort was
to subject the collected set of astrometric observations to tests in order to
extract a subset of high-accuracy observations.  To ensure that~this has
indeed been achieved requires an introduction of a cutoff or threshold of
choice for removing from the collection all individual observations whose
(observed minus computed) residuals in right ascension, \mbox{$(O\!-\!C)_{\rm
RA}$}, or declination, \mbox{$(O\!-\!C)_{\rm Decl}$}, exceed the limit.  In
practice, the task is made difficult by the fact that a residual exceeding
the limit does so for one of two fundamentally different reasons:\ either
because the astrometric position is of poor quality or because the comet's
motion is modeled inadequately.  The aim of the procedure is to reject the
data points with large residuals that are of the first category but not of
the second category; to discriminate between the two categories requires
an incorporation of an efficient filter.  Typically, the category into
which an observation with a large residual belongs is revealed by comparing,
with one another, observations over a limited orbital arc:\ if all or most
of them exhibit systematic residuals, the problem is in the comet's modeled
motion; if one or a few stand out in reference to the majority, it is the
poor quality of the individual observations.  Obviously, this filter fails
when there is only one or a very few observations available widely scattered
over a long orbital arc (Section 4).

Since we did not define what we meant by inadequate modeling of the comet's
motion at the start of the orbit-determination process, we used two tools to
filter out the observations of poor quality and at the same time to ensure
that no observation of required accuracy was removed.  One tool was a stepwise
approach, beginning with a cutoff at 10$^{\prime\prime}$ (Section~3) and
tightening it up; the other tool was to divide the whole observed orbital
arc into a number of segments of approximately equal and short enough length,
fitting an osculating gravitational solution through each of the segments.

In addition, throughout this work we consistently applied a self-sustaining
test of cutoff enforcement.  When a particular cutoff was introduced as a
limit for rejecting inaccurate observations, minor differences between two
consecutive solutions --- before and after elimination of the poor-quality
data --- could cause that some data points with the residuals slightly
exceeding the limit in the first solution (and therefore removed from the
set of used observations) just satisfied the limit in the second solution
(and were to be incorporated back into the set); while other data points
with the residuals just satisfying the limit in the first solution (and
therefore kept in the set) slightly exceeded the limit in the second
solution (and were now to be removed from the set).  In order to comply
with the cutoff rule, it was necessaary to iterate the process until the
number of astrometric observations that passed the test fully stabilized.

As described in Section 3 below, the single observation of 1993 is known to be
accurate to about 1$^{\prime\prime}$, while nearly a half of the observations
from 2005 and all from 2007-2010 were made with large telescopes; the more
advanced solutions (Sections 6.2--6.3) showed a posteriori that these data
points satisfied the tight limit and were not subjected to the aforementioned
test.
%
%

With these basic rules in mind, we applied an orbit-determination code {\it
EXORB7\/}, written by A.\,Vitagliano, both in its gravitational and
nongravitational modes.  Using the JPL DE421 ephemeris, the code accounts
for the perturbations by the eight planets, by Pluto, and~by the three
most massive asteroids, as well as for the relativistic effect.  The
nongravitational accelerations are incorporated directly into the equations
of motion (Section~4).  The code employs a least-squares optimization method
to derive the resulting elements and other parameters.

\begin{table*}[t] 
\vspace{-3.9cm}
\hspace{-0.52cm}
\centerline{
\scalebox{1}{
\includegraphics{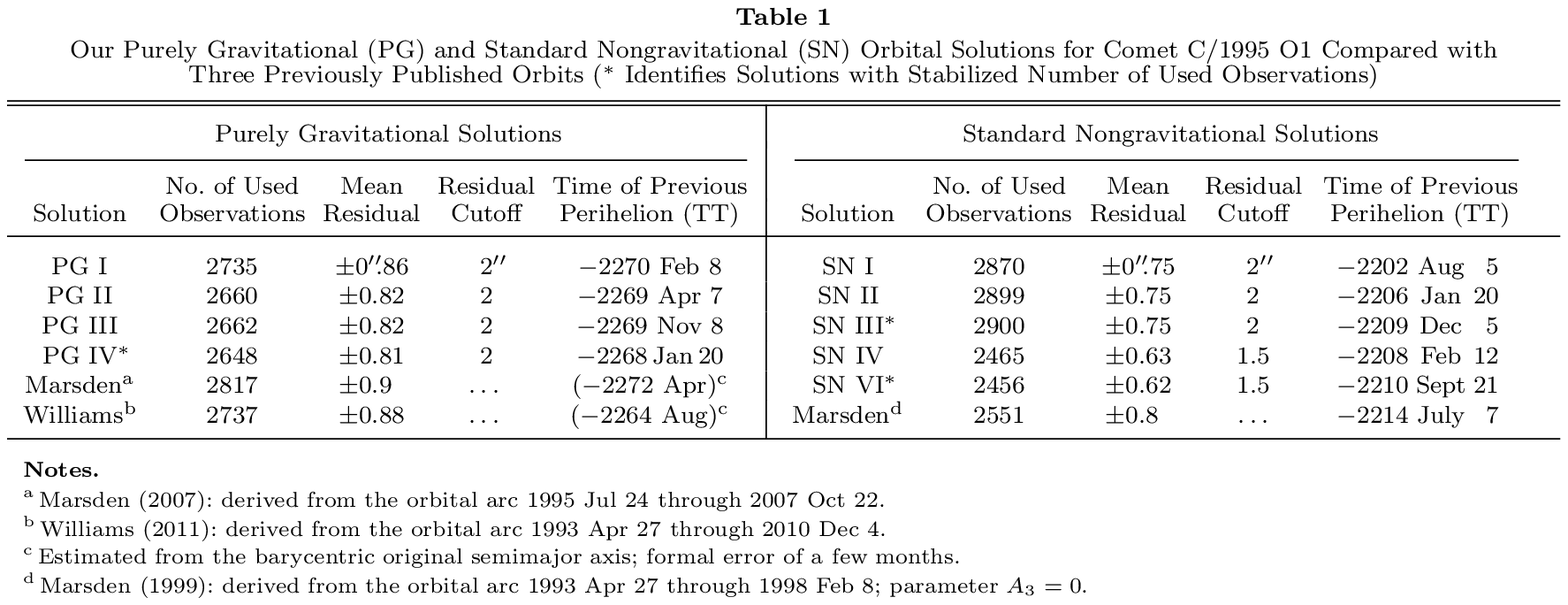}}} 
\vspace{-18.51cm}
\end{table*}

\section{Past Work and Purely Gravitational Solutions}
%
%
The comet's pre-discovery image from 1993, measured by McNaught (1995), is of
major significance, because it extends the observed orbital arc by 27~months.
Marsden (1999) commented on his (and other authors') difficulties with linking
this data point to the post-discovery, preperihelion observations and he
emphasized the magnitude of the problem at the time by pointing to the
possibility of ``a significant unmodeled perturbation [that] must have been
operative over a relatively short time interval,'' in a reference to
D.\,K.\,Yeomans.  The situation was not alleviated by McNaught's
remeasurement of the comet's 1993 image, which showed that its
astro\-metric position was accurate to within
1$^{\prime\prime}$.~Also,~\mbox{McNaught's}~\mbox{positional}~ver\-ification
of a known asteriod on the same plate confirmed the correct time.  The
problem was solved (temporarily) by assigning a greater weight to the 1993
data point.~New difficulties arose in an effort to link the preperihelion
observations with the post-perihelion observations~near~the end of 1997, and
the nongravitational terms had to be introduced into the equations of motion.
At the time of his paper's submission (1998 early February), the 1993
observation (assigned a unit weight) was noted by \mbox{Marsden} (1999)
to leave a residual of 1$^{\prime\prime}\!$.8 in right ascension even with
the nongravitational terms included.

Interestingly, Marsden's (2007) last published orbit\footnote{This set of
elements is also listed in Marsden \& Williams' (2008) most recent edition of
their catalog of cometary orbits.{\vspace{0.05cm}}} was gravitational, but
the 1993 pre-discovery position was left out and 2817 observations were
linked from only 1995 July 24 through 2007 October 22; the mean residual
equaled $\pm$0$^{\prime\prime}\!$.9.  On the other hand, the most recently
published orbit, by Williams (2011), is based on gravitationally linking
2737 observations from the entire period 1993--2010, leaving a mean residual
of $\pm$0$^{\prime\prime}\!$.88; no details are available on the quality of
fit.

The lesson learned from the obstacles encountered by Marsden (and others)
in trying to fit a 57-month-long arc of the comet's orbit leaves no hope that
a solution based on the gravitational law alone could adequately represent
a 211-month-long arc, nearly four times as long.  We also suspected that the
small number of observations available from the period of time after 2003 ---
not to mention the single 1993 pre-discovery observation --- would eventually
require their heavy weighting in comparison with the data from the interval of
1995--2003.

Given this history of orbit determination of C/1995~O1 and also in view of
the need of a reference standard for our further computations, we first
set to derive the best achievable purely gravitational solution.  To begin
with, we assigned a unit weight to each of the 3581 observations and
computed an initial gravitational orbit.  This step was necessary because
of the lack of a priori information on the quality of the individual
data points.  This solution, whose mean (rms) residual came out to be
$\pm$1$^{\prime\prime}\!$.56, served to obtain a complete set of initial
residuals.  We noticed the presence of a few residuals exceeding
10$^{\prime\prime}$ and a large number of residuals substantially
exceeding 2$^{\prime\prime}$; there were strong systematic trends in
right ascension, with the residuals exceeding 4$^{\prime\prime}$ in
2010.  The 1993 pre-discovery observation left a residual exceeding
3$^{\prime\prime}$ in declination.

Retaining an equal weight for all observations, these results prompted us
to set a residual cutoff at 2$^{\prime\prime}$ in either coordinate and to
iterate the gravitational least-squares fitting to the observations until
their employable number stabilized.  The left-hand side of Table~1 shows that
this task required four iterations to determine that the total number of
observations satisfying this condition settled down --- in the solution PG~IV
--- on 2648 and the mean residual on $\pm$0$^{\prime\prime}\!$.81.  This
solution places the comet's previous passage through perihelion early in the
year $-$2268.  Comparison with the orbits by Marsden (2007) and by Williams
(2011) shows that this perihelion time is close to an average of the two.
On the other hand, it differs by more than 50~yr from the perihelion time
predicted by Marsden's (1999) nongravitational solution.
%

The residuals \mbox{$O\!-\!C$} from the solution PG~IV are plotted in Figure 1.
The fit can by no means be considered satisfactory, so the predicted perihelion
passage in $-$2268 cannot be correct.  In right ascension the observation of
1993 leaves an acceptable residual, but a distinctly systematic trend starts
in 1998 from which time on the solution shows a progressive tendency toward
negative residuals, reaching $-$3$^{\prime\prime}$ in 2010.  In declination,
the pre-discovery observation leaves a residual greater in absolute value
than 2$^{\prime\prime}$.  Less prominent, short-term systematic trends are
also seen at other times in both coordinates.  Each residual that exceeds
2$^{\prime\prime}$ is legitimate, as it does so because of an imperfect model.
%
%

%
\begin{figure*}[t] 
\vspace{-3.82cm}
\hspace{-0.15cm}
\centerline{
\scalebox{0.8}{
\includegraphics{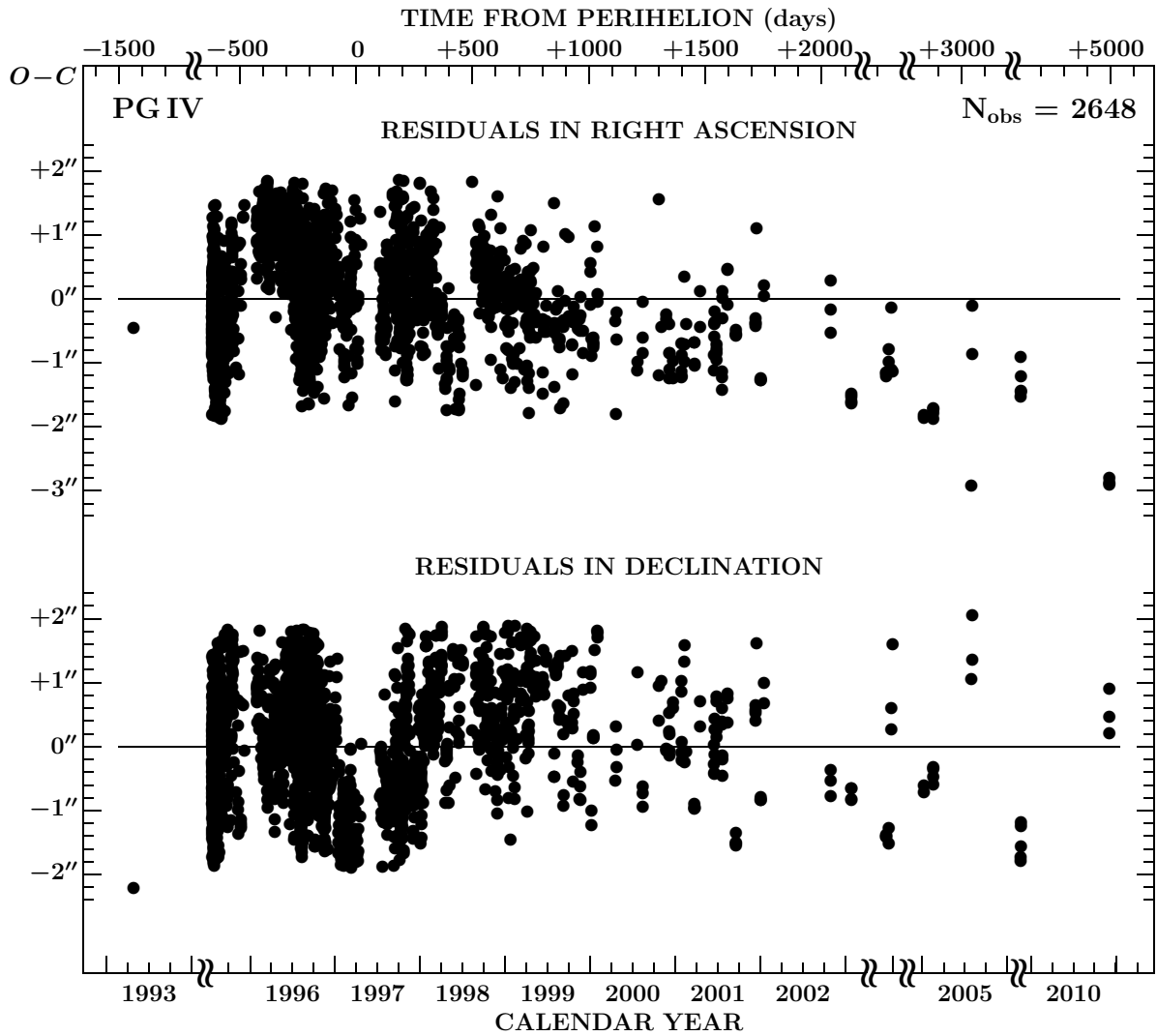}}} 
\vspace{-5.75cm}
\caption{Temporal distribution of residuals $O\!-\!C$ (observed minus
computed) in right ascension (top) and declination (bottom) from the
gravitational solution PG IV to 2648 accurate observations of comet
C/1995~O1 between 1993 and 2010.  The axis of abscissae is interrupted
in the years 1994--1995, 2003, 2004, 2006--2007, and 2007--2009 over
periods of time with no observations. A strong systematic trend is
apparent in right ascension starting in mid-1998 and reaching
3$^{\prime\prime}$ by 2010.  The pre-discovery observation leaves a
residual exceeding 2$^{\prime\prime}$ in declination.  Less prominent
systematic trends are apparent from 1995 to mid-1996 in right ascension
and during much of 1997 and 1998 through part of 1999 in declination.
All observations were given a unit weight.{\vspace{0.39cm}}}
\end{figure*}

\section{Standard Nongravitational Solutions}

Having confirmed that gravitational solutions are unacceptable, we next turned
to solutions based on the standard nongravitational formalism, proposed by
Marsden et al.\ (1973), in which the nongravitational terms were incorporated
directly into the equations of motion and the magnitude of the nongravitational
acceleration, as a function of heliocentric distance $r$, was expressed by an
empirical formula $g_{\rm ice}(r;r_0)$, designed to mimick the law of
momentum transfer driven by the sublimation of water ice,
\begin{equation}
g_{\rm ice}(r;r_0) = \psi \, \Lambda^{-m} (1 \!+\! \Lambda^n)^{-k},
\end{equation}
where \mbox{$\Lambda = r/r_0$}, $r_0$ is the scaling distance for water ice,
exponents $m$, $n$, and $k$ are constants, and $\psi$ is a normalization
coefficient, such that \mbox{$g_{\rm ice}(1\,{\rm AU};r_0) = 1$}.  Marsden et
al.'s (1973) formalism employs a so-called isothermal model, which averages
the Sun's incident radiation over a spherical nucleus' surface of constant
temperature, assuming an albedo of 0.1 in both the optical and thermal spectral
passbands.  For this model, the~\mbox{parameters} are \mbox{$r_0 = 2.808$ AU},
\mbox{$m = 2.15$}, \mbox{$n = 5.093$}, \mbox{$nk = 23.5$}, and \mbox{$\psi =
0.1113$}.  Even though the temperature of a cometary nucleus is known to vary
greatly over the surface, for the orbit-determination purposes the employed
simple model has served admirably over more than four decades and the formalism
is still used today.

Tested countless times in the past, this standard formalism stipulates a
constant orientation of the momentum-transfer vector in an {\small \bf
RTN} right-handed rotating Cartesian coordinate system.  Its cardinal
directions, referred to the comet's orbital plane, are the radial {\small
\bf R} (away from the Sun), transverse {\small \bf T}, and normal {\small
\bf N} components.  The components $j_{\bf C}(t)$ \mbox{({\small \bf C} =
{\small \bf R}, {\small \bf T}, {\small \bf N})} of the nongravitational
acceleration, to which the comet's nucleus is subjected, are, at time $t$
(when the heliocentric distance is $r$), equal to
\begin{equation}
\left[ \! \begin{array}{c}
j_{\bf R}(t) \\[0.07cm] j_{\bf T}(t) \\[0.07cm] j_{\bf N}(t)
\end{array}
\! \right] \!=\! \left[ \! \begin{array}{c}
A_1 \\[0.07cm] A_2 \\[0.07cm] A_3
\end{array}
\! \right] \cdot g_{\rm ice}(r;r_0),
\end{equation}
where $A_1$, $A_2$, and $A_3$ are the magnitudes of the nongravitational
acceleration's components in, respectively, the radial, transverse, and normal
directions at 1~AU from the Sun; they are the additonal parameters that are,
together with the orbital elements, determined from the employed observations
in the process of orbit determination by applying a least-squares optimization
procedure.

\begin{figure*}[t] 
\vspace{-3.75cm}
\hspace{-0.15cm}
\centerline{
\scalebox{0.79}{  
\includegraphics{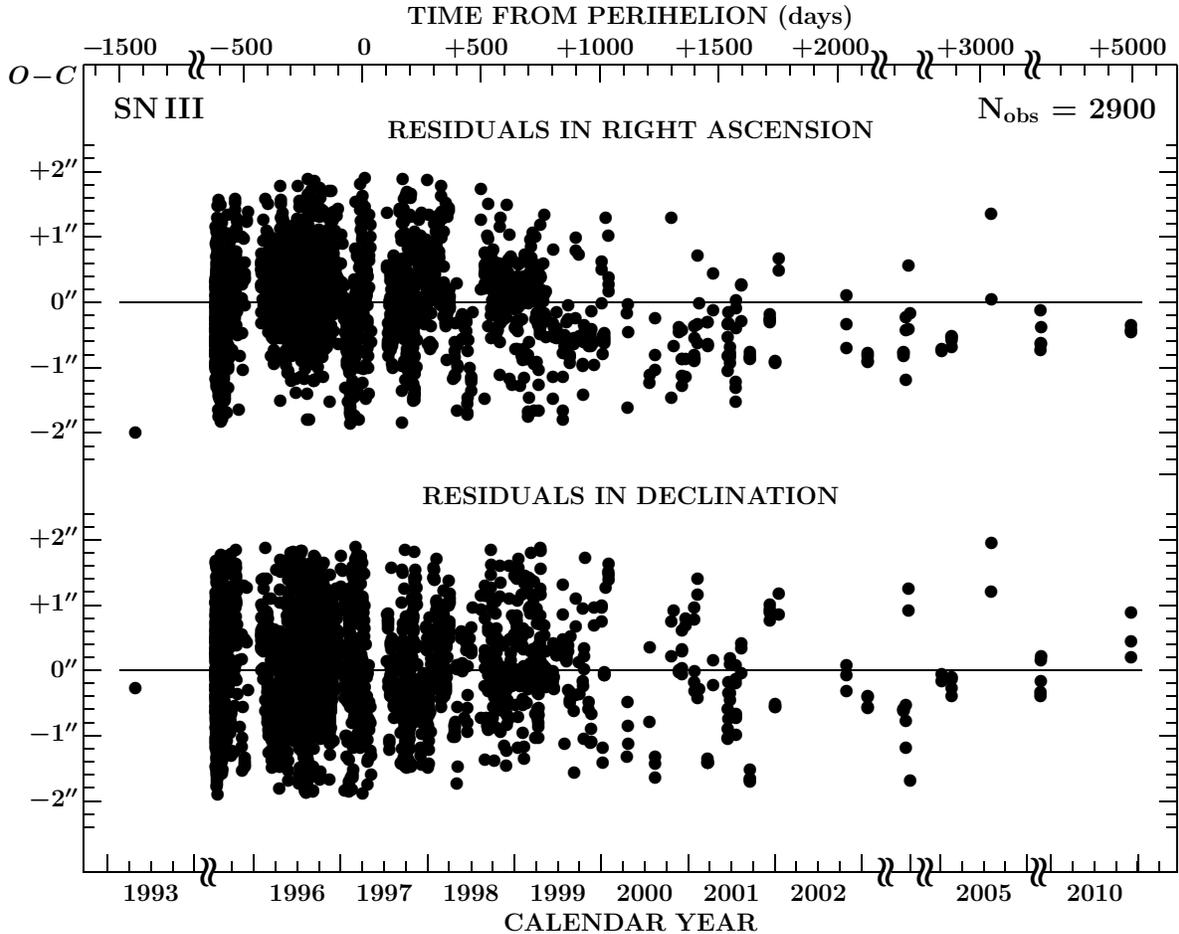}}} 
\vspace{-7.28cm}
\caption{Temporal distribution of residuals \mbox{$O\!-\!C$} (observed minus
computed) in right ascension (top) and declination (bottom) from the standard
nongravitational solution SN~III to 2900 accurate observations of comet C/1995~O1
between 1993 and 2010.  For more description, see the caption to Figure~1.
The fit is now much better, but not quite acceptable.  The pre-discovery
observation leaves a residual of 2$^{\prime\prime}$ in right ascension.  A
slight systematic trend in right ascension is seen between 1999 and 2010.  Only
in declination is the fit fairly satisfactory.{\vspace{0.4cm}}}
\end{figure*}

Retaining again an equal weight for all observations and a residual cutoff at
2$^{\prime\prime}$, we iterated the standard nongravitational least-squares
fitting to the observations until their number satisfying the conditions
stabilized.  Limited experimentation showed that the incorporation of the
normal component of the nongravitational acceleration in addition to the
usually included radial and transverse components improved the fit to a degree.
The right-hand side of Table~1 indicates that after three iterations the
number of employable observations leveled off at 2900, with the mean residual
of $\pm$0$^{\prime\prime}\!$.75 in the solution SN~III.  Thus, with somewhat
higher accuracy than the solution PG~IV, it accommodates $\sim$250 more
observations, suggesting that the incorporation of the nongravitational terms
into the equations of motion definitely had beneficial effects.

The solution SN III puts the comet's previous perihelion passage late in the year
$-$2209, within 6~yr of the time computed by Marsden (1999).  It thus appears
that, in terms of the previous perihelion time, there is a fair mutual agreement
among the various gravitational orbits on the one hand and among the standard
nongravitational orbits on the other hand, but that there is a systematic
difference of some 50--60 yr between the two types of orbital solutions, the
gravitational solutions predicting an earlier perihelion time.

%
The residuals \mbox{$O\!-\!C$} from the solution SN~III are plotted in Figure~2.
The 1993 pre-discovery observation is now fitted satisfactorily in declination,
but it leaves a residual of 2$^{\prime\prime}$ in right ascension.  The residuals
from the years 1999--2010 are also fairly satisfactory in declination, but they
have a tendency to being negative in right ascension.  We conclude that the
standard nongravitational law, although improving the match significantly, still
fails to offer a perfect fit and leaves thus room for further improvement.  In
terms of the predicted perihelion time at the previous return to the Sun, the
solution SN~III does not necessarily offer a better estimate than the purely
gravitational solution PG~IV.  This question also raises doubt about whether
the comet did indeed return to perihelion in the year $-$2214.

\section{Tightening the Cutoff for Residuals}
\vspace{0.08cm}

Before searching for a qualitatively superior approach in our quest for as perfect
a distribution of residuals as possible, we considered it appropriate to further
tighten the residual cutoff for the observations from the years 1995--2003 and
possibly even 2005.  We felt that the cutoff should not be tighter than, or
competitive with, the reported accuracy of {\lapeq}$\!\!1^{\prime\prime}$ of
{\vspace{-0.05cm}}the pre-discovery observation (Marsden 1999), which we were
determined to keep in the database under any circumstances.  We eventually
adopted a cutoff of 1$^{\prime\prime}\!$.5.

To check the effect of the tightened cutoff on the sample size, the predicted
previous perihelion time, and the nongravitational parameters under otherwise
identical conditions, we continued the investigation by means of the standard
nongravitational law.  We found that, with the new cutoff, three iterations
were needed to stabilize the number of employable observations; the runs were
SN~IV, SN~V, and SN~VI, the total numbers of used data points were, respectively,
2465, 2459, and 2456.  The first and last of these solutions are presented in
Table~1.  As expected, the distribution of residuals from SN~VI was, except for
the drop in their range, very similar to that from SN~III and, therefore, not
entirely satisfactory.

\begin{table}[t] 
\vspace{-4.18cm}
\hspace{4.23cm}
\centerline{
\scalebox{1}{
\includegraphics{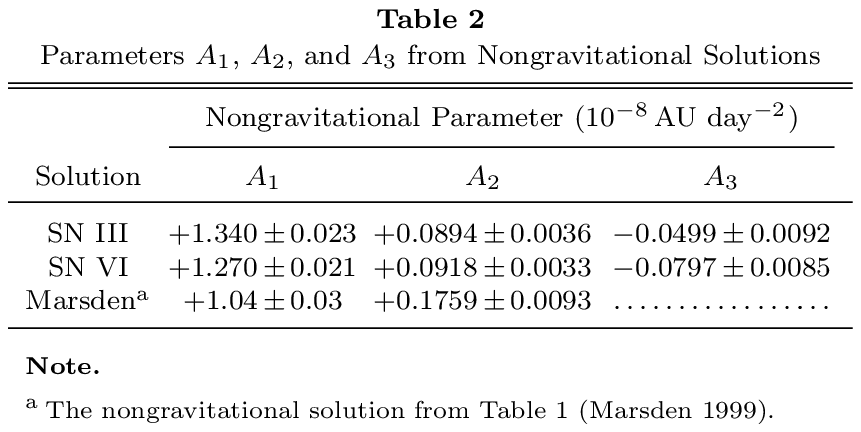}}} 
\vspace{-21cm}
\end{table}

Table 1 shows that the number of employable observations in the solution SN~VI
dropped by only about 15~percent compared with SN~III, a result that is a tribute
to good work by the astrometrists.  The predicted perihelion passage moved back
in time to September of $-$2210, by less than 15 months and still closer to
Marsden's (1999) prediction.  However, comparison of the nongravitational
parameters, in Table~2, suggests that the cutoff's  tightening made important
differences:\ the disparity between SN~VI and SN~III in the radial and normal
components' parameters, $A_1$ and $A_3$, just exceeded 3$\sigma$ of either run,
whereas the transverse component's parameters, $A_2$, of the two
solutions agreed to within 1$\sigma$.  The discrepancies between SN~VI and
Marsden's (1999) orbit, computed with an assumed $A_3$ equal to zero, are rather
startling, exceeding 7$\sigma$ in $A_1$ and even more in $A_2$.

\section{Modified Nongravitational Solutions}

In\,one of the previous papers\,(Sekanina\,\&\,Kracht~2015), we introduced a
generic {\it modified nongravitational law\/}, $g_{\rm mod}(r;r_0)$, and
applied it to much advantage in our study of the dwarf Kreutz sungrazing
comets, discovered in large numbers with the coronagraphs onboard the {\it
Solar and Heliospheric Observatory\/} ({\it SOHO\/}; Brueckner et al.\ 1995)
and, to a lesser degree, with the coronagraphs onboard the two spacecraft of
the {\it Solar Terrestrial Relations Observatory\/} ({\it STEREO\/}; Howard
et al.\ 2008).

The modified nongravitational (MN) law retains the values of the exponents
$m$, $n$, and $k$ of the standard law (1), but varies the scaling distance
$r_0$ and the constant $\psi$.  This intervention is fully justified on
account of Marsden et al.'s (1973) finding that the shapes of the normalized
sublimation-rate curves against heliocentric distance for a variety of
species are fairly similar in a log-log plot except for major horizontal
shifts, which means that they are relatively insensitive to the exponents
$m$, $n$, and $k$, but critically dependent on the scaling distance $r_0$.
For a given absorptivity and emissivity of the nuclear surface, $r_0$
measures essentially an effective latent heat of sublimation, $L_{\rm sub}$,
of the observed mix of outgassing species, varying to a first approximation
inversely as its square,
\begin{equation}
r_0 \simeq \left( \! \frac{\rm const}{L_{\rm sub}} \! \right)^{\!\!2} \! ,
\end{equation}
where a calibration by water ice gives{\vspace{-0.05cm}} for the constant a value
of 19\,100 AU$^{\frac{1}{2}}$\,cal\,mol$^{-1}$\,in the case of an isothermal model
but 27\,000 AU$^{\frac{1}{2}}$\,cal\,mol$^{-1}$\,for outgassing from a subsolar
region only, the two extreme values of the scaling distance.  In the following,
I employ the first model for a reason discussed briefly at the end of Section 9.4.

For the dwarf comets of the Kreutz sungrazing system the application of this
modified version of the nongravitational law was instrumental because the
species that were sublimating profusely from their disintegrating nuclei
in close proximity of the Sun were {\it much less\/} volatile than water
ice (Sekanina \& Kracht 2015); these refractory materials have \mbox{$r_0
\ll 2.8$ AU}.  For C/1995~O1, on the other hand, the modified law is
necessary because a large body of data (e.g., Crovisier et al.\ 1996,
1997; Jewitt et al.\ 1996; Biver et al.\ 1997, 1999; Bockel\'ee-Morvan
\& Rickman 1999; Crovisier 1999; Despois 1999; DiSanti et al.\ 1999)
provides undisputable evidence for abundant release of species {\it much
more\/} volatile than water ice; for these ices, in particular for carbon
monoxide and carbon dioxide, \mbox{$r_0 \gg 2.8$ AU}.

\subsection{Assigning Greater Weights to Critical Observations}
A number of initial orbital runs based on the modified nongravitational law
consistently showed that the fit to the observations could not be improved
significantly and the systematic trends in the residuals at the ends of the
orbital arc removed, unless these first and last observations --- that we
herefater call {\it critical\/} --- were assigned substantially greater
weights than the observations near the arc's middle.  We first allotted
weight 40 to the pre-discovery observation in 1993, weight 20 to each of
the three observations on 2010 December 4, and weight 10 to each of the
five observations on 2007 October 20--22, so that we had at that point a
total of 9 critical observations.  All other employed observations, from
the years 1995--2005, were allotted a unit weight.

We searched for orbital solutions based on a set of the observations weighted
this way assuming several scaling distances $r_0$ and compared the mean residual
left by all the observations used with the mean residual left only by the critical
observations.  In the subsequent runs we changed the weight assignment and
included the 2005 observations among the critical ones.  Some of the solutions
obtained as we tried to stabilize the number of observations satisfying the
residual cutoff of 1$^{\prime\prime}\!$.5 (similarly to the solutions PG~I
through PG~IV and SN~I through SN~III in Table~1), are presented in Table~3.
They show the assigned weights and a dramatic impact of the stabilization
process on the number of retained observations.

\begin{table*} 
\vspace{-4.15cm}
\hspace{-0.52cm}
\centerline{
\scalebox{1}{
\includegraphics{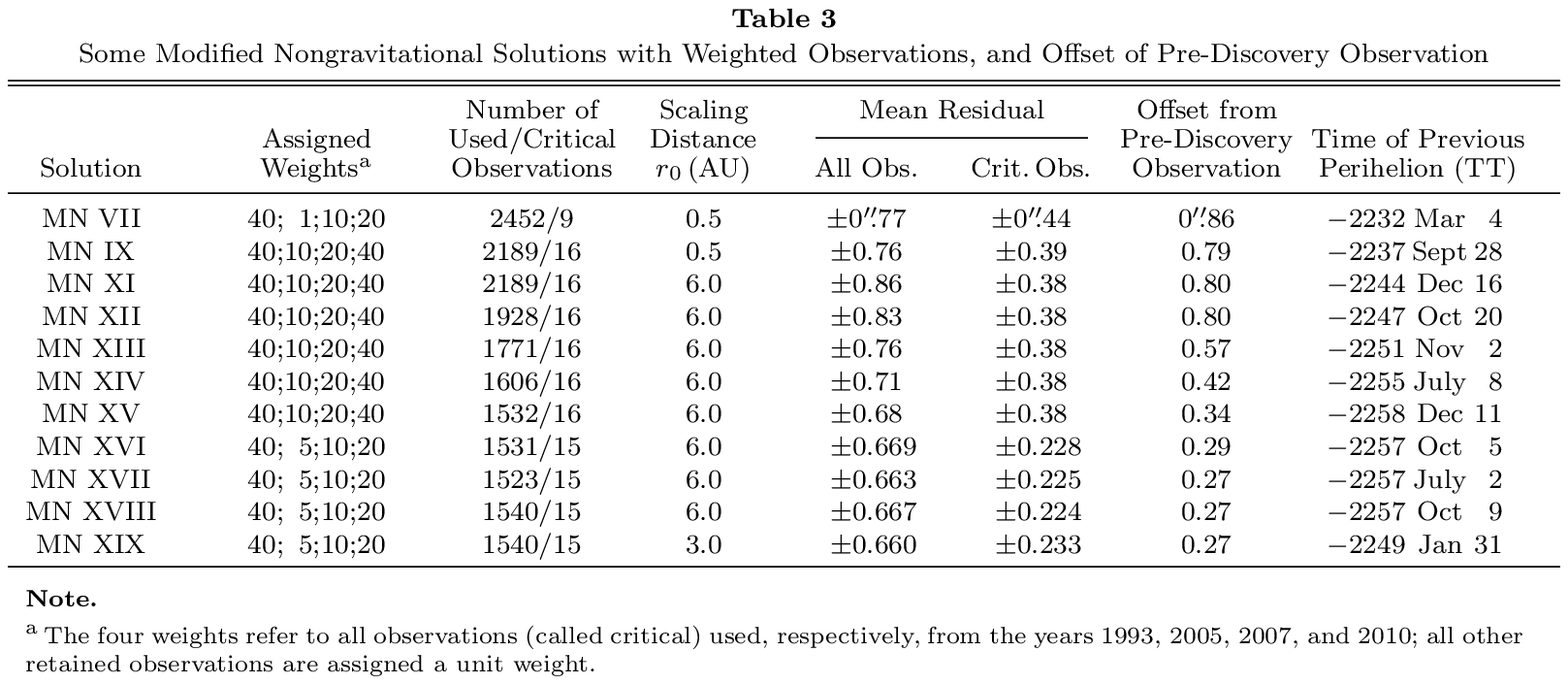}}} 
\vspace{-17.9cm}
\end{table*}

These results provide an answer to the stubborn problem that Marsden (1999)
was struggling with in his effort to accommodate the pre-discovery observation.
The bottom line is that this could readily be done (with a residual of $<\!\!
0^{\prime\prime}\!$.5) by weighting the observation heavily enough; however,
the resulting orbital solutions demanded rejection of almost one half of all
observations that we started with ($\sim$2900), because their residuals failed
to satisfy the tight cutoff.  We investigated the temporal distribution of the
rejected observations and found that nearly 500 (or one third) of them came
from the period of time between mid-March 1997 and early March 1998 (two weeks
before perihelion to 11 months after perihelion) and that only very few
acceptable observations from this time remained in the set.  A vast majority
of the rejected observations left residuals that were positive in right
ascension and negative in declination.  For example, the 69 rejected
observations during 1997 April left an average residual of
\mbox{+3$^{\prime\prime}\!.0 \pm 1^{\prime\prime}\!.4$} in right ascension
and \mbox{$-4^{\prime\prime}\!.4 \pm 1^{\prime\prime}\!.9$} in declination;
similarly, the average residuals amounted, for example, to
\mbox{+2$^{\prime\prime}\!.5 \pm 1^{\prime\prime}\!.2$} and \mbox{$-2^{\prime
\prime}\!.3 \pm 1^ {\prime\prime}\!.3$}, respectively, from the 88 rejected
observations during 1997 October, and to \mbox{+2$^{\prime\prime}\!.2 \pm
0^{\prime \prime}\!.9$} and \mbox{$-1^{\prime\prime}\!.2 \pm 0^{\prime\prime}
\!.5$}, respectively, from the 40 rejected observations during 1998 February.
These offsets do not correlate with any cardinal direction, such as the
antisolar direction.  From Table~3 we note the~lack of correlation between
the mean residual from all the used observations and the mean residual from
the critical observations.  For example, comparison of the solutions MN~IX
and MN~XI suggests that the fit with a scaling distance $r_0$ of 0.5~AU is
clearly better than the fit with \mbox{$r_0 = 6$ AU} in terms of all the
employed observations, but slightly worse in terms of the critical
observations.  The solutions MN~XI through MN~XV indicate, on the other
hand, that for a given scaling distance the quality of fit does not change
among the critical observations, but it improves among all the used
observations as their number drops dramatically.  The highly weighted
1993 pre-discovery observation is nonetheless fitted the better the smaller
is the mean residual from all the used observations.  We further note
from Table~3 that, as the number of retained observations drops, the previous
perihelion passage moves systematically by many years to earlier times; the
perihelion time predicted from the solution MN~XV is within 11 years of the
time predicted in Table~1 from the gravitational solution PG~IV!

The last four entries of Table 3, whose weights for~the critical data ---
40\,(1993),\,5\,(2005),\,10\,(2007),\,20\,(2010) --- are hereafter called the
{\it Weight System~I\/}, are the most important ones.  The last observation of
2005 was now rejected, leaving just 15 critical observations, with the total
number of observations eventually stabilized at 1540.  MN~XVIII and MN~XIX,
which fit the data points essentially equally well, are the basis for deriving
the optimum solution, the final step in our orbit-determination effort; it is
dealt with in the next subsection.

\begin{table}[hb] 
\vspace{-3.85cm}
\hspace{4.23cm}
\centerline{
\scalebox{1}{
\includegraphics{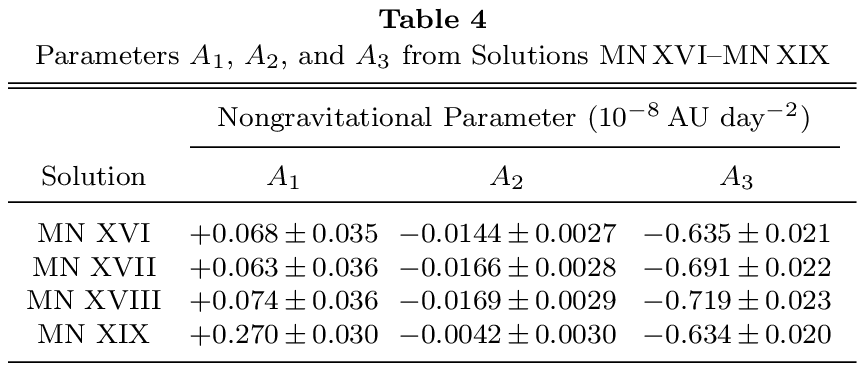}}} 
\vspace{-22.08cm}
\end{table}

Before addressing that issue, we should comment on the properties and the
implications of the most important solutions in Table~3, MN~XVI through MN~XIX.
These solutions require, similarly to MN~XV, that nearly all the observations
between mid-March 1997 and early March 1998 be rejected.  The nongravitational
parameters from MN~XVI through MN~XIX, listed in Table~4, reveal some interesting
features.  The most surprising is that by far the highest acceleration is in
the direction normal to the orbital plane, $A_3$.  We will return to this result
in the section in which we confront orbital evidence with information on the
nuclear rotation of the comet.  Table~4 also shows that the ratio of
\mbox{$|A_3|/A_1$} varies strongly with the scaling distance $r_0$, dropping
from near 10 at \mbox{$r_0 = 6$ AU} to a little over 2 at \mbox{$r_0 = 3$ AU},
when the transverse component $A_2$ is very small and poorly defined.

Comparison with the solutions SN~III and SN~VI from Table~2 also offers some
rather unexpected conclusions.  The total magnitude{\vspace{-0.04cm}}
of the nongravitational acceleration, $\sqrt{A_1^2 \!+\! A_2^2 \!+\! A_3^2}$,
is now lower by a factor of $\sim$2.  For the solutions MN~XVI through MN~XVIII
this could be understood as an effect of the scaling distance (which of course
equaled 2.8 AU for the SN solutions), because a greater scaling distance means a
greater contribution to the integrated effect from larger heliocentric distances.
However, this does not explain the factor of 2 between the SN solutions and
MN~XIX, whose $r_0$ was nearly the same.  The incompatibility of the SN and
MN results is also underscored by the opposite signs of the transverse component.
It thus appears that assigning weights to the critical observations and relaxing
the scaling distance had a major effect on the results.
%
%
%

\begin{table}[b] 
\vspace{-4.05cm}
\hspace{4.3cm}
\centerline{
\scalebox{1}{
\includegraphics{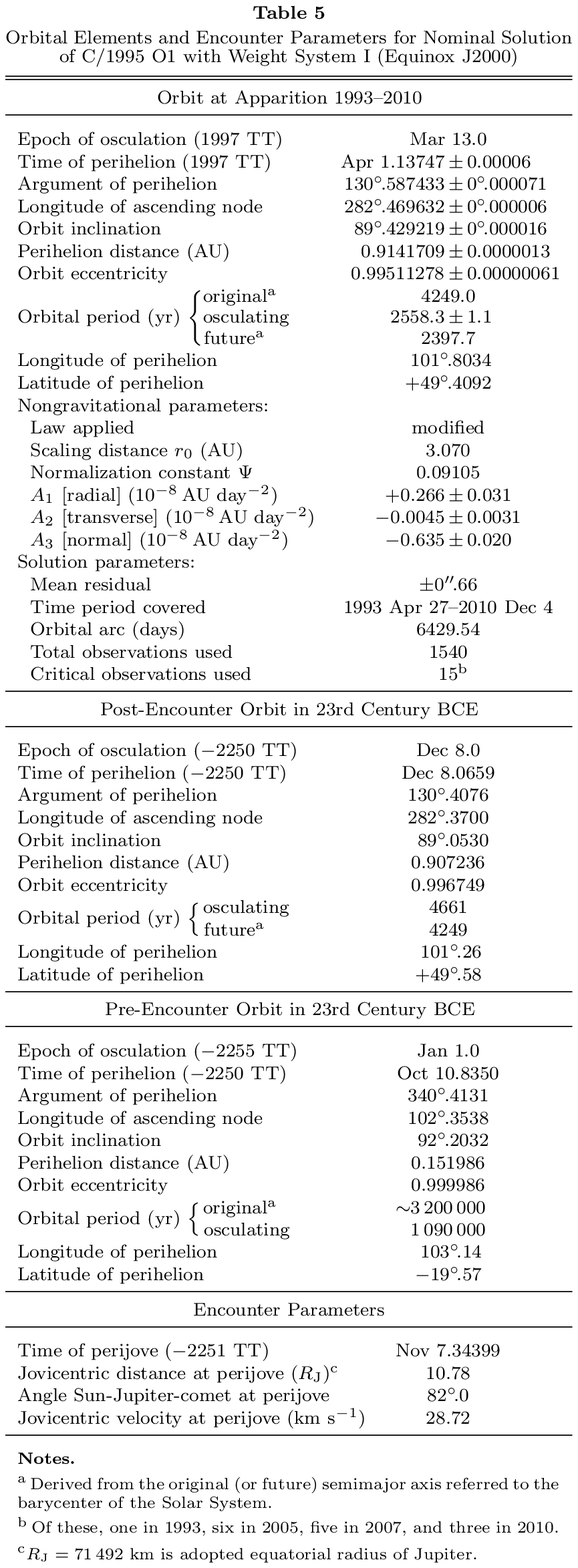}}} 
\vspace{-2.4cm}
\end{table}

\subsection{Weight System I:\ Final Optimization of Orbit}

Table 3 shows that the uncertainty in the scaling distance $r_0$ implies an
uncertainty of at least several years in the time of perihelion passage in the
23~century BCE.  Because of this indeterminacy, it could never be proven that
C/1995~O1 did in fact experience a close encounter with Jupiter.  However,
pursuing Marsden's (1999) suggestion that ``{\it it is not entirely improbable
that {\rm [}the comet underwent\/{\rm ]} a recent dramatic approach\/}'' to
the planet,~we exploited the tabulated dependence of the previous perihelion
time on the scaling distance.  By slightly adjusting $r_0$, we were able to
determine the time of perijove and the corresponding Jovicentric distance that
were compatible with plausible pre-encounter orbital constraints.  We found
that the comet had to pass through perihelion in early December of $-$2250 in
order to approach closely Jupiter.  This was the planet's only revolution that
was consistent with the close-encounter requirement.

To achieve a similarly favorable configuration during the previous Jovian
revolution, the comet would have to have passed through perihelion in late
January of $-$2261.  However, this time requires \mbox{$r_0 \approx 8$ AU},
for which the radial parameter \mbox{$A_1 < 0$}, in contradiction to the
physical model of nongravitational forces.

On the other hand, a favorable configuration one Jovian revolution later
would require that the comet's perihelion occurred in mid-October of
$-$2238.  A set of computer runs showed that, unfortunately, there was no
solution with a perihelion time later than near the end of $-$2247, and
even that implied an unrealistically small scaling distance of \mbox{$r_0
< 1$ AU}.

A fine tuning of the orbital elements, starting with the set MN~XIX in Table~3
and based, as described below, on the constraint that C/1995~O1 was in the
previous return a dynamically new comet, resulted in a nominal solution with
Weight System~I, which is presented in full detail~in Table~5.  The perihelion
time on $-$2250 December 8 suggests that Marsden's (1999) choice for the
encounter time was three Jovian revolutions much too late.

\begin{table}[b]
\vspace{-4cm}
\hspace{4.23cm}
\centerline{
\scalebox{1}{
\includegraphics{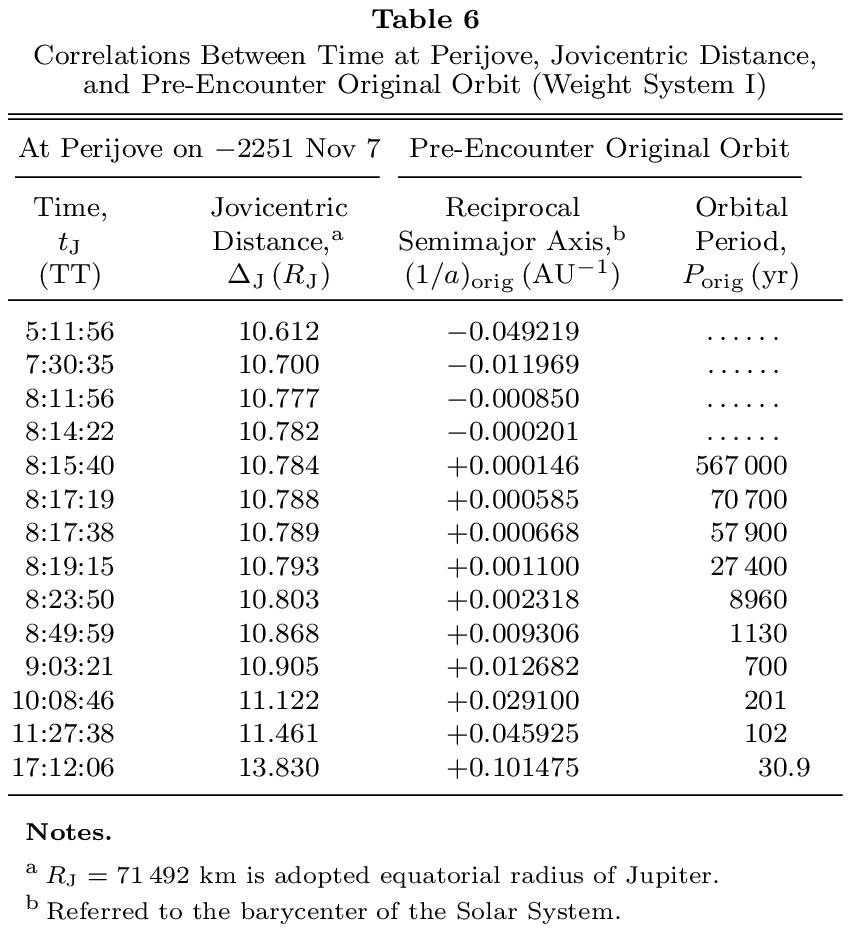}}} 
\vspace{-16.4cm}
\end{table}

Table 6 provides insight into the optimization procedure that was employed
to select the solution in Table~5.  The time of perijove, $t_{\rm J}$,
correlates tightly with both the Jovicentric distance at the time,
$\Delta_{\rm J}(t_{\rm J})$, and with the pre-encounter orbit's original
semimajor axis, $(1/a)_{\rm orig}$ (referred to the barycenter of the
Solar System), or the equivalent original orbital period, $P_{\rm orig}$.
The interval of $t_{\rm J}$ in Table~6, covering 12~hours on November 7 of
the year $-$2251, corresponds to an interval of 0.001~AU in the scaling
distance $r_0$ and shows that the minimum perijove distance $\Delta_{\rm J}$
was attained certainly before 7:00.  Although we will estimate the exact time
below, it is outside the range of interest because the comet would then have
arrived along a strongly hyperbolic orbit, an unacceptable scenario.

Realistically, the comet should have passed through perijove after 8:15:11,
when its pre-encounter barycentric orbit was elliptical.  If C/1995~O1 was
a dynamically new comet arriving from the ``core'' region of the Oort Cloud
--- for which we adopt with Marsden et al.\ (1978) a heliocentric distance
of $\sim$43\,000~AU) --- it would have arrived at perijove just 10~seconds
later, at 8:15:21.  The comet's original orbital period would have then
amounted to some 3\,200\,000~yr, its Jovicentric distance at perijove
would have been near 10.8~$R_{\rm J}$, and its Jovicentric velocity about
29~km~s$^{-1}$, as the nominal solution in Table~5 indicates. 

Of course, C/1995 O1 may have been a long-period comet before the encounter,
but not dynamically new.  If so, it would have arrived later still, perhaps
as late as 8:30 or even 9:00.  Its perijove distance would then have been
only slightly greater. For two related reasons it is, however, less likely
that the comet had moved in an orbit of a relatively short period before the
encounter:\ (i)~it would have had a high probability of approaching Jupiter
on a number of occasions at times before this encounter and (ii)~having made
many revolutions about the Sun with a very small perihelion distance, it
would long ago have gotten depleted of highly volatile species, in conflict
with the observations (Section~9).

\begin{figure}[t] 
\vspace{-9.83cm}
\hspace{2.39cm}
\centerline{
\scalebox{0.76}{
\includegraphics{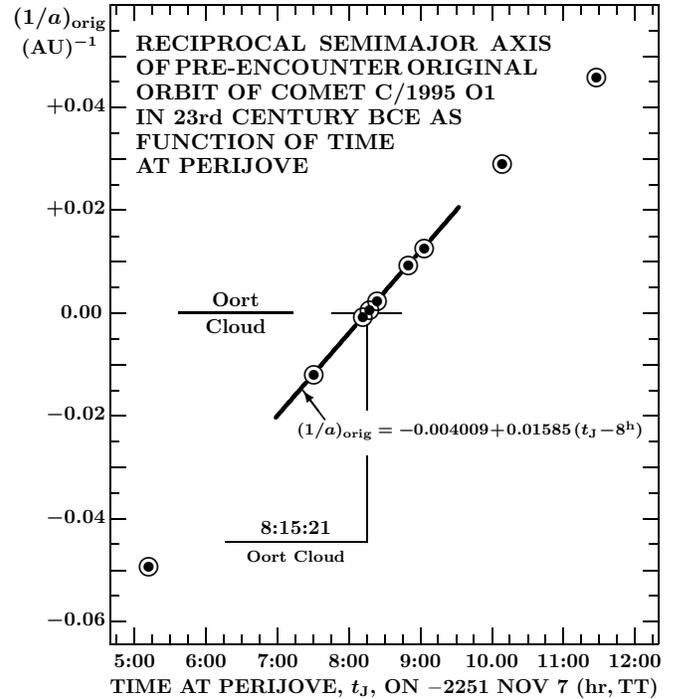}}} 
\vspace{-3.63cm}
\caption{Plot of the reciprocal semimajor axis of the comet's pre-encounter
original barycentric orbit, $(1/a)_{\rm orig}$, against the time at perijove,
$t_{\rm J}$, on $-$2251 November 7.  To prevent overcrowding, only some of the
14 entries from Table~6 are displayed.  They~all are distributed along a curve
that over intervals shorter than 2~hr can with high accuracy be approximated
with a straight line, as shown.  An arrival from the Oort Cloud is depicted
with a horizontal line at \mbox{$(1/a)_{\rm orig} = +0.000046$\,(AU)$^{-1}$}
and described with a corresponding perijove time of 8:15:21 TT.{\vspace{0.5cm}}}
\end{figure}

The steep dependence of the original barycentric semimajor axis, $(1/a)_{\rm
orig}$, on the time of perijove, $t_{\rm J}$, during the day of $-$2251
November 7 is illustrated in Figure~3.  Although the relationship is more
complex over wider intervals of time, a straight-line approximation is
satisfactory over intervals shorter in $t_{\rm J}$ than about
2~hr.~A~least-squares fit to the data points between $\sim$7:30 and
$\sim$9:00, which was used above to pinpoint the time of perijove for
C/1995~O1 as an Oort Cloud comet, can be expressed as follows:
\begin{eqnarray}
(1/a)_{\rm orig} & = & -0.004009 + 0.015850\,(t_{\rm J}\!-\!8^{\rm h}),
 \nonumber \\[-0.05cm]
                 &   & \pm 0.000011 \, \pm \! 0.000024
\end{eqnarray}
where $(1/a)_{\rm orig}$ is in (AU)$^{-1}$ and $t_{\rm J}$ in hr.

Similarly, the relationship between the time of perijove and the perijove
distance, $\Delta_{\rm J}$, is over the entire interval of time from 5:10 to
11:30 approximated by a least-squares polynomial
\begin{eqnarray}
\Delta_{\rm J} & \!\!\:=\!\!\: & 10.751826 + 0.1191931(t_{\rm J}\!-\!8^{\rm h})
 + 0.024733(t_{\rm J}\!-\!8^{\rm h})^2 \!, \nonumber \\[-0.05cm]
 & & \!\!\pm 0.000006 \,\pm\! 0.0000002 {\hspace{1cm}} \;\;\;\pm\! 0.000001
\end{eqnarray}
where $\Delta_{\rm J}$ is expressed in units of the Jovian equatorial radius
$R_{\rm J}$ (1\,$R_{\rm J} = 71\,492$ km) and $t_{\rm J}$ is again~in~hr.  A search
for a minimum provides \mbox{($\Delta_{\rm J})_{\rm min} = 10.608 \, R_{\rm J}$}
at \mbox{$(t_{\rm J})_{\rm min} = 5$:35:25} or 2:39:56 before the perijove
time for C/1995~O1 as an Oort Cloud comet.

The orbit in Table 5, referred to the osculation epoch of 1997 Mar 13, is
generally in accord with the published gravitational orbits.  For example, it
agrees with Marsden's (2007) orbit to better than 90 seconds in the peri\-helion
time, to better than 4$^{\prime\prime}$ in the angular elements, to better than
1000~km in the perihelion distance, and to better than 20~yr in the osculating
orbital period.  However, with the exception of the arument of perihelion, our
formal errors are smaller than the differences, in some of the elements by two
orders of magnitude.  The original values of the orbital period derived by us
and by Marsden deviate by 20~yr, the future period by 8~yr.  Our result predicts
for the previous perihelion time the year $-$2252, Marsden's orbit the year
$-$2272 (cf.\ Table~1).

\begin{figure*}[t] 
\vspace{-3.75cm}
\hspace{-0.15cm}
\centerline{
\scalebox{0.79}{
\includegraphics{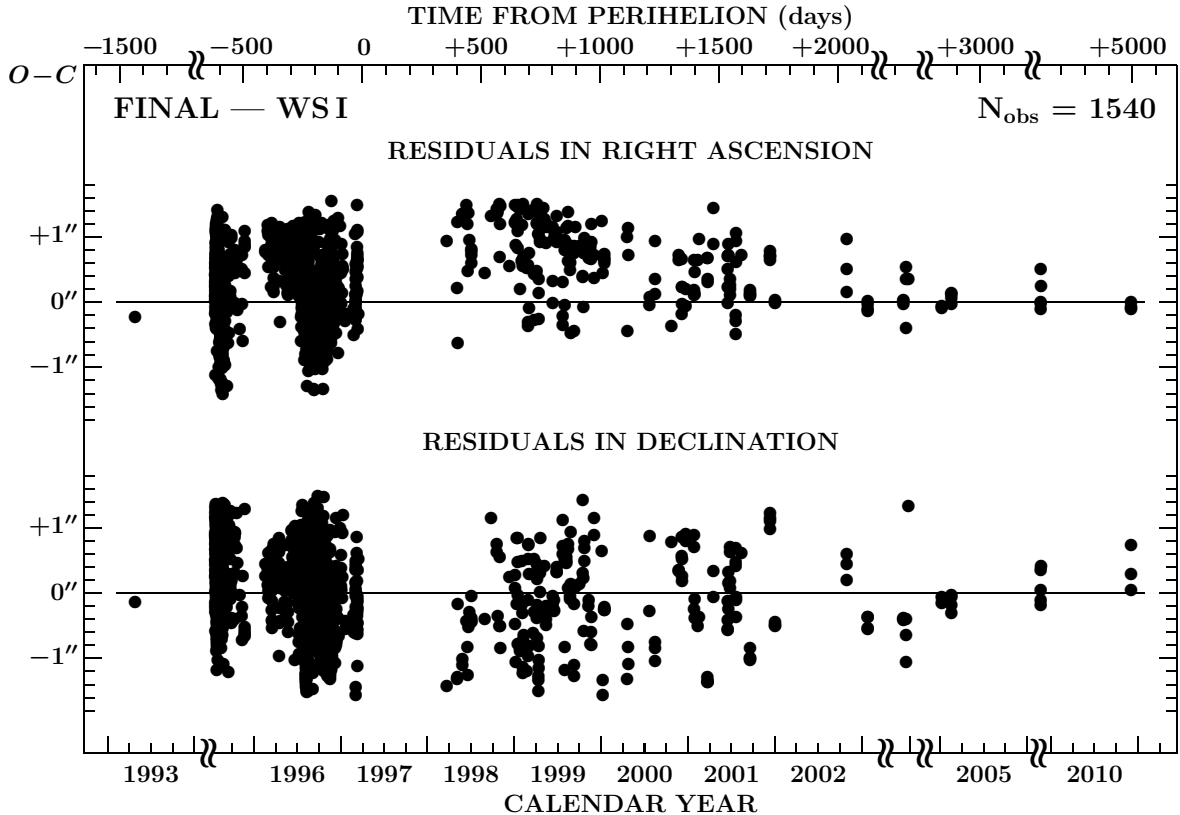}}} 
\vspace{-8.8cm}
\caption{Temporal distribution of residuals \mbox{$O \!-\! C$} (observed minus
computed) in right ascension (top) and declination (bottom) from the nominal
nongravitational solution listed in Table~5, based on 1540 accurate observations
of comet C/1995~O1 between 1993 and 2010, which left residuals not exceeding
1$^{\prime\prime}\!$.5 in either coordinate.  The fit is judged perfect in
declination, but many observations from 1998--2000 still leave mostly positive
residuals.{\vspace{0.5cm}}} 
\end{figure*}

A more recent gravitational orbit, by Williams (2011), differs from our nominal
solution by 45~seconds in the perihelion time, by 0$^{\prime\prime}\!$.1 in the
argument of perihelion, by 3$^{\prime\prime}\!$.6 in the longitude of the node,
by 3$^{\prime\prime}\!$.0 in the inclination, by 450~km in the perihelion
distance, and by 12~yr in the osculation orbital period.  It predicts that the
previous perihelion time occurred in the year $-$2264 (cf.\ Table~1).  In
general, Marsden's (2007) and Williams' (2011) orbits agree with each other
better than either of them with our orbit.  We suspect that this is so in part
because of our incorporation of the nongravitational terms in the equations of
motion, although the weighting of the critical observations may also have an
effect.  As already noted in Sections~4 and 6.1, the normal component of the
nongravitational acceleration --- which has hardly ever been employed in
orbital computations of comets --- not only improves the fit, but turns out
to be the most prominent component, more than twice the magnitude of the
radial component.  On the other hand, the transverse component is very small
and poorly defined.  The overall magnitude of the nongravitational
acceleration at 1~AU from the Sun is 0.69\,$\pm$\,0.02~AU~day$^{-2}$, or
about one half the acceleration that resulted from the standard
nongravitational solutions (Table~2).  This is noteworthy because
the magnitude of the nongravitational acceleration derived earlier was
judged to be too high to be compatible with the presumably large nucleus
of the comet (Marsden 1999, Sosa \& Fern\'andez 2011).

The scaling distance $r_0$ of the nongravitational law in the nominal solution
in Table~5 is only slightly greater than the scaling distance for water ice in 
Equation~(1).  This result can be interpreted to mean that the nongravitational
effect was due largely to water ice, with only relatively minor contributions
from more volatile species.  We return to this issue in Section~9.

Somewhat surprising is the very small perihelion distance, of only about
0.15~AU, of the pre-encounter orbit.  Two other prominent dynamically new
comets approached the Sun to a similar distance, C/1973~E1 (Kohoutek) and
C/2006~P1 (McNaught), but in terms of the inclination the pre-encounter orbit
of C/1995~O1 is much more similar to the latter of the two comets.  And while
the orbital plane's nodal line and inclination were affected very little by
the Jovian encounter (by 1$^\prime$ and~a little more than 3$^\circ$,
respectively), the line of apsides moved by 69$^\circ$ ({\it sic!\/}), the
nodes got interchanged, and the perihelion distance increased by a factor of
six.  C/1995~O1 passed through perihelion two months later than it would have
in its pre-encounter orbit.


\begin{table*}
\vspace{-3.85cm}
\hspace{-0.52cm}
\centerline{
\scalebox{1}{
\includegraphics{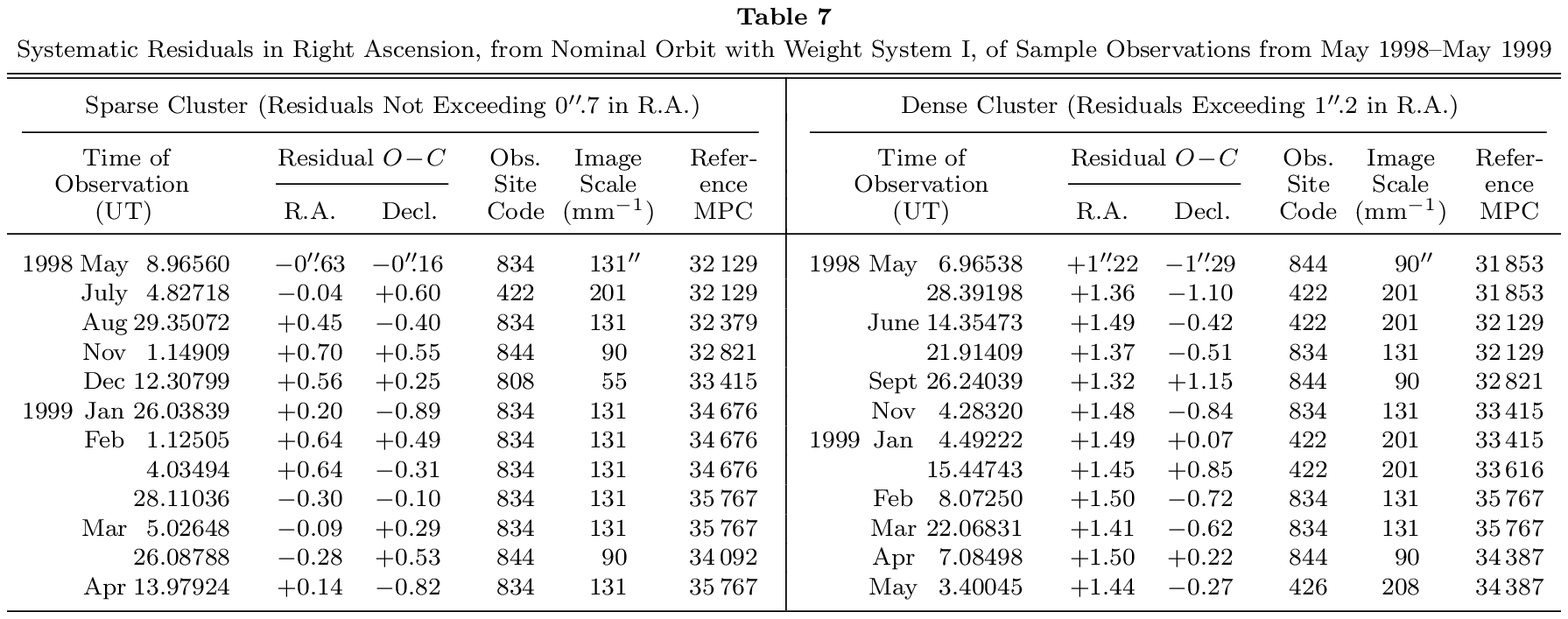}}} 
\vspace{-17.82cm}
\end{table*}

The distribution of residuals from the nominal solution in Table~5 left by
the 1540 retained observations~is~dis\-played in Figure~4.  The distribution
is satisfactory in declination, but there is a significant excess of positive
residuals in right ascension in the course of 1998 and 1999.  This is a tail
of the set of much more strongly positive residuals left by the observations
made between mid-March 1997 and early March 1998, all of which had to be
discarded from the nominal solution.  There are two clusters of residuals in
R.A.\ in Figure 4 in the 1998--1999 time slot --- an upper dense one and a
lower sparse one --- separated by a tilted strait with no data points.~The
lower, lightly populated cluster contains observations that are more
consistent with the solution in Table~5.

To investigate this peculiar double-cluster distribution in greater detail, we
present in Table~7 a dozen observations from either cluster, made in the period
of time from early May 1998 through early May 1999.  Each observation with the
residuals is identified by the observing site code and the reference to a
{\it Minor Planet Circular\/} (MPC), in which it was published.  From the list
of observing sites that always precedes the list of reported observations in
the MPCs, we extracted the information on each instrument's aperture and the
{\it f\/} ratio and converted it into the image scale presented in column~5 of
Table~7.  The aim was to check whether there was any systematic difference
between the image scales (CCD pixel sizes were unfortunately unavailable) of
the instruments employed to make the seemingly more accurate observations on
the left and the image scales of the instruments used to make the observations
leaving systematic residuals in R.A.\ on the right.  As seen from the table,
it turned out that the same instrument sometimes provided acceptable
residuals, while at other times it did not.  The average image scale for the
observations on the left is better than that on the right, but only by a small
margin of 124$^{\prime\prime}$\,mm$^{-1}$ vs 151$^{\prime\prime}$\,mm$^{-1}$.
This result means that the problem with the residuals in R.A.\ either lies
elsewhere or the cutoff of $\pm$1$^{\prime\prime}\!$.5 is tighter than the
level of accuracy achievable by the category of telescopes (with apertures
of, generally, 25--50~cm) used by most observers at the times the comet was
bright enough to be within their reach.  We note that on the one hand, the
systematic trend in R.A. is limited to the time interval of high activity
(associated with a complex coma morphology), but it fails to correlate with
the comet's orientation relative to the Sun, so that the offsets cannot
readily be attributed to sunward or antisunward condensations.

\subsection{Introduction of New Weight System}
The distribution of residuals might be sensitive to the weight assigned to the
pre-discovery observation of 1993.  It is recalled that it indeed was this
observation's residual in R.A.\ that caused problems when attempts were made
in 1998 to link it with the post-discovery observations (Marsden 1999).  The
weight system applied in Section 6.2, which assigned the 1993 observation
weight 40, resulted in an excellent fit to this data point by the nominal
solution in Table~5, leaving a residual of $-$0$^{\prime\prime}\!$.23 in
R.A.\ and $-$0$^{\prime\prime}\!$.14 in decl.\ (Figure~4).   Given the
position's uncertainty ``within 1~arcsec'' (Marsden 1999) --- the 124~cm UK
Schmidt telescope, with which the observation was made, has an image scale
of $\sim$67$^{\prime\prime}$\,mm$^{-1}$ --- one should allow these residuals
to become quite a bit greater and still consistent with the stated astrometric
error.  The implication is that the weight assigned to the 1993 observation
was unnecessarily high and that a lower weight might be more appropriate.

The weight assignment to the critical observations, not only to the pre-discovery
observation, also has broader ramifications in terms of the orbital results and
predictions of the comet's motion in the previous return to the Sun.  It is
therefore desirable to address this issue in greater detail.  To examine the
impact, we chose a set of different weights, called hereafter Weight System II,
as follows:\ the observation of 1993 was assigned weight 20 (instead of 40), the
2010 observations weight 15 (instead of 20), while the weights of the 2005 and
2007 observations were left unchanged at 5 and 10, respectively.

We then proceeded in a way similar to that in Section~6.1, with the nominal
solution in Table~5 used as a starting set of elements in the iterative
procedure of stabilizing the number of retained observations.  However, unlike
before (Table~3), at each step of the process we optimized the scaling
distance $r_0$ of the modified nongravitational law to fit the encounter time.
Accordingly, we were successively deriving the solutions MN~XXI~to~MN~XXVII
listed in Table~8.  

\begin{table*} 
\vspace{-4.15cm}
\hspace{-0.52cm}
\centerline{
\scalebox{1}{
\includegraphics{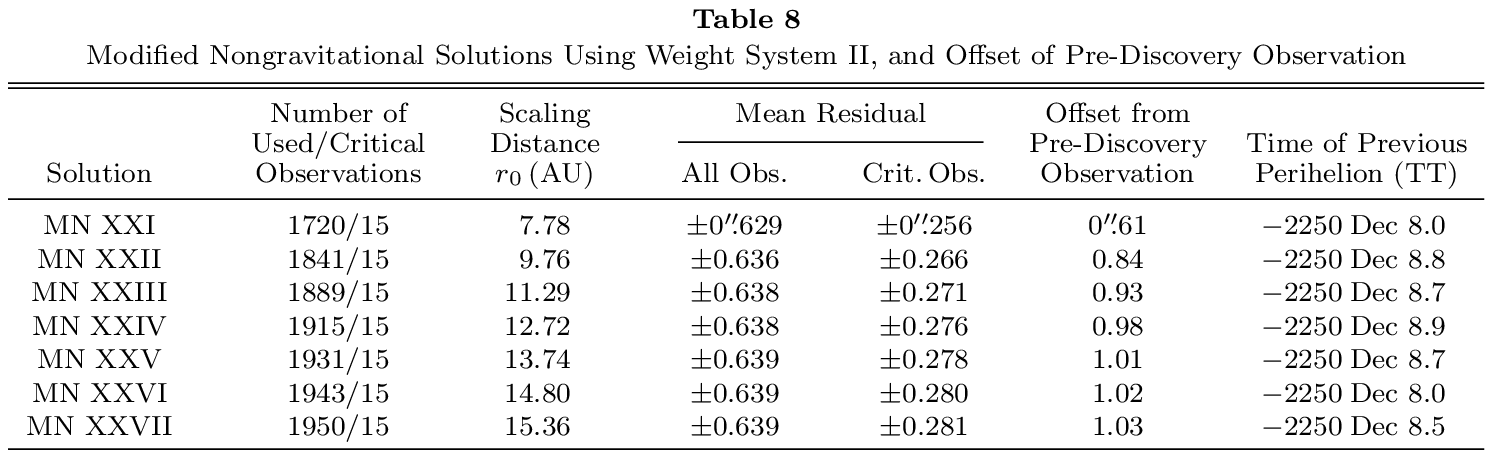}}} 
\vspace{-20.3cm}
\end{table*}

The results, which again allow a close approach only in the year $-$2251 but
not in $-$2262 or $-$2239, are rather striking.  The first iteration with
Weight System II immediately indicated that 180 more observations, whose
residuals exceeded the cutoff of $\pm$1$^{\prime\prime}\!$.5 before, would
now fit.  Simultaneously the scaling distance increased dramatically by
almost 5~AU, while the offset of the 1993 pre-discovery observation grew to
about twice the offset of the nominal solution in Table~5 (Figure~4), still
well within the reported uncertainty.  The first solution was far from
stabilizing the number of retained observations, and the trend continued:\
121 additional observations were accommodated by the second iteration, which
required a scaling distance greater by a yet another 2~AU, while the offset of
the pre-discovery observation now increased to more than 0$^{\prime\prime}\!$.8
and was slowly approaching the limit of the measured error.  After five more
iterations, the number of retained observations finally stabilized at 1950, an
increase by more than 500 relative to the solution in Table~5.  Interestingly,
in spite of these additional accommodated data points the mean residual dropped
slightly in comparison with the nominal solution in Table~5.  However, the mean
residual of the 15 critical observations increased a little, unquestinably owing
to the greatly increased offset of the pre-discovery observation.  This offset
was now at the limit of uncertainty, with the residuals, in the sense ``observed
minus computed'', amounting to $-$0$^{\prime\prime}\!$.86 in right ascension and
$-$0$^{\prime\prime}\!$.57 in declination.

\begin{table}[b] 
\vspace{-3.85cm}
\hspace{4.23cm}
\centerline{
\scalebox{1}{
\includegraphics{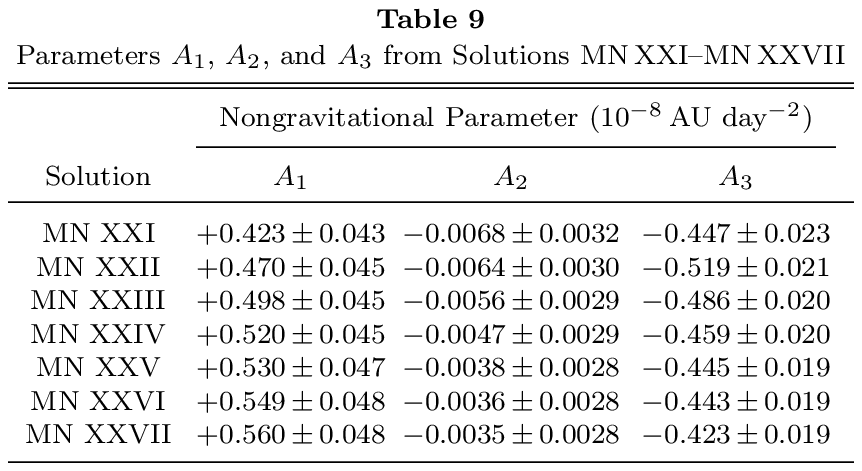}}} 
\vspace{-21.05cm}
\end{table}

The nongravitaional parameters derived from the solutions MN~XXI through
MN~XXVII likewise differ significantly from those for the nominal orbit in
Table~5, as illustrated by comparing Table~9 with Table~4.  While the total
magnitude is approximately the same, it is now the radial component of the
nongravitational acceleration whose magnitude dominates that of the normal
component, while the magnitude of the transverse component is again
insignificant and poorly defined.

In summary, there is evidence that the strong systematic trends in the
residuals in R.A., which necessitated rejection of all observations made
between mid-March 1997 and early March 1998 and showed rather prominently
in the residuals of the retained observations throughout May 1999, were
caused by an excess weight assigned to the 1993 pre-discovery observation.
When the weight was reduced by a factor of two, the orbital solution
accommodated more than 400 observations that had to be rejected before,
increased the scaling distance of the modified nongravitational law by a
factor of five, and decreased the magnitude of the normal component of the
nongravitational acceleration, so that the radial component now dominates;
all this at the expense of a fit to the pre-discovery position, with the
total residual increasing from less than 0$^{\prime\prime}\!$.3 to
$\sim$1$^{\prime\prime}$, the estimated uncertainty of the astrometric
measurement.

\subsection{Weight System II:\ Final Optimization of Orbit}
To determine the comet's nominal orbit and its arrival time from the Oort
Cloud under the constraints of Weight System~II, we fine-tuned the solution
MN~XXVII in Table~8.  The results are listed in detail as a nominal solution
in Table~10, whose format is identical with that of Table~5 to allow direct
comparison with the nominal solution with Weight System I.  The agreement
between the two sets of orbital elements is excellent, but there are major
differences in the nongravitational parameters that will be addressed in
Section~9.  As for the encounter parameters, the time of perijove is now
predicted 24 minutes earlier and the Jovicentric distance at perijove is
merely 0.05 Jupiter's equatorial radius, or 3600~km, smaller.  An earlier
time for the encounter is consistent with a greater scaling distance and a
higher value of $A_1$, both of which mean that under Weight System~II the
orbital motion was subjected to higher integrated nongravitational effects,
so the temporal gap between the perijove in the year $-$2251 and the
perihelion in 1997 must be slightly longer.

\begin{table}[b]
\vspace{-4.0cm}
\hspace{4.3cm}
\centerline{
\scalebox{1}{
\includegraphics{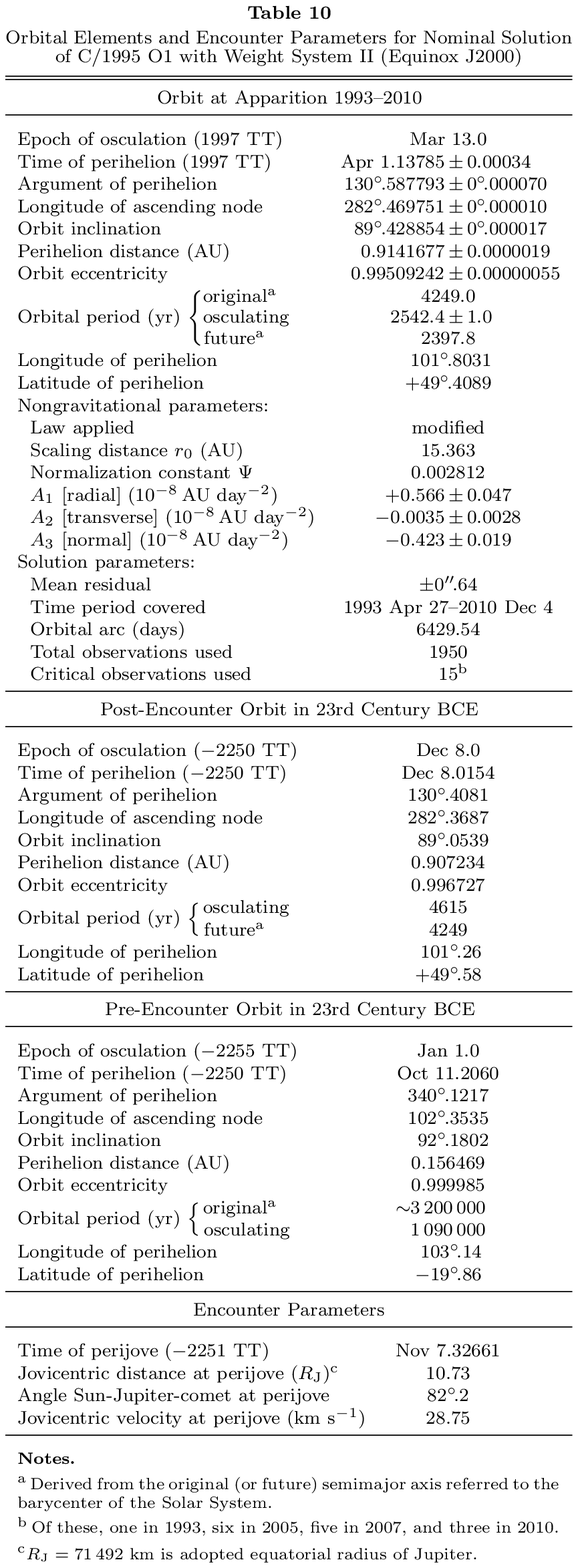}}} 
\vspace{-2.43cm}
\end{table}

The expressions for the reciprocal semimajor axis of the pre-encounter,
original barycentric orbit, $(1/a)_{\rm orig}$ (or the orbital period
$P_{\rm orig}$), and for the Jovicentric distance at perijove,
$\Delta_{\rm J}$, are similar to Equations~(4) and (5), respectively.
A dozen solutions, spanning more than 10~hr in the perijove time
$t_{\rm J}$ and listed in Table~11, illustrate the relationships.
A least-squares fit to the data points covering less than one hour of
the relevant range of $t_{\rm J}$ provides the following relation:
\begin{eqnarray}
(1/a)_{\rm orig} & = & +0.002658 + 0.016180\,(t_{\rm J}\!-\!8^{\rm h}),
 \nonumber \\[-0.05cm]
                 &   & \pm 0.000003  \, \pm \!0.000012
\end{eqnarray}
where $(1/a)_{\rm orig}$ and $t_{\rm J}$ are again in (AU)$^{-1}$ and hr,
respectively.  We do not show its plot, because it looks
very much like Figure~3:\ the slope is only 2\% steeper and the fitted
straight line is shifted, as already pointed out, to an earlier time by
25 minutes.

From a wider interval of $t_{\rm J}$, spanning about 7.5~hr, the perijove
distance (in units of Jupiter's equatorial radius) is given by an expression:
\begin{eqnarray}
\Delta_{\rm J} & = & 10.74861 + 0.139837\,(t_{\rm J}\!-\!8^{\rm h})
 + 0.024681\, (t_{\rm J}\!-\!8^{\rm h})^2 \!. \nonumber \\[-0.05cm]
 & & \!\!\pm 0.00004 \, \pm \! 0.000021 {\hspace{1.4cm}} \!\pm\! 0.000006
\end{eqnarray}
The minimum perijove distance of 10.551 Jupiter's equatorial radii took
place at 5:10:02, or 160~minutes before the perijove time of C/1995~O1 as
an Oort Cloud comet (Table~10).

\begin{table}[b] 
\vspace{-4.0cm}
\hspace{4.23cm}
\centerline{
\scalebox{1}{
\includegraphics{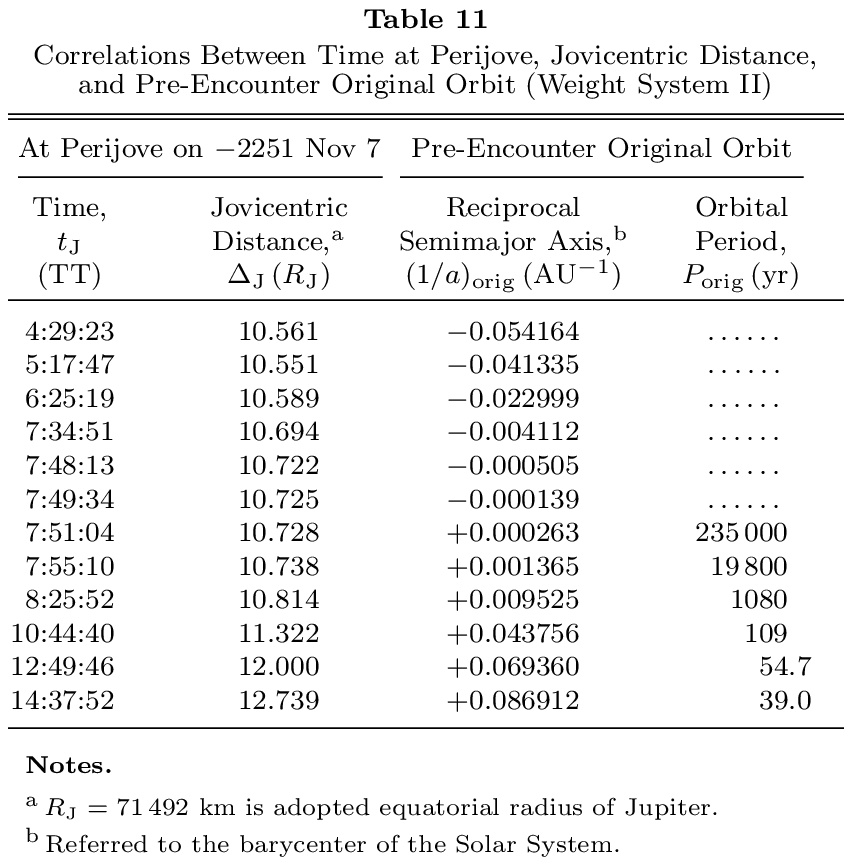}}} 
\vspace{-17.1cm}
\end{table}

The distribution of residuals from the nominal orbit in Table 10, left by the
1950 observations, is presented in Figure~5.  It depicts the $\sim$400
reinstated data points between mid-March 1997 and early March 1998 that
replace the prominent gap in Figure~4.  A residual gap of a few months
centered on mid-June 1997 has to do with the comet's persistently small
elongations from the Sun (with a minimum of 21$^\circ\!$.5 on June 10), which
was responsible for no astrometric data (with one exception) between May~12
and July~14 and made the observations over a few adhering months difficult,
less frequent, and of lower quality.  The tail of systematically positive
residuals (in excess of 1$^{\prime\prime}$) in R.A.\ in 1998--1999 has
likewise disappeared.  The strong systematic effect in R.A.\ in Figure~4 may
be slightly overcorrected in Figure~5,~but there is no truly worrisome effect
of the kind.  A greatly improved fit in R.A.\ to the observations from the
denser 1997--1998 cluster, which is apparent by comparing Table~12 with
Table~7, corroborates a superior match to the observations by the solution
from Table~10.  In declination, the distribution of residuals is considered
rather satisfactory, with no prominent long-term systematic trends.

\begin{figure*}[t] 
\vspace{-3.85cm}
\hspace{-0.15cm}
\centerline{
\scalebox{0.79}{
\includegraphics{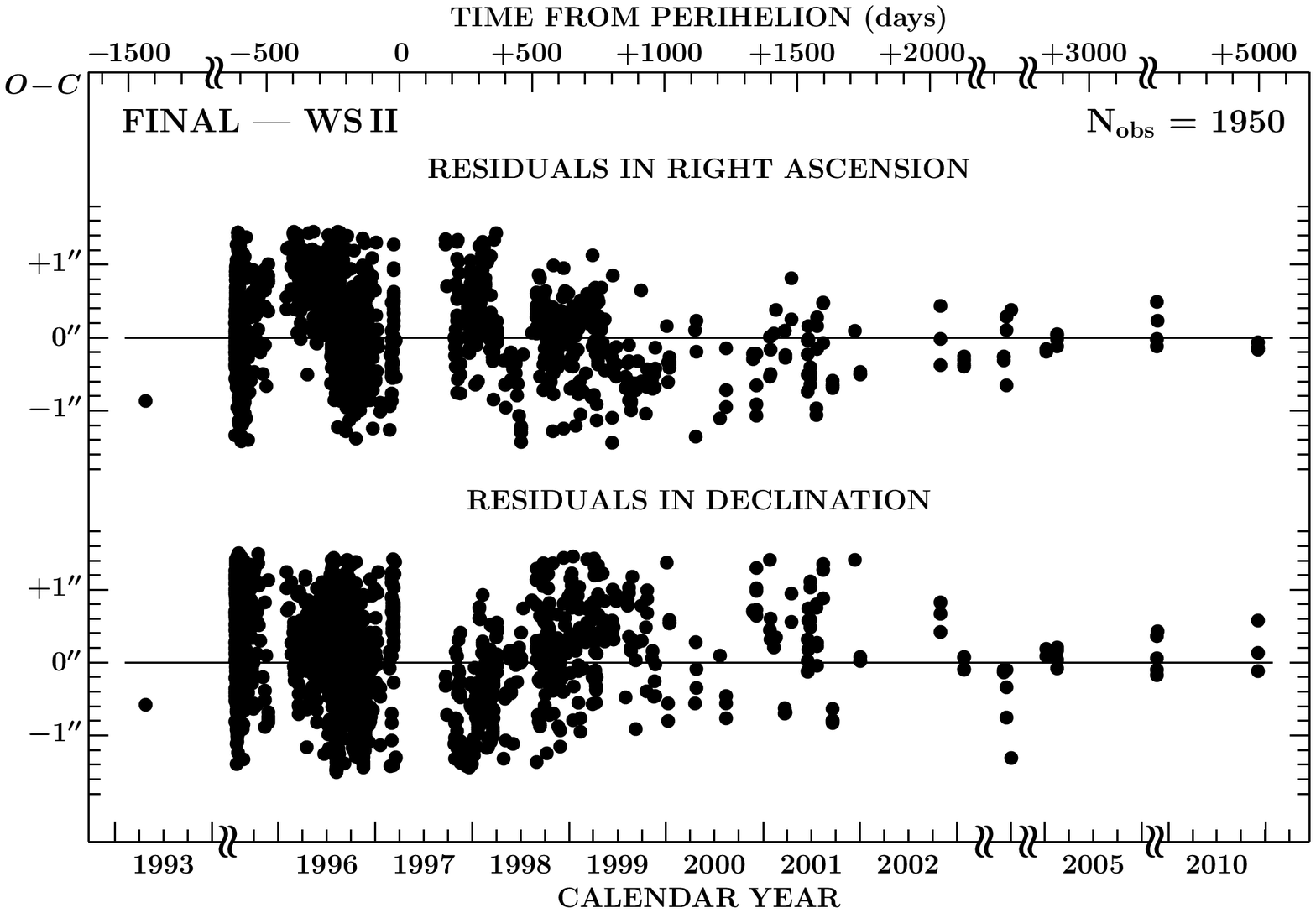}}} 
\vspace{-8.9cm}
\caption{Temporal distribution of residuals \mbox{$O\!-\!C$} (observed minus
computed) in right ascension (top) and declination (bottom) from the nominal
nongravitational solution listed in Table~10, based on 1950 accurate
observations of comet C/1995~O1 between 1993 and 2010, which left residuals
not exceeding 1$^{\prime\prime}\!$.5 in either coordinate.  The fit is overall
judged to be quite satisfactory, implying for the 1993 pre-discovery
observation residuals that reach the limit of its estimated
uncertainty.{\vspace{0.4cm}}}
%
\vspace{-4.1cm}
\hspace{-0.52cm}
\centerline{
\scalebox{1}{
\includegraphics{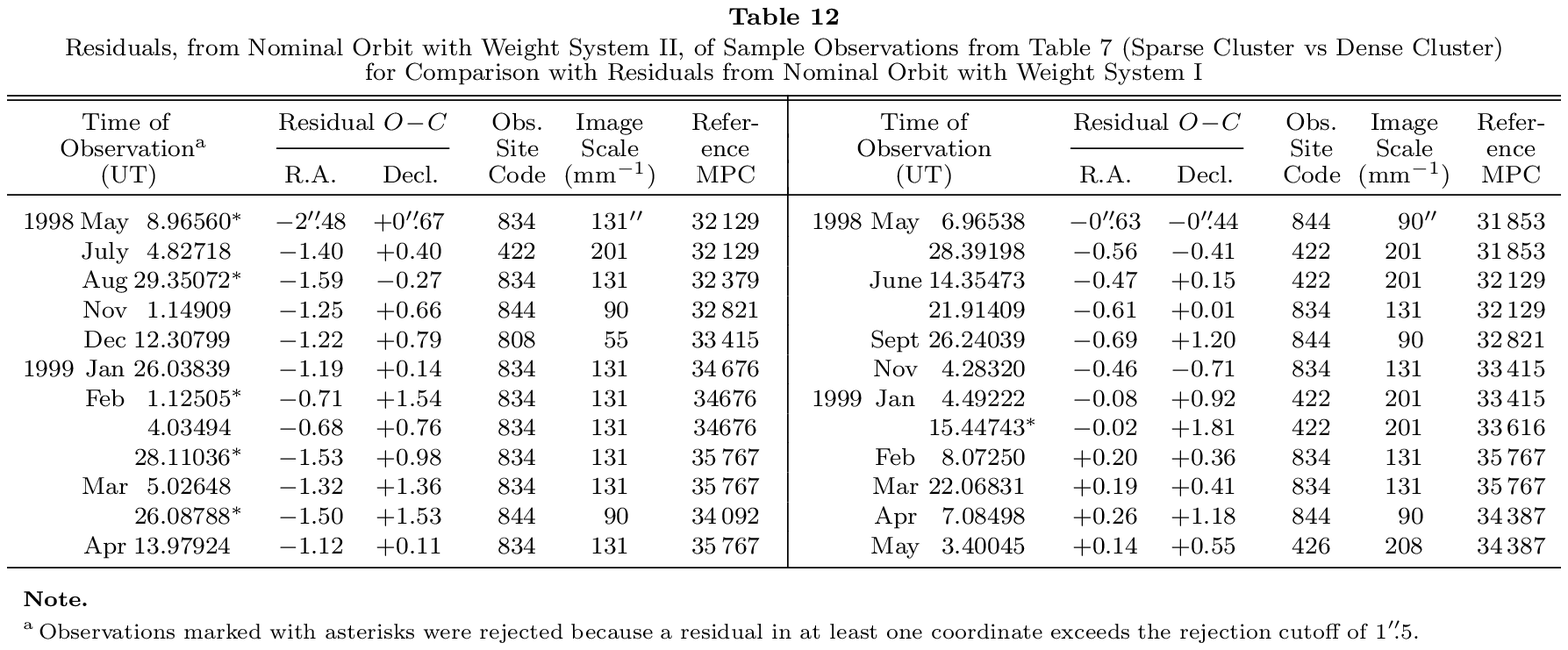}}} 
\vspace{-17.96cm} 
\end{figure*}

Concluding this section, we note that the nominal orbit with Weight System~II
offers --- in terms of orbital quality --- a solution superior to the nominal
orbit with Weight System I:\ it accommodates 410~more observations to within
1$^{\prime\prime\!}$.5, while still fitting the 1993 pre-discovery observation
to $\sim$1$^{\prime\prime}$.  Both scenarios (Tables~5 and 10) allow for an
approach of C/1995~O1 to less than 11 equatorial radii of Jupiter on $-$2251
November 7, 396 days before perihelion, and, if the comet originated in the
Oort Cloud, for its arrival at perijove at times that agree to within one half
hour.  Moreover, the perijove in both scenarios depends only weakly on the
pre-encounter orbit; with Weight System~II, for the original orbital periods
of 100\,000~yr, 10\,000~yr, and 1000~yr, the time of perijove would shift
forward by merely 1.6, 7.8, and 36.9~minutes and the perijove distance would
increase by 0.003, 0.018, and 0.090~Jupiter's equatorial radii, respectively.

As illustrated in Figure~6, the comet was approaching Jupiter from above the
ecliptic plane and was diverted by the planet's gravity back above the
ecliptic plane in a trajectory that was strongly hyperbolic relative to
Jupiter, with an eccentricity of 4.0.  The comet was launched toward the
Sun along a new, very different orbit, whose plane though agreed with the
original one to 3$^\circ$.  At the time of encounter the comet's position
was at close proximity of the descending node of the original orbit
and the ascending node of the new orbit.  After reaching perihelion about
two months later and at a distance from the Sun six times greater than it
would have in the absence of the encounter, the comet began to recede from
the Sun, once again approaching the ecliptical plane.  Some 35~days after
perihelion it crossed the descending node (now on the other side of the
Sun) for the second time and continued its journey to aphelion, diving
almost perpendicularly to the ecliptic.

\begin{figure*}[ht] 
\vspace{-12.75cm}
\hspace{0.44cm}
\centerline{
\scalebox{0.84}{
\includegraphics{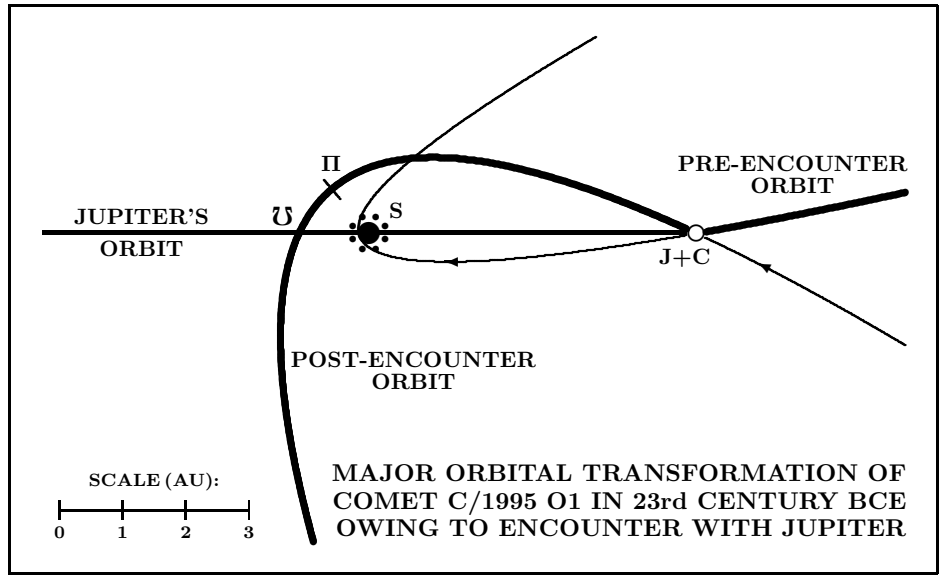}}} 
\vspace{-4.7cm}
\caption{Major orbital transformation of C/1995 O1 on $-$2251 November 7, at
the time of its close encounter with Jupiter, in projection onto the comet's
orbital plane, which was nearly exactly perpendicular to Jupiter's orbital
plane.  The comet's actual path is depicted~by~the thick curve, the
extrapolated portions of the pre-encounter and post-encounter orbits by
the thin curve, with the scale and direction of motion indicated.  The
Sun is marked by S, Jupiter and the comet at the time of encounter by J+C,
the comet's perihelion point by $\Pi$, and its second descending node, on
the post-encounter orbit, by $\mho$.  The first descending node, on the
pre-encounter orbit, and the ascending node on the post-encounter orbit
coincide on the scale of the plot with the comet's and Jupiter's position
at the time of encounter.{\vspace{0.4cm}}}
\end{figure*}

\section{Predicted Appearance of Comet C/1995 O1\\in the 23rd Century BCE and
Search\\for Possible Historical Records}
Searching for historical records would be pointless, if the comet's appearance
was affected by unfavorable observing geometry.  While the viewing conditions
for a comet moving in an orbit perpendicular to the plane of the ecliptic
cannot remain inferior for long, we are primarily concerned with a short
orbital arc near perihelion, when the comet is intrinsically the brightest.
Given the perihelion distance near 0.9~AU and the argument of perihelion at
130$^\circ$, the comet would at perihelion project to the terrestrial observer
at an angular distance of less than 24$^\circ$ from the Sun (at a geocentric
distance of more than 1.7~AU), if the Earth should then be crossing the
comet's orbital plane on the side of the ascending node.  On the other hand,
at the descending node, at a heliocentric distance of 1.1~AU, the comet would
be, under the most favorable conditions, at opposition with the Sun~and only
0.1~AU from the Earth.

An ephemeris, based on the post-encounter orbital elements from the 23rd
century BCE in Table~10, shows that the viewing circumstances in this
configuration were rather favorable.  In Figure~7 we present the comet's
predicted path in the sky over a period of 4~months, from $-$2250 October
23 (46~days before perihelion, position 1) through $-$2249 February 20
(74~days after perihelion, position 13).  This time interval was dictated
by the comet's predicted brightness.  Assuming, conservatively, that in the
23rd century BCE the comet was as bright intrinsically as in the 1997
apparition, we calculated its apparent brightness from the light-curve fits
published by Kidger (1999).  Over the orbital arc in Figure~7 the comet was
predicted to have remained brighter than apparent magnitude +2, reaching
a peak magnitude of $-$1.9, rivaling Jupiter, near position 7, about two
weeks past perihelion and shortly after entering the constellation Andromeda.
As seen with the naked eye near perihelion, the coma was predicted to have
amounted to 32$^\prime$ in diameter, practically the angular size of the
Moon.\footnote{This prediction is based on 433 naked-eye estimates of the
coma diameter of C/1995~O1 made in 1997 between 10 days before and 10 days
after perihelion, and reported to the {\it International Comet Quarterly\/}
(Green 1997, 1998, 1999).  The data average, a diameter of 1\,150\,000~km,
was equivalent to 19$^\prime\!$.6 at the 1997 perihelion.}

The predicted viewing geometry is described in a self-explanatory Table 13.
A minimum geocentric distance of 0.644~AU was found to have taken place on
$-$2250 December 23, while the elongation reached a minimum of 41$^\circ\!$.2
on $-$2250 October 16 and a maximum of 71$^\circ\!$.8 on $-$2249 January 9.
The phase angle reached a peak of 74$^\circ\!$.5 on $-$2250 December 19, so
no forward-scattering effect could be expected.  The comet was predicted to
have stayed brighter than magnitude 0 for 67~days and brighter than magnitude
$-$1 for 40 days.

We conclude that in this scenario the comet must have been a spectacular
object quite favorably placed~in the sky to the northern-hemisphere observer
for a long enough period of time that it could not have been overlooked.
Nonetheless, the probability that the comet's apparition was recorded is
low, because historical records of comets from this very distant past are
extremely fragmentary.  Even if it was, we probably would not be able to
recognize it because of uncertainties in dating the celestial phenomena at
these times.

\begin{figure*}[t] 
\vspace{-5.45cm} 
\hspace{-0.77cm}
\centerline{
\scalebox{1.0}{ 
\includegraphics{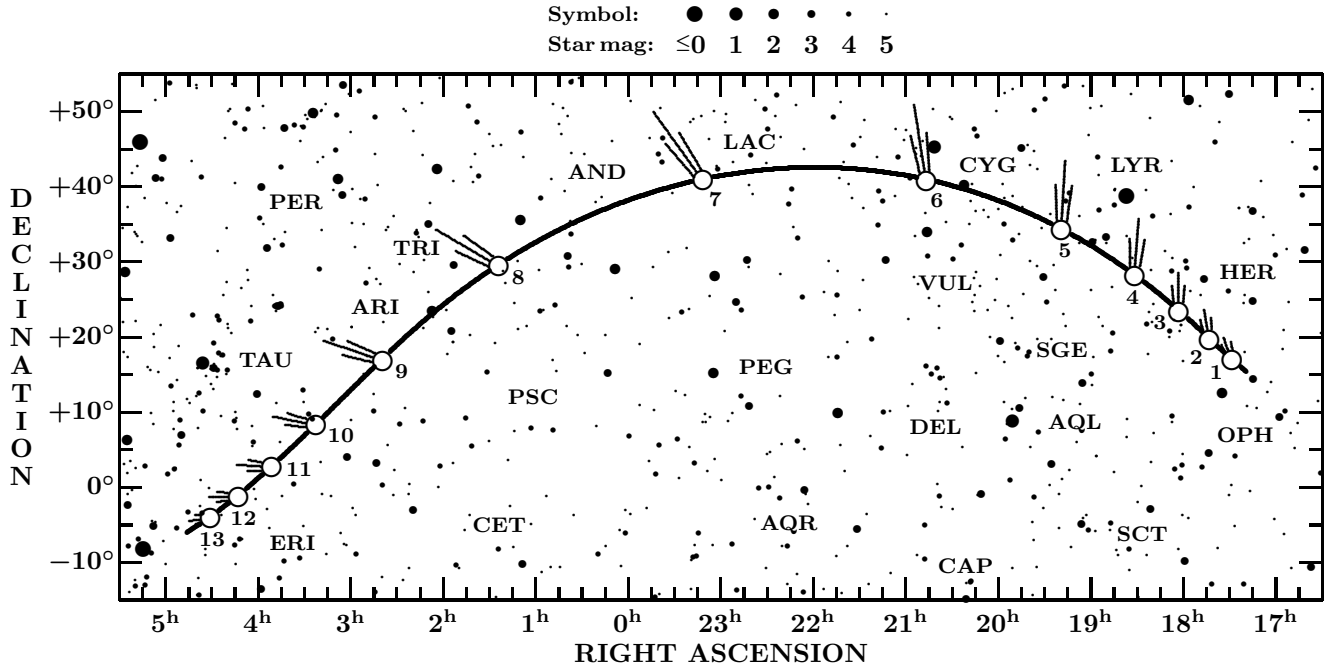}}} 
\vspace{-15.35cm} 
\caption{Predicted path of comet C/1995 O1 over the sky during its arrival from
the Oort Cloud in the year $-$2250.  The 13 positions are for times between
$-$2250 October 23 and $-$2249 February 20, spaced at 10 days intervals.  The
numbers 1--13 refer to the emtries in Table~13.  The tail's orientation shows
the antisolar direction, while its length measures approximately the comet's
apparent brightness, which varied from +1.8 at position 1 to $-$1.9 at position
7, back to +2.0 at position 13..  The equinox is J2000.{\vspace{0.5cm}}}
\end{figure*}

It is accepted that reliably determined times of events in Chinese historical
records extend to only about $-$840.  Information from earlier times is vague
and sporadic, because there are no contemporaneous sources; all known events
from the early period were compiled from sources dated after $-$770 (such as
{\it Ch\={u}nqi\={u}\/} or {\it Ch'un-ch'iu\/}, the {\it Spring and Autumn
Annals\/}; e.g., Liu et al.\ 2003).

Equally or more uncertain are the time assignments for events in other
ancient civilizations.  In partcular, the currently accepted chronology of
the Egyptian history, introduced less than two decades ago (Shaw 2000),
differs in the 3rd millennium BCE from a previous chronology (Breasted
1906) by almost 300~yr.  It thus comes as no surprise that most modern
catalogs of historical records of comets, summarizing information collected
from the ancient sources, do not reach the 3rd millennium BCE, and they
especially do not list the orbital elements.  Indeed, Williams (1871) begins
in the year $-$610, Galle (1894) in $-$371, Baldet \& de Obaldia (1952)
and Vsekhsvyatsky (1958) in $-$466 (presumably 1P/Halley), Porter (1961)
and Marsden \& Williams (2008) in $-$239 (undoubtedly 1P/Halley), Ho (1962)
in the 14th century BCE, and Kronk (1999) in $-$674.  Only Baldet's (1950)
{\it liste g\'en\'erale\/} and Pingr\'e's (1783) {\it com\`etographie\/}
(one of Baldet's sources) start, respectively, in the 24th and 23rd
centuries BCE.  Pingr\'e's, Williams', Baldet's, and Ho's lists and many
additional sources were incorporated by Hasegawa (1980) into his catalogue
of ancient and naked-eye comets, but with some notes and comments left out.

A reference to C/1995 O1 in $-$2250 could potentially be one of the first
six entries common to Baldet's (1950) and Hasegawa's (1980) lists:\ $-$2315,
$-$2296 (or $-$2284), $-$2287, $-$2254, $-$2241 (or $-$2269 or $-$2226/5),
and $-$2191 (or $-$2024).  The origin of the first five is Chinese, while the
last record is from Egypt.  The objects from the years $-$2315, $-$2287, and
$-$2254 are listed~by Baldet in reference to Stratton (1928), whose subject
was the records of novae.  However, our inspection of Stratton's paper fails
to confirm Baldet's objects of $-$2315~and $-$2287 in the constellation
Crater.  Stratton does indeed point at the encyclopedia {\it T'u shu chi
ch'-\^eng\/} as the source of the earliest record of a bright object in the
sky, but the year is identified as $-$2678, not $-$2315.  Furthermore, we
note Stratton's remark on the star of $-$2254 in Scorpius, but also his
reference to a yellow star in $-$2237, which Baldet omits. Stratton regards
both objects as likely novae.

\begin{table}[b] 
\vspace{-3.85cm}
\hspace{4.23cm}
\centerline{
\scalebox{1}{
\includegraphics{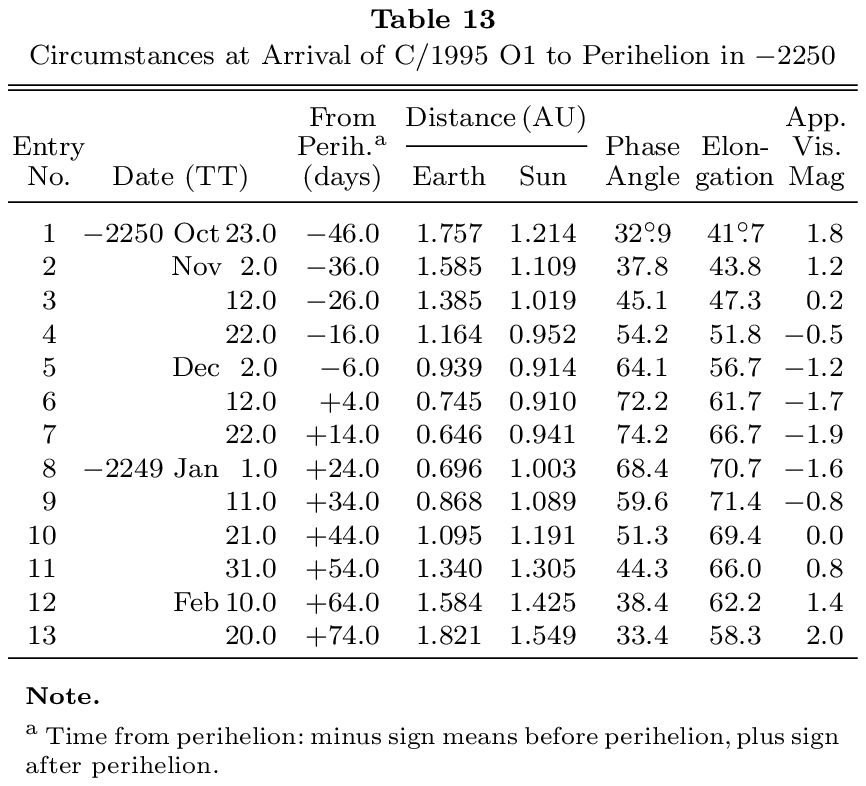}}} 
\vspace{-17.67cm}
\end{table}

The events of $-$2296 (etc.), $-$2241 (etc.), and $-$2191 (etc.) were copied
by both Baldet (1950) and Hasegawa (1980) from Pingr\'e (1783), who provides
a fair amount of detail.  He explains that the first of the three events was
related to the birth of Emperor Yu the Great, the first ruler of the Xia
Dynasty.  But Yu's historical existence has been a matter of controversy,
because the time of his presumed reign preceded the oldest known written
documents by about a millennium.  Doubts have also persisted  on the subject
of an elaborate irrigation system that Yu as the ruler is believed to have
overseen to control floods that for generations had inundated the vast
plains of the country along the Yellow River.  Pingr\'e (1783) adopted that
the reign of Yu began in $-$2223 and that he died at the age of 100 in
$-$2196 or $-$2184.  Then the star, which is said to have appeared as an omen
of good fortune during his mother's pregnancy, would have lit up in the sky
in $-$2296 or $-$2284.  The modern interpretation of the historical records
places the beginning of the Xia Dynasty to about $-$2070,\footnote{This date
for the beginning of the Xia Dynasty was fixed by the {\it Xia-Shang-Zhou
Chronology Project\/}, a multi-disciplinary project commisioned by the
People's Republic of China in 1996 to determine with accuracy the location
and time frame of the Xia, Shang, and Zhou Dynasties.  This Project provides
dates that are generally later than given by the numerous variations of the
traditional historiography, based mainly on the work by Sima Qian (died
$-$85), the legendary historian of the Han Dynasty; for more details,
see e.g.  {\scriptsize \tt
https://en.wikipedia.org/wiki/Xia\_Shang\_Zhou\_Chronology\_Project.}}
which would mean that the date listed by Pingr\'e is too early by some
150~yr.\footnote{To set a frame for dating records of comets, Pingr\'e often
refers to major historical events (e.g., the death of emperors), for whose
timing he usually relies on information from secondary sources, such as the
works of the French Jesuits A.\ Gaubil (1689-1759) or J.-A.-M.\ de Moyriac
de Mailla (1669-1748), or the Belgian Jesuit P.\ Couplet (1623-1693), all
of whom were missionaries to China.}  Moreover, Pang \& Yau (1996) argue that
a statement in the {\it Bamboo Annals\/} that during Yu's reign `` \ldots the
Sun disappeared by day and reappeared at night \ldots'' is a reference to
a ``double sunset'' eclipse of the Sun on $-$1911 September 24.  This timing
is in excellent agreement with the result from a very recent discovery by Wu
et al.\ (2016) of a catastrophic-flood event that was caused by breaching a
natural dam created by an earthquake-driven landslide, all of which has by
radiocarbon dating been pegged to around $-$1920, the time that is expected
to correlate with the beginning of Yu's reign.  It is thus likely that
Pingr\'e's time for the bright star associated with Yu's birth was at least
300~yr too early and could not refer to C/1995~O1.

The event of $-$2241 (etc.) happened during the reign of Emperor Shun, who was
Yu's predecessor.  Referring to Father Couplet (see footnote 6), Pingr\'e says
that a new star, equaling a half Moon, was seen in $-$2241, the 16th year of
Shun's reign.  However, based on independent sources for the date of Shun's
ascension to the throne, Pingr\'e also finds as possible the years $-$2269
or $-$2226/5.  Given the shift of Pingr\'e's time frame, the event probably
occurred in the 20th century BCE and, likewise, could not be considered a
reference to C/1995~O1.

There is hardly any point in commenting on the event of $-$2191 (or $-$2024),
not only because it is even more recent, but also because Pingr\'e expresses
so much skepticism about this case that one wonders why he bothered to include
it in his cometography in the first place.

The only remaining object that as yet has not been discredited, is Stratton's
(1928) yellow star of $-$2237.  There appears to be no additional information
available, but the time-frame contraction mentioned above is probably also
applicable in this case, even if Stratton should be incorrect in his belief
that this was a nova.  The inevitable conclusion is that any effort to
identify a historical record with C/1995~O1 is doomed to failure.

\section{Predicted Motion of C/1995 O1 Over Several Future Revolutions about
the Sun}
The constraint that C/1995 O1 had a close encounter with Jupiter and was a
dynamically new comet in the previous return to the Sun fixes the comet's
motion over more than one full revolution about the Sun.  With this wide
enough base as an initial condition, it is plausible to investigate the
comet's expected orbital evolution over a limited number of future returns
to perihelion.  This limit is determined by the rate of error propagation
in the orbital elements, which is in this case a function of the semimajor
axis of the comet's pre-encounter, original barycentric orbit.  By prescribing
an (unknown) error to the value of $(1/a)_{\rm orig}$, we can estimate the
propagated errors in the elements, which, as shown below, increase in general
with time (or with the number of returns) exponentially.  The most sensitive
element is by far the perihelion time, to which we pay particular attention.

\begin{figure}[b] 
\vspace{-9.2cm}
\hspace{2.62cm}
\centerline{
\scalebox{0.825}{
\includegraphics{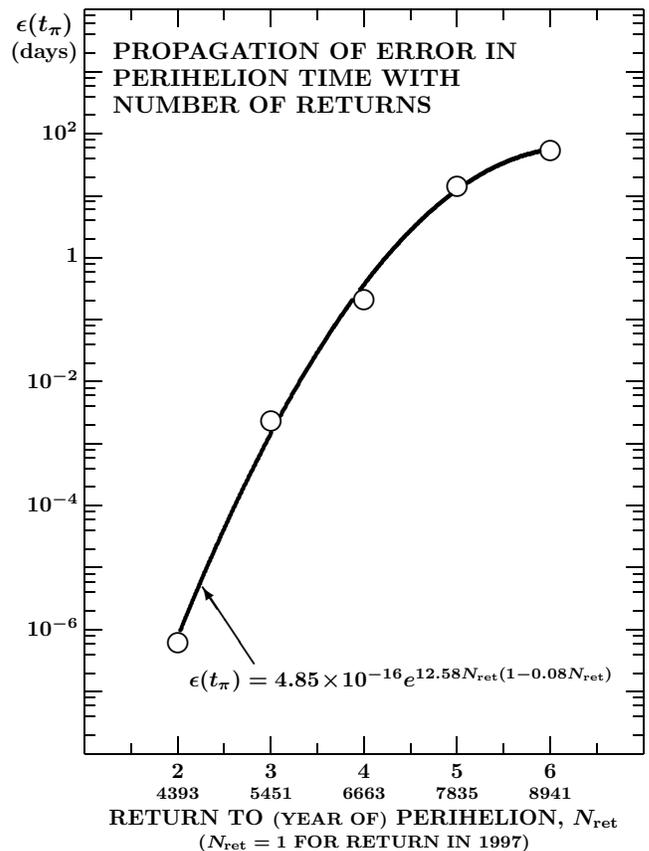}}} 
\vspace{-3.75cm}
\caption{Propagation of the error in the perihelion time, equivalent to an
error of 0.000001 AU$^{-1}$, as a function of time, expressed in terms of
a number returns to perihelion, $N_{\rm ret}$, since $-$2250.  The empirical
fit is the best we were able to achieve in the given range of returns and
is not to be extrapolated outside this range.{\vspace{-0.09cm}}}
\end{figure}

\begin{table*}[t] 
\vspace{-3.8cm}
\hspace{-0.53cm}
\centerline{
\scalebox{1}{
\includegraphics{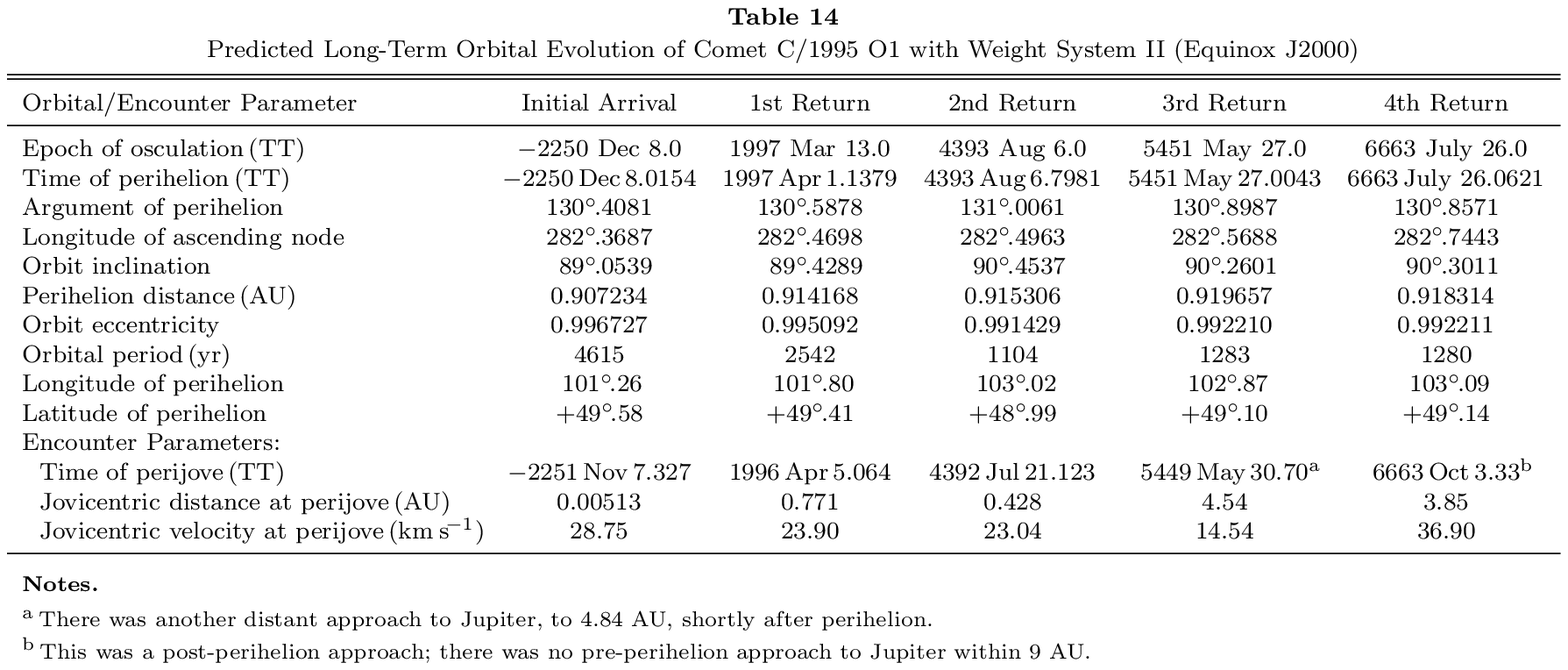}}} 
\vspace{-17.6cm}
\end{table*}

For the purpose of determining the rate of error propagation, we began the
orbit integration forward in time from the osculating orbit in 1997 listed
in Table~10,~which was assumed to be independent of the error in $(1/a)_{\rm
orig}$.  [It should be the comet's post-encounter orbit in $-$2250 as well
as the time and distance at perijove that would change slightly, if
$(1/a)_{\rm orig}$ differed {\vspace{-0.035cm}}from the adopted value of
+0.000046~AU$^{-1}$ (Section~6.2).] The results~of the computations are
shown in Figure~8, which suggests that an error of 0.000001~AU$^{-1}$
transforms to a negligibly small error of 0.05 second in the perihelion time
at the next return, to which we assign \mbox{$N_{\rm ret} = 2$}.  However, at
the following return (\mbox{$N_{\rm ret} = 3$}), the propagated error already
increases to 3.4~minutes and at the subsequent returns to, respectively,
0.21~day (\mbox{$N_{\rm ret} = 4$}), 14.4~days (\mbox{$N_{\rm ret} = 5$}),
and 53.1~days (\mbox{$N_{\rm ret} = 6$}).  We terminated the computations
at this point, because given that a realistic uncertainty of $(1/a)_{\rm
orig}$ of{\vspace{-0.04cm}} C/1995~O1, even if it was a dynamically new
comet, should be at least 0.000010~AU$^{-1}$ and perhaps still greater,
the corresponding uncertainty of the perihelion time at the subsequent
returns could become comparable with Jupiter's orbital period, rendering
the comet's encounters with the planet (and the resulting orbital
transformations) unpredictable.  

The complete sets of orbital elements with Weight System~II, including the
initial arrival and the current return (both from Table~10) are presented in
Table~14, in which the future returns are terminated in the 7th millennium.
To our surprise, the table shows that as long as the comet experienced the
very close approach to Jupiter in $-$2251, it undergoes encounters to less
than 1~AU from the planet at {\it three consecutive\/} apparitions:\ a
moderate approach in 1996 will be followed by a yet another preperihelion
encounter at the next return, in the year 4392!  

For the scenario based on Weight System I, the results are rather similar for
the expected return in 4393, the perihelion time taking place about 24~days
earlier and the Jovicentric distance at perijove amounting to 0.58~AU, but
the predictions by the two scenarios of the next perihelion time, in the 6th
millennium, already differ by more than 200~yr.  Yet, the fundamental features
of the orbital evolution, especially the triple Jovian encounter at three
consecutive returns, are common to both scenarios and thus independent of the
weight system for the critical observations.

\begin{table}[b] 
\vspace{-3.9cm}
\hspace{4.23cm}
\centerline{
\scalebox{1}{
\includegraphics{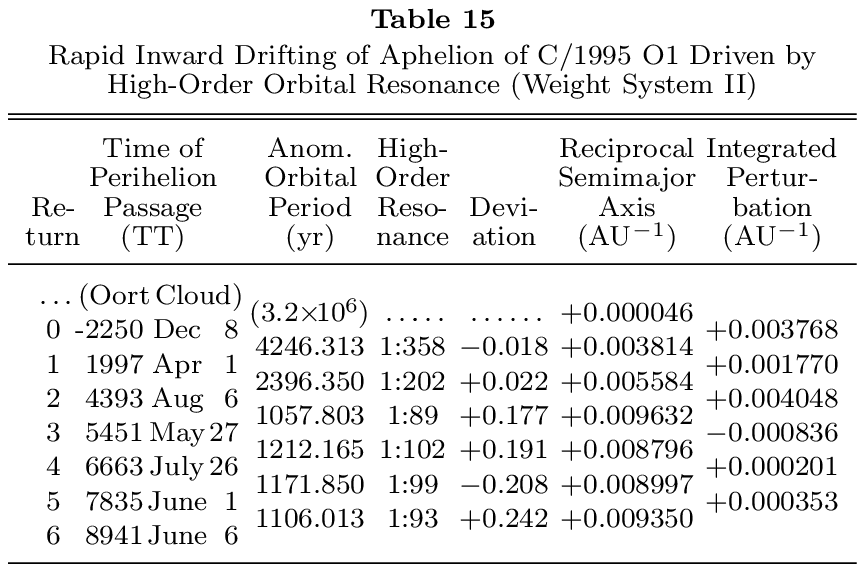}}} 
\vspace{-19.93cm}
\end{table}

In a recent paper (Sekanina \& Kracht 2016), we called attention to the fact
that Jovian perturbations exerted on a comet in the course of recurring close
or moderate encounters, especially when they take place at several consecutive
returns (in a scenario that we described as a high-order orbital-cascade
resonance), can provoke rapid inward drifting of the comet's aphelion.  In
magnitude, these effects rival those triggered by a single extremely close
encounter.  When we wrote that 2016 paper, we did not expect that C/1995~O1
--- then the next object of our interest --- would provide us with such a
nice example of the process of orbital-cascade resonance, ending up with
an aphelion distance of $\sim$200~AU after only three revolutions about the
Sun.

Table~15 presents the details of this remarkable orbital evolution:\ from the
perihelion times in column~2 we determined the anomalistic orbital periods in
column~3, which between $-$2250 and 1997 and between 1997 and 4393 turned out
to be, as column~4 shows, practically commensurable with the sidereal orbital
period of Jupiter (as the minor deviations in column~5 confirm); column~6
converts the orbital periods from column~3 into the reciprocal values, $1/a$,
of the semimajor axis, while the last column lists the differences between
the neighboring values of $1/a$, that is, the perturbations integrated over
one revolution about the Sun.  The table shows that in merely three revolutions
about the Sun, by 5451, the comet will reduce its orbital period from some
3~million~yr to only slightly more than 1000~yr.  An equally surprising result
is that the peak integrated perturbation is not the one associated with the
close approach of 0.005~AU to the planet in $-$2251, but with an encounter
to 0.43~AU in 4392!  The effect of the Jovian encounters is illustrated by an
average integrated perturbation of $1/a$:\ between $-$2250 and 5451 it amounts
to about 0.0032~AU$^{-1}$rev$^{-1}$, being very systematic and always positive;
 whereas during the three revolutions after 5451 it equals, in absolute value,
only 0.00046~AU$^{-1}$rev$^{-1}$ and is random.

\section{Correlation among Nongravitational Law,\\Orbital History, and Activity\\of
 Comet C/1995~O1}
In Section 6.4 we expressed our preference for the orbital solution in
Table~10 (Weight System II) over the solution from Table~5 (Weight System I)
solely on the grounds of orbital quality (the distribution of residuals) and
the number of accommodated observations.  We noticed that by far the most
striking distinction between the parameters of the two solutions was the
nongravitational law.  The solution from Table~5 required a scaling distance
of \mbox{$r_0 = 3.07$ AU}, implying the prevalence of water ice in the
effects of the nongravitational acceleration, and, unexpectedly, the
dominance of the acceleration's normal component over the radial component.
On the other hand, the preferred solution from Table~10 is in line with a
scaling distance of more than 15~AU, suggesting that the contribution to
the detected nongravitational acceleration by ices much more volatile
than water ice was important.  In addition, the radial component of~the
nongravitational acceleration exceeded the normal component, even though
by a factor of $\sim$4/3 only.  The transverse component came out to be
small and poorly determined from both solutions.

In Section 6 we already remarked on a sizable body of data that points to a
large abundance of highly-volatile species sublimating from the nucleus of
C/1995~O1.  The question is to what extent is the nongravitational law,
preferred for its orbital prerogatives, consistent with evidence based on
the physical observations.  Because of a record post-perihelion span of
available data, extending to 32~AU from the Sun, it is this branch of the
orbit that we examine first.

\subsection{Nongravitational Law and Major Contributors to the Comet's
 Post-Perihelion Activity}
Our aim is now twofold: (i) to compare the two applied versions of the
nongravitational law (Weight Systems I and II) with the sublimation curves
of common ices released from the nucleus and (ii)~to examine the correlations
between the laws and the observed production rates of these species.  To
proceed with the first task, we describe the sublimation curves of three major
compounds:\ water ice, carbon dioxide, and carbon monoxide in the order of
increasing volatility.

The isothermal approximation to the nongravitational (or sublimation) law
for water ice, extensively tested on the orbital motions of a large number
of comets, is taken from Marsden et al.'s (1973) standard Style II
nongravitational model, as already mentioned in Section~4.  For carbon
dioxide the critical data on the saturated vapor pressure and the latent
heat of sublimation were taken from Azreg-A\"{\i}nou (2005), whereas for
carbon monoxide from an extensive compilation by Wylie (1958).

\begin{table}[b] 
\vspace{-3.85cm}
\hspace{4.23cm}
\centerline{
\scalebox{1}{
\includegraphics{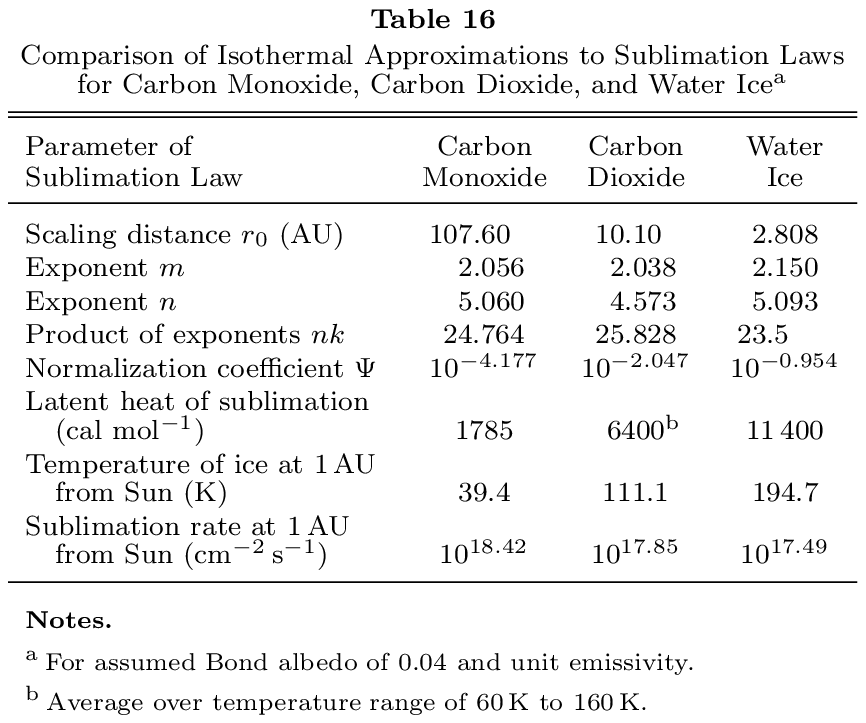}}} 
\vspace{-18.53cm}
\end{table}

The parameters of an empirical fit, of the type expressed by Equation~(1),
to the normalized sublimation (or momentum-transfer) rates of the three
species are listed in Table~16, together with the latent heat of sublimation
and the ice temperature and sublimation rate at 1~AU from the Sun.  These
rates, per cm$^2$ per s, are, respectively for carbon monoxide and carbon
dioxide, 8.5 times and 2.3 times greater than for water ice.  Masswise the
ratios are still higher:\ 13.2 for carbon monoxide and 5.6 for carbon dioxide.
An approximation to the scaling distance given by Equation~(3) provides the
values of 114.5~AU for carbon monoxide and 8.9~AU for carbon dioxide, in
reasonable agreement with the tabulated numbers obtained by fitting the
sublimation rates derived directly from the rela\-tions for vapor pressure
as a function of temperature.

To proceed with the second task, we next collected the data on the production
rates of water, carbon dioxide, and carbon monoxide from the nucleus of
C/1995~O1 after perihelion.  For water ice the production rates were
determined by Dello Ruso et al.\ (2000) directly from the ground-based
high-resolution 2--5~$\mu$m infrared spectroscopic observations and by
Crovisier et al.\ (1999) from observations with two instruments on board
the {\it Infrared Space Observatory\/} (ISO); and, furthermore, by Combi
et al.\ (2000) from the images of the hydrogen Lyman-alpha coma taken with
the SWAN all-sky camera on board~the SOHO spacecraft; by Weaver et al.\
(1999b) via OH production rates from ultraviolet observations with the
Hubble Space Telescope (HST); by Stern et al.\ (1999) and Harris et al.\
(2002), both groups using in principle the same technique to analyze their
observations taken, respectively, with a mid-UV/visible imager on board the
Space Shuttle and with a wide-field ground-based instrument; and by Biver et
al.\ (1999, 2002) and Colom et al.\ (1999), who employed a radio telescope to
obtain measurements of the 18-cm emission of OH.

Since the radiation from carbon dioxide cannot be observed from the ground
(e.g., Crovisier 1999), the only existing post-perihelion data on its
production rate were derived from the ISO observations of the 4.25-micron
band (Crovisier et al.\ 1999).

For carbon monoxide our primary source of information was the paper by
Gunnarsson et al.\ (2003), specifically the sum of the production rates
from their model of the nucleus' subsolar and isotropic outgassing, with
the data taken at a wavelength of 1.3~mm at ESO, La Silla.   They cover a
range of heliocentric distances from 2.87~AU to 10.75~AU.  This set was
supplemented with the results from the radio observations by Biver et al.\
(1999, 2002) at several discrete wavelengths between 0.65~mm and 2.6~mm;
from the infrared observations by DiSanti et al.\ (1999); and from the ISO
infrared observations by Crovisier et al.\ (1999).  From the common data
points by Gunnarsson et al.\ (2003) and Biver et al.\ (1999, 2002) we found
that the latter required a minor correction of $-$0.05 in log(production
rate) to match the former on the average.  In accordance with the
conclusions by DiSanti et al.\ (1999) we plotted the ``native'' production
rates at heliocentric distances smaller than 2~AU, but the ``total'' rates
at larger distances.

\begin{table}[b] 
\vspace{-3.85cm}
\hspace{4.23cm}
\centerline{
\scalebox{1}{
\includegraphics{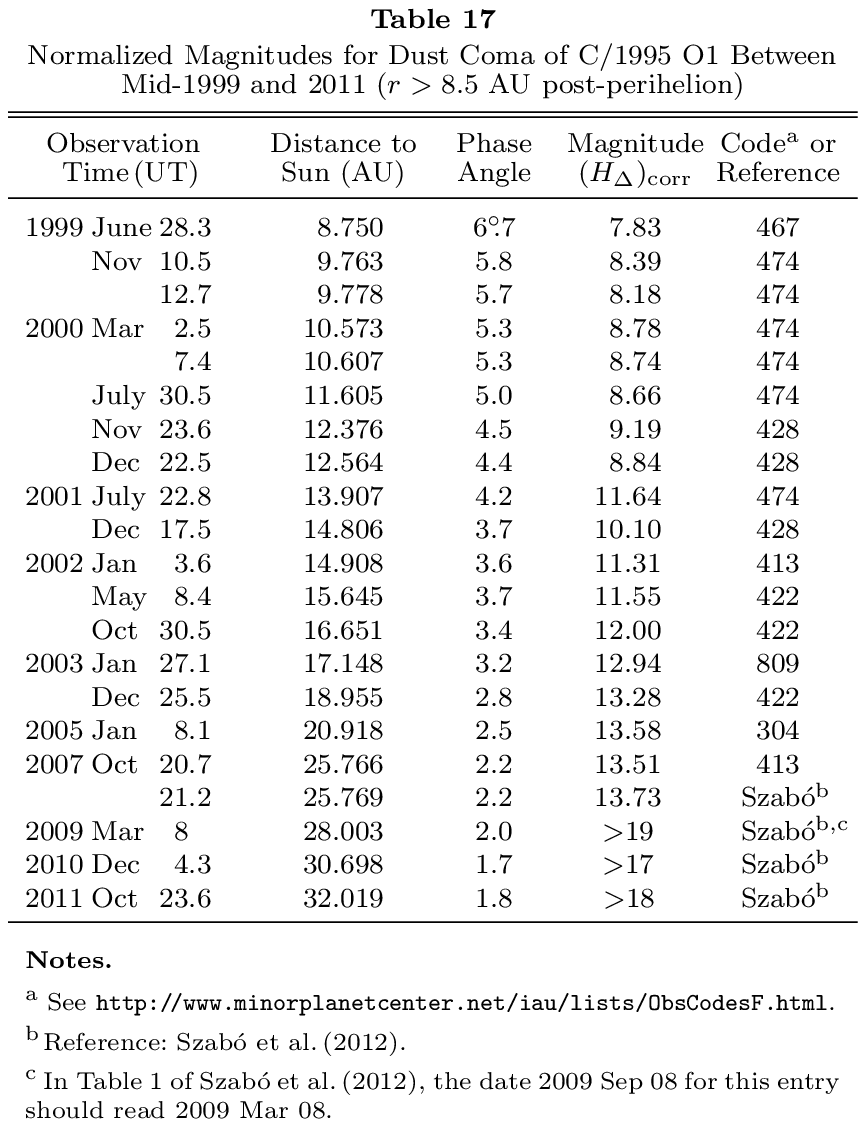}}}  
\vspace{-14.12cm}
\end{table}

The production of carbon monoxide was systematically monitored up to 14 AU
from the Sun.  The activity at still larger distances could only be assessed
from the magnitudes of the dust present in the coma.  As shown in Table~17,
for the period from mid-1999 to late 2007 we derived these data from the
total magnitudes reported to the {\it Minor Planet Center\/} (see footnote
1 in Section 2) by the observers from the selected sites; for the period
from late 2007 on from the magnitudes published by Szab\'o et al.\ (2008,
2011, 2012).  Each resulting magnitude $(H_\Delta)_{\rm corr}$ was obtained
by normalizing the observed magnitude to a unit geocentric distance, by
subtracting the contribution from the nucleus, and by referring the residual
brightness to a zero phase angle using the ``compound'' Henyey-Greenstein
law, as modified by Marcus (2007).  Table~17 indicates that in 2007 the
comet, at 25.8~AU from the Sun, was definitely active (Szab\'{o} et al.\
2008), but by 2009, at 28~AU from the Sun, the nucleus became inert
(Szab\'{o} et al.\ 2012).  This development is also supported by the Spitzer
Space Telescope thermal-infrared observations made in 2005 and 2008 by
Kramer et al.\ (2014).  As the comet's heliocentric distance between the
two dates increased from 21.6~AU to 27.2~AU, the flux dropped much more
steeply than required by an inverse square power law, with the dust in the
coma having been substantially depleted over the period of the three years.

\begin{figure*}[t]  
\vspace{-7.29cm}
\hspace{-0.2cm}
\centerline{
\scalebox{0.78}{
\includegraphics{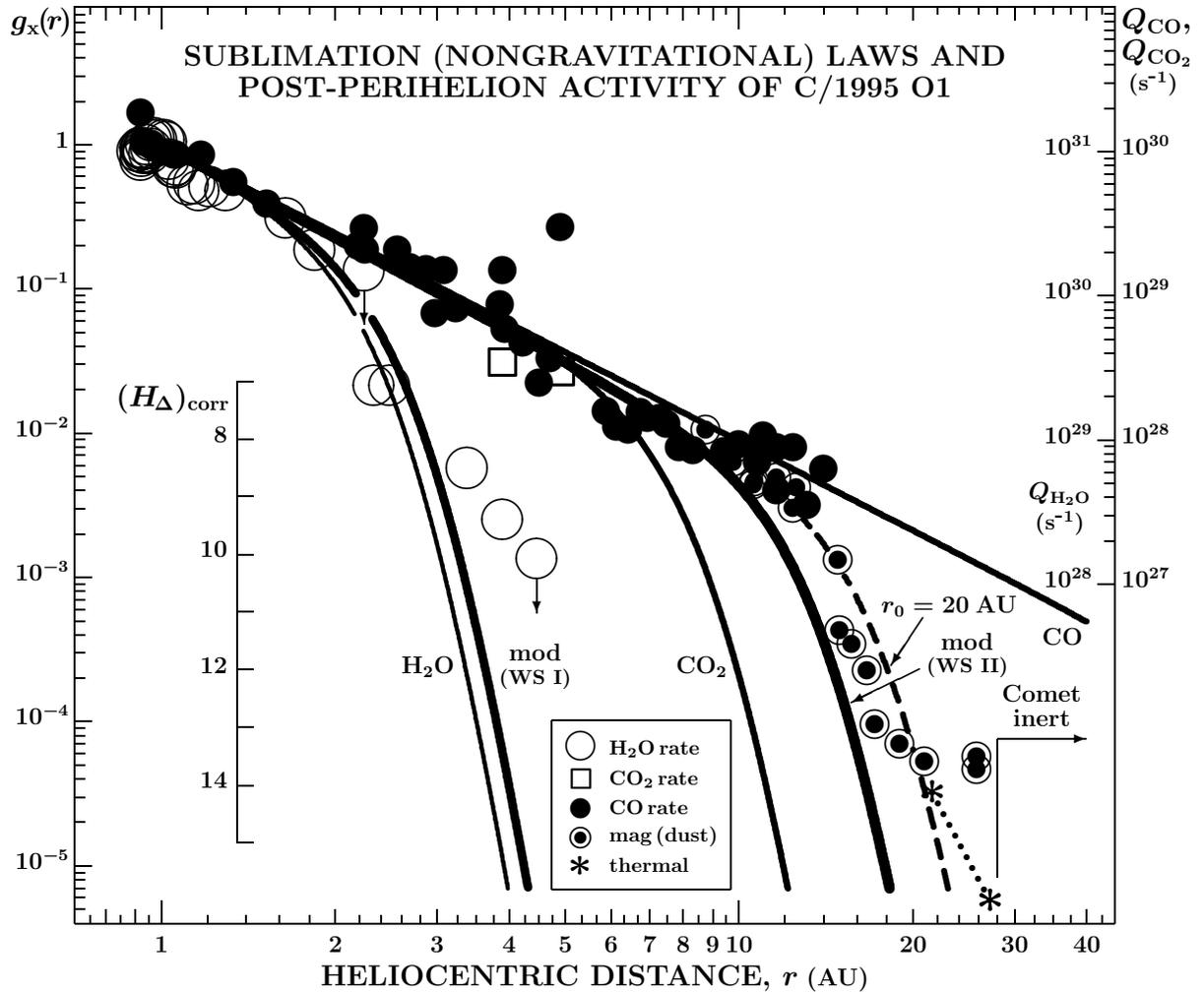}}} 
\vspace{-2.38cm}
\caption{Post-perihelion activity of C/1995 O1 and comparison of the
isothermal approximations to the nongravitational and sublimation laws,
$g_{\rm mod}(r;r_0)$.  The preferred nongravitational law has a scaling
distance \mbox{$r_0 = 15.36$ AU} (Weight System II or WS\,II), the other
has \mbox{$r_0 = 3.07$ AU} (Weight System I or WS\,I).  Also shown are
the sublimation laws for water ice, carbon dioxide, and carbon monoxide
In addition, we plot the observed data collected from various sources
(see the text):\ the production rates~of water $Q_{{\rm H}_2{\rm O}}$
{\vspace{-0.03cm}}(large open circles; including two 3$\sigma$ upper
limits --- circles with arrows), carbon dioxide $Q_{{\rm CO}_2}$
(squares), and carbon {\vspace{-0.03cm}}monoxide $Q_{\rm CO}$ (solid
circles; the anomalously high rate is an ISO data point, Crovisier et
al.\ 1999); the total magnitudes of the dust coma at helio\-centric
distances greater than 8~AU (the CCD data, obtained by a selected set
of observers and extracted from the {\it Minor Planet Center\/}'s
Observations Database, corrected for the contribution from the nucleus,
and normalized to a unit geocentric distance and zero phase,
($H_\Delta)_{\rm corr}$; circled dots); and, in relative units, the
dust coma's thermal flux between 21.6~AU and 27.2~AU (connected
asterisks), based on the Spitzer results by Kramer et al.\ (2014).
A crude fit to the dust magnitude data far from the Sun by a $g_{\rm
mod}(r;r_0)$ law is depicted by a dashed line to show that they are
approximated by \mbox{$r_0 \approx 20$ AU}.  Szab\'o et al.\ (2012)
argue that the comet became inactive beyond a heliocentric distance
of $\sim$28~AU.{\vspace{0.7cm}}}
\end{figure*}

The collected data on the comet's post-perihelion activity in Figure~9
show a number of important features.  Comparison of the two modified
nongravitational laws with the sublimation curves for water ice, carbon
dioxide, and carbon monoxide --- all normalized to 1~AU --- suggests
that, the law with Weight System~I is, as expected, very close to the
water-ice sublimation curve, while the law with Weight System~II extends
to larger heliocentric distances than the carbon-dioxide sublimation curve.
Thus, no mix of H$_2$O and CO$_2$ could explain the law with Weight System~II;
a contribution from CO is necessary.

The water-production rates are consistent with the theoretical water-ice
sublimation curve up to about 2.5~AU from the Sun.  Beyond that point only two
positive observations exist, which suggest that the rate decrease proceeded
more slowly than predicted by the simple theory.

Only two measurements exist in the post-perihelion period of time for the
production rate of carbon dioxide (obtained with the ISO); they are fairly
{\vspace{-0.03cm}}compatible with an expected $\sim \! r^{-2}$ sublimation
curve between 3.5~AU and 5~AU from the Sun.  At 3.89~AU, the production rates
of H$_2$O and CO$_2$ were comparable, about \mbox{$3 \times\!10^{28}$\,s$
^{-1}$}, but masswise the rate of CO$_2$ was more than twice greater.

The production rate of carbon monoxide follows the approximately inverse-square
power law from perihelion all the way to 14~AU.  Among more than 40 collected
measurements only one is strongly out of line; it is the second (and last)
post-perihelion measurement obtained with the ISO on 1998 April 6 at 4.9~AU
from the Sun.  In Figure~9 this anomalous rate exceeds the level of the
nucleus outgassing rate from 1998 March 18--19 (at 4.7~AU from the Sun), as
determined by Gunnarsson et al.\ (2003), by a factor of 8.  Crovisier et
al.\ (1999) speculated on a possible outburst, but related that no
simultaneous flare-up was reported in the light curve.  We may add that
no excessive production of CO$_2$ is seen in the plot to have accompanied
the CO event either.

The normalized magnitudes $(H_\Delta)_{\rm corr}$ of the dust coma are drawn
to overlap the CO production rates in Figure~9 in the range of heliocentric
distances from 8.7~AU through $\sim$14~AU.  A satisfactory correlation between
both appears from the plot, suggesting that \mbox{$(H_\Delta)_{\rm corr} =
10$}~is equivalent to \mbox{$Q_{\rm CO} \simeq 1.6\times\!10^{27}$\,s$^{-1}$},
even though a correspondence between magnitudes and any other activity index
is only approximate because of the dust ejecta's finite residence time in the
coma.  The magnitudes begin to progressively deviate from the inverse-square
power law downward at about the time of termination of the CO production-rate
data; the magnitudes follow, very approximately, a law similar to that with
Weight System~II, but with a greater scaling distance $r_0$, estimated at
$\sim$20~AU.  The last three data points from Table~17~are not plotted in
Figure~9; they would be located way~\mbox{below} the plot's foot line,
referring, in conformity with Szab\'{o} et al.'s (2012) conclusion, to a
vanishing dust coma.

Two important inferences on the post-perihelion activity of C/1995~O1 are
(i)~the production of carbon monoxide did not apparently follow an
inverse-square power law beyond $\sim$15~AU from the Sun, as it should~(up
to $\sim$100~AU) if free CO were available on the nucleus surface, but a
steep drop in the dust-coma brightness~set in at $\sim$15~AU and no dust
could be detected~in~the~coma from 28~AU on; and (ii)~the dynamically
determined nongravitational law of Weight System II approximates the comet's
post-perihelion activity curve fairly successfully up to almost 20~AU from
the Sun, but to achieve the best possible correspondence, the law's scaling
distance needs to be increased by some 30\%.

A\,peculiar post-perihelion~\mbox{evolution}~of~the~\mbox{inner}\,coma's
brightness, observed extensively by Liller (2001), will be addressed in
Part~II of this investigation.  At this point we are interested in a
potential perihelion asymmetry of the comet's activity, for which purpose
we now describe the developments along the incoming branch of the orbit.

\begin{table*} 
\vspace{-4.2cm}
\hspace{-0.53cm}
\centerline{
\scalebox{1}{
\includegraphics{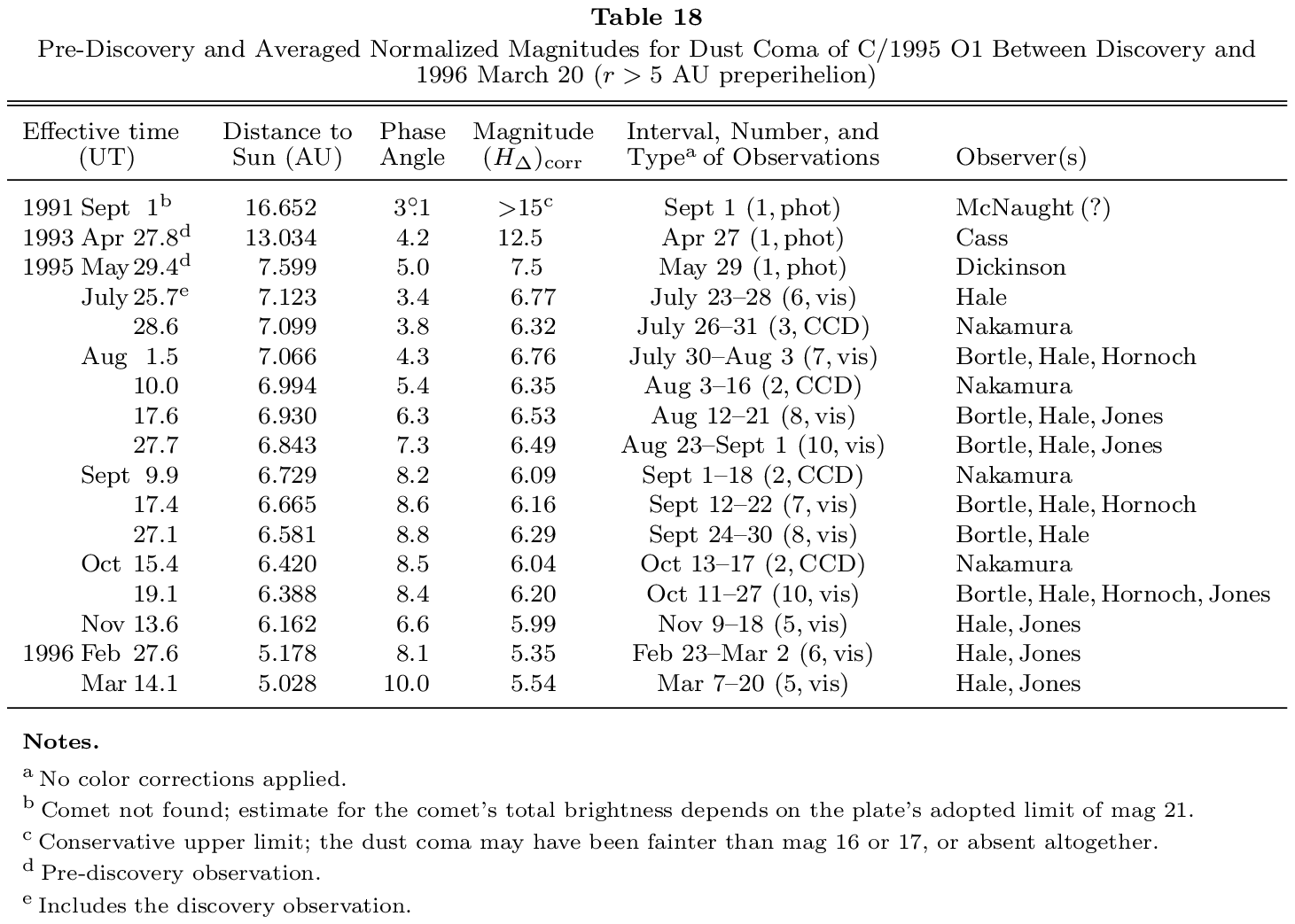}}} 
\vspace{-14.3cm}
\end{table*}

\subsection{The Comet's Preperihelion Activity}
The procedure we now followed was the same as for the post-perihelion activity
in Section 9.1.  The data on the water production rate were assembled, as
before, from the works by Dello Russo et al.\ (2000), Crovisier et al.\
(1999), Combi et al.\ (2000), Weaver et al.\ (1997, 1999b), Harris et al.\
(2002), Biver et al.\ (1997, 1999), and Colom et al.\ (1999); in addition,
we used Weaver et al.'s (1999a) result derived from a ground-based observation
of the 4.65-$\mu$m band.

For carbon dioxide, the preperihelion list has six entries, two ISO
measurements by Crovisier et al.\ (1999) and four HST upper limits by
Weaver et al.\ (1999b).  On the average, the CO$_2$ production rate is
about a factor of two or so lower than the CO production rate at the same
heliocentric distance (between 2.7 and 4.6~AU).  Relative to the production
of water, the CO$_2$ production is lower by a factor of 4--5 at 2.9~AU, but
they are comparable to one another at 4.6~AU.

The preperihelion production of carbon monoxide was monitored at millimeter
and submillimeter wavelengths, starting soon after discovery, by Biver et
al.\ (1997, 1999).  It was also measured by Jewitt et al.\ (1996) in 1995,
by Womack et al.\ (1997) in 1995-1996, by Crovisier et al.\ (1999) in 1996,
by DiSanti et al.\ (1999) in 1996--1997, and by Weaver et al.\ (1999a) in
1997.

Normalized dust-coma magnitudes would provide little additional information
on the comet's preperihelion activity, if there were no pre-discovery
observations.  Because there were, these magnitudes do provide --- their
low accuracy notwithstanding --- some fairly tight constraints.  The
1993 observation with the UK Schmidt (McNaught 1995), made by C.\ P.\
Cass and already referred to, shows, when combined with the discovery
observation, that the rate of brightening between 13~AU and 7~AU was
much steeper than $r^{-2}$.  The activity is further constrained by the
failed effort by McNaught (1995) to detect the comet on a plate obtained
on 1991 September 1.  Assuming with McNaught that the limiting magnitude
of the UK Schmidt plate was 21, the nondetection of the comet in 1991
may indicate that its nucleus was nearly or completely inactive, because
Szab\'{o} et al.'s (2012) results imply that the nucleus should have had
a visual magnitude of about 22 at the time of the 1991 observation,
equivalent to a normalized magnitude of about 16.  The upper limit on
the dust-coma magnitude in Table~18 is therefore very conservative.  On
the other hand, the estimated normalized magnitude of the nucleus at the
time of the 1993 detection is about 15, so that the dust coma was then much
brighter than the nucleus.  Even McNaught's (1995) estimate of 19 for the
so-called nuclear magnitude was at least 1.5~mag brighter than the bare
nucleus.

\begin{figure*} 
\vspace{-7.3cm}
\hspace{1.35cm}
\centerline{
\scalebox{0.78}{
\includegraphics{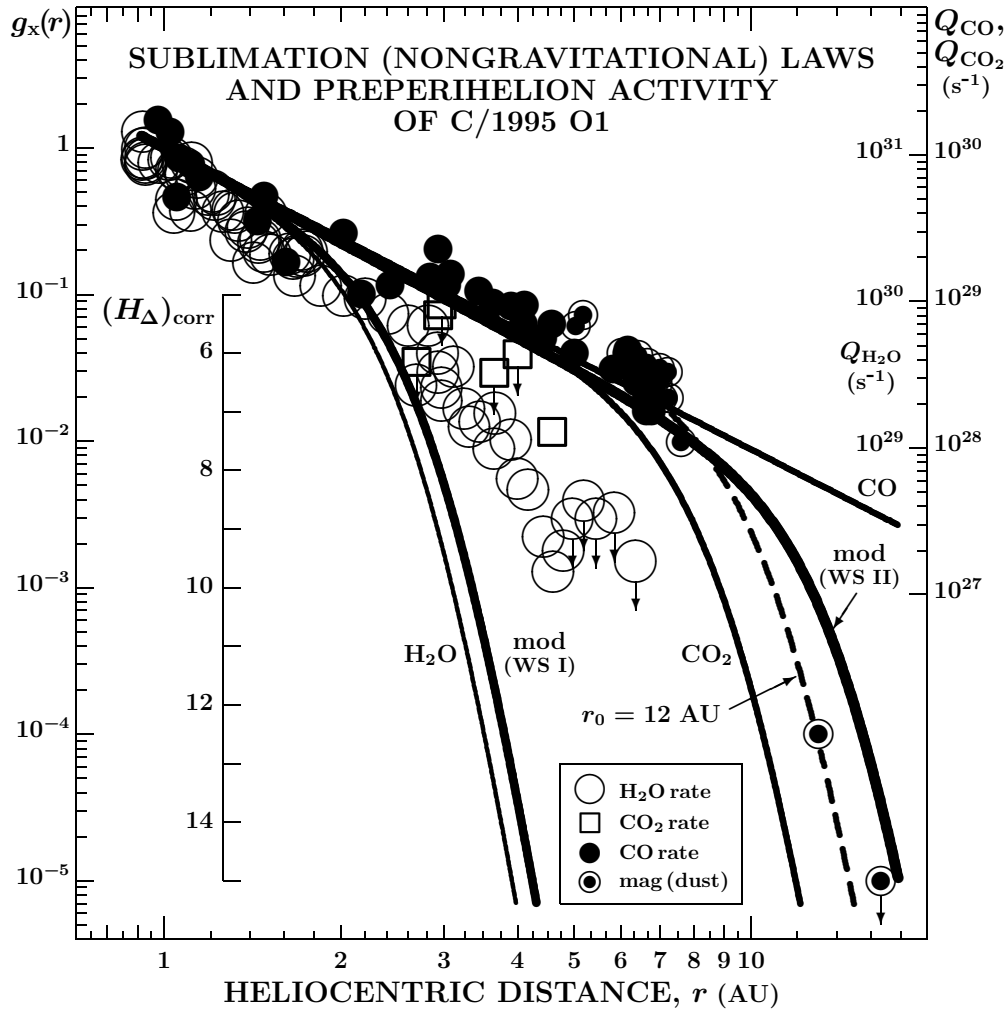}}} 
\vspace{-2.68cm}
\caption{Preperihelion activity of C/1995 O1 and comparison of the isothermal
approximations to the nongravitational and sublimation laws, $g_{\rm
mod}(r;r_0)$, which are normalized the same way as in the plot of
post-perihelion activity in Figure~9.  Similarly, the scales for the
observed production rates of water, carbon monoxide, and carbon dioxide
have not been changed.  Note the increased scatter in the CO data below
3~AU from the Sun.  The pre-discovery dust magnitude from 1993 is fitted
by a $g_{\rm mod}(r;r_0)$ law with \mbox{$r_0 \approx 12$ AU}.  Comparison
with Figure~9 suggests a striking asymmetry relative to perihelion.  For a
description of the symbols used, see the caption to Figure~9.{\vspace{0.6cm}}}
\end{figure*}

In response to the request by the {\it Central Bureau for Astronomical
Telegrams\/} for fortuitous pre-discovery observations of the comet (Marsden
1995), another message --- besides that by McNaught --- was sent by George
(1995), who reported an apparent image on a photograph taken with an 8.5-cm
f/1.7 lens on a Kodak Royal Gold 400 film by T. Dickinson on 1995 May 29.40
UT, when the comet was of mag 11.7.  Although this observation was made only
some 8 weeks before discovery, it suggests that the comet was still
brightening very rapidly.

Once the comet was discovered, the number of visual observations of its total
brightness became overwhelming, and in Table~18 we list averages of estimates
by four experienced observers, all of whom used reflectors with apertures
between 31~cm and 41~cm:\ J.\,Bortle, A.\,Hale, K.\,Hornoch, and A.\,F.\,Jones.
To allow a sufficient overlap in heliocentric distance with the observations of
carbon monoxide, we covered a wide enough period of time from discovery to 1996
March, during which the comet's heliocentric distance dropped from 7.1~AU to
5~AU.  These visual brightness estimates are compared in Table~18 with a series
of CCD magnitudes by A.\,Nakamura, who used a 60-cm f/6.0 Ritchey-Chr\'etien
reflector.  Both the visual and CCD data were taken from several issues of
the {\it International Comet Quarterly\/} (Green 1995, 1996).

The collected information on the preperihelion activity of C/1995~O1 is
presented in Figure~10.  Although there are similarities between this plot
and Figure~9 in general, differences are readily perceived in detail.  The
water production rates are higher than the post-perihelion ones farther
from the Sun, but slightly lower near perihelion, so that the production
of H$_2$O increases less steeply with decreasing heliocentric distance
before perihelion than it decreases with increasing distance on the way
out.  The preperihelion variations in the water production rate can be
fitted with a modified nongravitational law (see the beginning of
Section~6) with a scaling distance of \mbox{$r_0 \simeq 5$ AU}.  This is
not puzzling because a theoretical water sublimation curve for a subsolar
point requires \mbox{$r_0 = 5.6$ AU}; the observed production curve is
therefore still compatible with a simple sublimation model.

The measured preperihelion production curve of carbon monoxide exhibits some
inconsistencies.  At heliocentric distances exceeding $\sim$3~AU, the data by
Biver et al.\ (1997, 1999), Jewitt et al.\ (1996), DiSanti et al.\ (1999),
and Crovisier et al.\ (1999) are compatible, showing a systematic increase
in the production at a rate slightly steeper than an inverse-square power
of decreasing heliocentric distance, strongly resembling the post-perihelion
behavior.  On the other hand, the production rates by Womack et al.\ (1997)
are for some reason systematically lower by a factor of several, and they
are not plotted in Figure~10.

At smaller heliocentric distances the situation is still more precarious.
The CO production curve published by Biver et al.\ (1997, 1999) is stalling
or slightly dropping during the approach to the Sun between 3~AU and 2~AU,
then grows steeply between 2~AU and perihelion.  In contradiction, DiSanti
et al.'s (1999) results suggest a smooth, systematic increase in the
production rate all the way to perihelion.  This conflict obscures the
merit of the points that Bockel\'ee-Morvan \& Rickman (1999) make in their
comparing the degree of success achieved by three independent models of
comet interior proposed, respectively, by Capria et al.'s (1997),\footnote{An
updated fit to the observations of C/1995~O1 by this model appears in Capria
et al.\ (2002).} by Prialnik (1997), and by Enzian et al.\ (1998).

The normalized preperihelion magnitudes $(H_\Delta)_{\rm corr}$ are linked
with the CO production curve by assuming mag 10 to be equivalent to
\mbox{$1.1 \!\times\!  10^{27}$\,s$^{-1}$}, which differs slightly from the
post-perihelion equivalent,~apparently~because of a somewhat different
dust-to-CO mass ratio in the coma (Section~9.5).  When comparing Figures~9 and
10, the most striking feature is a major \mbox{\it perihelion\,asymmetry\/}
at heliocentric distances larger than $\sim$7~AU, in the sense that the
activity is systematically lower before than after perihelion:\ whereas the
post-perihelion dust magnitudes crudely fit a $g_{\rm mod}(r;r_0)$ law with
\mbox{$r_0 \approx 20$ AU}, the 1993 pre-discovery magnitude and the 1991
upper limit are both consistent with \mbox{$r_0 \approx 12$ AU}.  The
dynamically determined scaling distance of $\sim$15.4~AU appears to be a fair
compromise, which --- the asymmetry notwithstanding --- implies a remarkable
{\small \bf correlation between~the nongravitational acceleration exerted on
the comet's nucleus and its observed activity}.

\subsection{Nongravitational Laws Incorporating\\Perihelion Asymmetry}

The apparent detection of perihelion asymmetry in the activity of C/1995~O1
leads up to the issue of an asymmetric nongravitational law, which,
unfortunately, the employed software package does not allow us to apply at
present.  This tool has over the past several decades been debated in the
literature fairly extensively, even though the modern, sophisticated methods
of accounting for a nongravitational acceleration, by incorporating it
directly in the equations of motion, focused from the beginning on the
symmetrical models (Marsden 1969).  Tested on 2P/Encke's comet, the early
orbital solutions based on an asymmetric law failed (Marsden 1970) for
reasons that have not been fully understood.  Similarly, employing an
empirical nongravitational law that was consistent with 1P/Halley's
asymmetric water production-rate  curve, \mbox{Yeomans} (1984) was unable
to improve upon the results obtained with a symmetrical law.  Rickman \&
Froeschl\'e (1983) [see also Froeschl\'e \& Rickman (1986) and Rickmam \&
Froeschl\'e (1986)] calculated the expected nongravitational parameters
for Halley's comet from their thermal model of comets, but Landgraf (1986),
applying them in his orbital solutions, concluded that, in spite of an
improvement, a further refinement was desirable.  Subsequently, Sitarski
(1990) developed a hybrid model, employing the symmetrical nongravitational
law by Marsden et al.\ (1973), to derive precessional parameters ---
inherently requiring perihelion asymmetry --- for short-period comets.

Returning to the problem of 2P/Encke and motivated by a perception of major
seasonal effects and nonrandom distribution of active regions on the nucleus
surface as well as by the observed statistical correlation between the
perihelion asymmetry of the light curves of comets and the sense of the
nongravitational effect, Sekanina (1988b) suggested that, at the risk of
a potentially significant loss of generality, the form of an asymmetric
law can be prescribed by applying the standard symmetrial law (Marsden
et al.\ 1973), $g_{\rm ice}(r;r_0)$ [see Equation~(1)], in which the
heliocentric distance $r$ is taken not at the time of observation, $t$,
but at a time \mbox{$t \!-\! \tau$}, hence $r(t\!-\!\tau)$, where $\tau$
is a constant ``offset'' time.  The nongravitational acceleration then
peaks at a time \mbox{$t_\pi \!+ \tau$} rather than at the time of
perihelion, $t_\pi$; that is, before perihelion if \mbox{$\tau < 0$},
or after perihelion if \mbox{$\tau > 0$}.

In orbital applications of this asymmetric law, the offset time $\tau$ is to
be determined by optimizing an orbital solution that fits the astrometric
observations of an examined comet.  \mbox{Yeomans} \& Chodas (1989) employed
this paradigm to develop an orbit-determination procedure that allows one to
establish the optimum offset time~$\tau$~as the value used in a solution with
a minimum rms residual.  Applying this approach to eight short-period comets
(including 1P/Halley but not 2P/Encke), they found that in a number of orbital
runs this new law improved the fit to astrometric observations and indicated
that the radial, rather than transverse, component of the nongravitational
acceleration accounted for practically the whole effect (the normal component
having been ignored); in most cases, the peak acceleration was offset from
perihelion a few weeks in one direction or the other, both~the magnitude and
the sign of the offset time correlating with those of the offset of the
outgassing peak (usually represented by the light curve) from perihelion.

Sitarski (1994a, 1994b) went even further in his application of this type of
asymmetric law.  He started from his hybrid model, improving it markedly
by incorporating the offset time $\tau$ and its rate of variation directly
into the least-squares procedure, which included all three components of the
nongravitational acceleration.  Under certain assumptions, he succeeded in
fitting --- in a single run --- the orbital motion of comet 22/P Kopff over
13~returns to perihelion between 1906 and 1990 with an rms residual of
$\pm$1$^{\prime\prime}\!$.56; and by rigorous extrapolation in satisfying,
to better than $\pm$1$^{\prime\prime}$, the discovery positions of the comet
at its next apparition.

While~an~asymmetric~\mbox{nongravitational}~law~might~improve a fit to the~\mbox{orbital}~\mbox{motion}~of~C/1995~O1~over~the~solution~with Weight~\mbox{System}~II,~the~paradigm~that~is~cus\-tomized for applications~to~\mbox{short-period}~comets~should be modified.~A~\mbox{constant}~\mbox{offset}~time~$\tau$,~which --- based \mbox{on Combi et al.'s\,(2000)~extensive~observations of}~near-perihelion
water production rates --- should equal about +19~days for this comet, could
not explain the striking difference between the heliocentric distances of the
very steep preperihelion increase of activity in Figure~10 (from 13~AU down to
8~AU) and its equally steep post-perihelion drop in Figure~9 (from 12~AU up
to 28~AU).  In terms of the scaling distance $r_0$, the difference between
the post-perihelion and preperihelion activity curves is \mbox{$20 \!-\! 12
= 8$}~AU, whereas a time difference of 19~days is equivalent to merely
$\sim$0.13~AU at 12~AU and just about 0.10~AU at 20~AU.  Measuring the
perihelion asymmetry at a level of, for example, \mbox{$g_{\rm mod} = 0.0001$}
(see Figures~9 and 10), the preperihelion and post-perihelion heliocentric
distances become, respectively, \mbox{$r_{\rm pre} \!=\! 12.56$\,AU} and
\mbox{$r_{\rm post} \!=\! 19.37$\,AU} and the equivalent times from perihelion
\mbox{$t_{\rm pre} \!-\! t_\pi \!=\! -1360$ days} and \mbox{$t_{\rm post}\!-\!
t_\pi \!=\! +2538$ days}, so in absolute value their difference is 1178~days.
For the $g_{\rm mod}$ law to peak at \mbox{$t_\pi \!+\! 19$ days} and at the
same time satisfy \mbox{$g_{\rm mod}(12.56\,{\rm AU};12\,{\rm AU}) = g_{\rm
mod}(19.37\,{\rm AU};20\,{\rm AU})$},~it~must apply \mbox{$r(t_{\rm pre} \!-\!
\tau_\ast) = r(t_{\rm post} \!-\! \tau_\ast) = r(t_\ast)$}~and,~accordingly,
\mbox{$\tau_\ast = 589$ days}, \mbox{$t_\ast = t_\pi \, \mp$\,1949 days}, and
\mbox{$r_\ast = 16.14$ AU}.  In order that $g_{\rm mod}(r_\ast; r_0) = 0.0001$
at 16.14~AU, the scaling distance $r_0$ ought to equal
\begin{equation}
r_0 = \left\{\! \frac{1 - r_\ast^{n + m/k} [g_{\rm mod}(r_\ast; r_0)]^{1/k}}
 {r_\ast^{m/k} [g_{\rm mod}(r_\ast; r_0)]^{1/k} - 1} \! \right\}^{\!\!1/n}
 \!\!\! = 16.10 \:{\rm AU}.
\end{equation}
We note that the scaling distance of the law for Weight System~II differs
from this value by only 0.74~AU.

The offset times at peak outgassing and at $\sim$16~AU from the Sun represent
two critical check points of the asymmetric law $g_{\rm mod}(r;16.10$\,{\rm
AU}).  Since the offset time at $\sim$16~AU is considerably greater --- by a
factor of 31 --- than the offset time near perihelion, the law $g_{\rm
mod}(r;\tau;r_0)$ with a constant offset time $\tau$ in the argument of
heliocentric distance, $r(t\!-\!\tau)$, could not obviously fit the observed
perihelion asymmetry of C/1995~O1.  It is legitimate to seek a generalized
form of $\tau$, called now $\tau_\star$, as a product of the offset time at
peak outgassing, $\tau$, and a function $h[r_\star]$ of heliocentric distance
\mbox{$r_\star \!=\! r(t\!-\!  tau_\star)$}.  This function should be
increasing with $r_\star$, rather than $r$ (\mbox{$dh/dr_\star > 0$}), and
it should equal unity, \mbox{$h = 1$}, at the time of peak outgassing,
$t_{\rm peak}$ [when \mbox{$t_{\rm peak} \!-\! \tau = t_\pi$} and
\mbox{$r_\star = r(t_{\rm peak} \!-\! \tau) = q\:\!$}]:
\begin{equation}
r(t \!-\! \tau_\star) = r_\star = r(t \!-\! \tau \, h[r_\star]).
\end{equation}
We consider two general forms of $h[r_\star]$:\ one is a power function, the other
an exponential.  In either case, it is necessary to make sure that $h[r]$ does not
overcorrect the asymmetry near aphelion.  The argument of $r$ should then read
either as
\begin{equation}
r_\star = r(t\!-\!\min\{\tau [r_\star/q]^\nu,\tau_{\rm max}\}),
\end{equation}
or as
\begin{equation}
r_\star = r(t\!-\!\min\{\tau \exp[\mu(r_\star \!-\! q)^\kappa], \tau_{\rm max}\}),
\end{equation}
where $\nu$, $\mu$, $\kappa$, and $\tau_{\rm max}$ are constants.  In practice,
either equation is readily solved by successive, rapidly converging iterations,
starting with \mbox{$r_\star \!=\! r$} in the expressions $(r_\star/q)^\nu$ or
$\exp[\mu(r_\star \!-\! q)^\kappa]$.  The two nongravitational laws should be
referred to as, respectively, $g_{\rm mod}(r;\tau,\nu;r_0)$ and
$g_{\rm mod}(r;\tau,\mu,\kappa;r_0)$, but when describing them, as well as
$g_{\rm mod}(r;\tau;r_0)$, summarily, we use the same notation as for the
symmetrical law, $g_{\rm mod}(r;r_0)$.  We note that any of the asymmetric laws
reduce to the symmetrical law when \mbox{$\tau = 0$}, regardless of the values
of $\nu$, $\mu$, or $\kappa$.  Similarly, the variable offset-time laws reduce
to the constant offset-time law when \mbox{$\nu = 0$} or \mbox{$\mu = 0$}.

For some comets, a better fit should be obtained with expression (10), for
others with expression (11).  The main difference between the functions
(10) and (11) is that the former allows the asymmetry to increase more
gradually.  The exponential, on the other hand, fits the cases where the
perihelion asymmetry increases only insignificantly at small to moderate
heliocentric distances, but rather dramatically farther from the Sun.  As
a practical procedure, we suggest that the overall trend in the perihelion
asymmetry of the gas production law between the two end points be employed
to choose the type of the function $h[r_\star]$, whereas the magnitudes of
the offset time at peak outgassing and far from the Sun be used to fix the
function's constants.

Insight into the issue of selecting the function $h[r_\star]$ is provided by
determining the dependence of the difference in $\log g_{\rm mod}(r;r_0)$
between a preperihelion and post-perihelion nongravitational effect (or a
production rate), ${\Delta_{\sf asym}}\log g_{\rm mod}(r;r_0)$, on the offset
time $\tau_\star$ as a function of heliocentric distance.  Differentiating
the relation between the time from perihelion and heliocentric distance
for parabolic motion (thus obtaining a tight upper limit on the derivative
$dr/dt$) and equating the differential $dt$ with $\tau_\star$, we
find\\[-0.2cm]
\begin{equation}
\Delta_{\sf asym} \log g_{\rm mod}(r;r_0) = 0.02113\,\frac{\sqrt{r_\star
 \!\!-\! q}}{r_\star^2} \! \left( \! m \!+\! \frac{nk \Lambda_\star^n}{1 \!+\!
 \Lambda_\star^n} \! \right) \! \tau_\star,
\end{equation}
where $r_\star$ and $q$ are in AU, $\tau_\star$ in days, and
\mbox{$\Lambda_\star = r_\star/r_0$}.  The difference $\Delta_{\sf
asym}\log g_{\rm mod}(r;r_0)$ has the same sign as $\tau_\star$; when
positive, $g_{\rm mod}(r;r_0)$ is greater after perihelion, and vice versa.

Comparing Figures~9 and 10 we note that the primary features of the perihelion
asymmetry of C/1995~O1's activity could be fitted much better by the offset
times that follow an exponential law rather than a power law.  The asymmetry
is distinctly apparent in close proximity of perihelion (particularly in the
variations of the production rate of water) than it is between 3~AU and 7~AU
from the Sun (in the CO production).  At still larger distances, the
differences between preperihelion and post-perihelion rates of outgassing
increase dramatically.  When \mbox{$\kappa = 1$} in Equation~(11), the two
{\vspace{-0.04cm}}check points require that \mbox{$\mu = 0.2255$ AU$^{-1}$},
so that \mbox{$\tau_\star = 75$ days} at \mbox{$r = 7$ AU}.  When
\mbox{$\kappa = 2$}, then one gets \mbox{$\mu = 0.0148$ AU$^{-2}$} and
\mbox{$\tau_\star = 32.9$ days} at \mbox{$r = 7$ AU}.  And when \mbox{$\kappa
= 3$}, it follows that \mbox{$\mu = 0.000973$ AU$^{-3}$} and \mbox{$\tau_\star
= 23.7$ days}.  On the other hand, with the power law from Equation~(10), one
finds \mbox{$\nu = 1.20$}, so that \mbox{$\tau_\star = 219$ days} at \mbox{$r
= 7$ AU}.  Inserting each of these values of $\tau_\star$ into Equation~(12),
the perihelion asymmetry in the nongravitational law $g_{\rm mod}(r;r_0)$ at
\mbox{$r = 7$ AU} becomes \mbox{$\Delta_{\sf asym} \log g_{\rm mod} =
0.198$}, 0.087, and 0.063, respectively, for the three exponential cases,
but 0.58 for the power-law case.  The case with \mbox{$\kappa = 3$} is
especially satisfactory, as the asymmetry mimicks day-to-day variations
in the production rate of carbon monoxide, while the power-law case is
rather unacceptable.

The superior quality of the exponential law is also apparent from the following
figures, in which three asymmetric nongravitational laws are compared with a
symmetrical one within 200~days of perihelion (Figure~11) and on another time
scale, spanning 18~years (Figure~12).  The constant-offset law displays
nicely an asymmetry near perihelion, but gradually less so farther from
the Sun; it grossly underrates the asymmetry, seen in Figures~9 and 10, at
very large heliocentric distances, where it is virtually equivalent to the
symmetrical law.  The power law fits the large perihelion asymmetry very
far from the Sun, but underrates the nongravitational effect before
perihelion and overrates it after perihelion on a time scale of months
from perihelion.  A remarkable quality of the exponential law is that it
mimicks the behavior of the constant-offset law near perihelion, but of
the power law far from the Sun.

\begin{figure}[t]  
\vspace{-4.1cm}
\hspace{-0.15cm}
\centerline{
\scalebox{0.43}{
\includegraphics{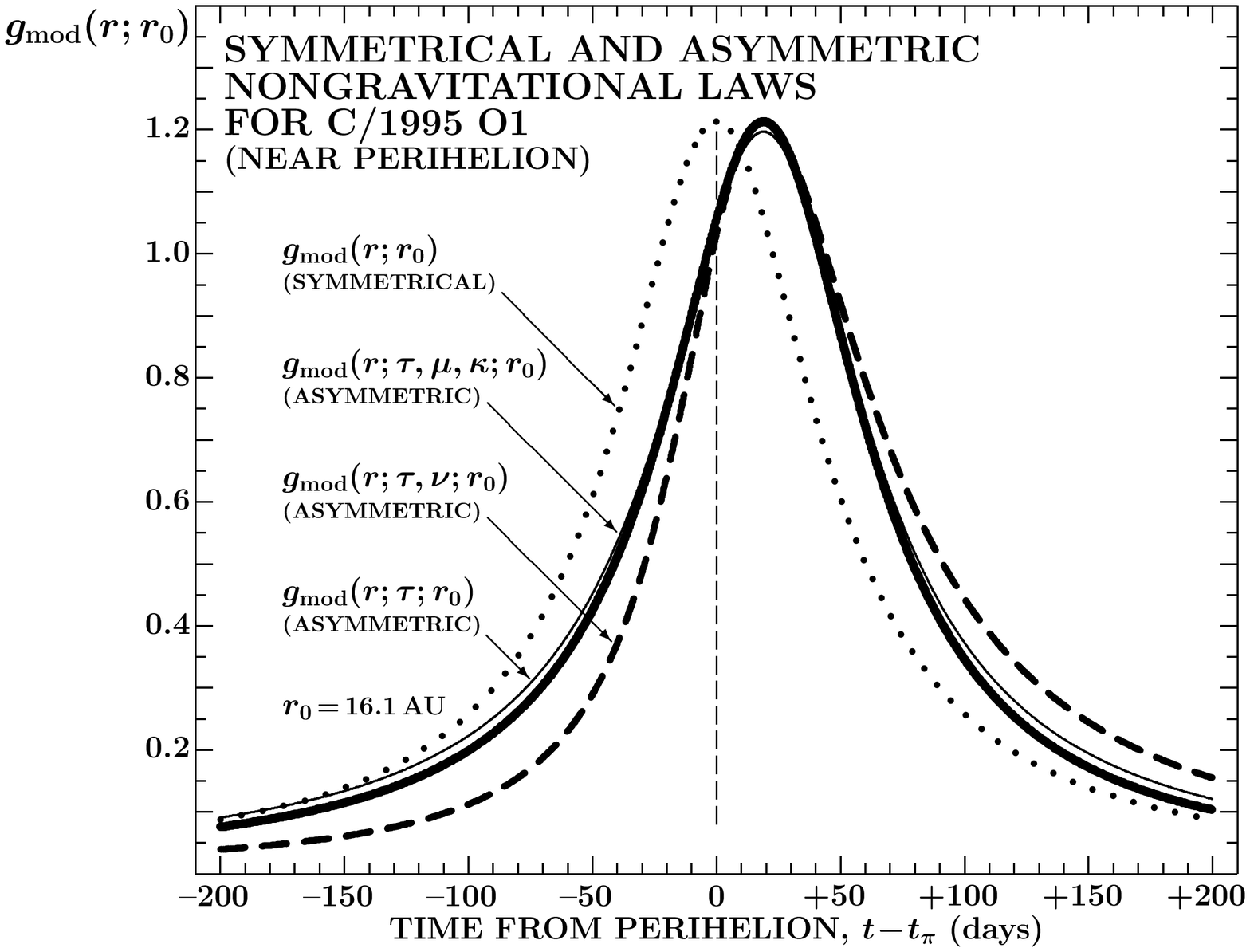}}}  
\vspace{-1.95cm}
\caption{Comparsion, within 200~days of perihelion, of four modified
nongravitational laws, $g_{\rm mod}(r;r_0)$, for comet C/1995~O1, all of them
having the same scaling distance of \mbox{$r_0 = 16.10$ AU}:\ (i)~$g_{\rm
mod}(r;r_0)$, a symmetrical law with respect to perihelion; (ii)~$g_{\rm
mod}(r;\tau; r_0)$, an asymmetric, law with a constant offset time of
\mbox{$\tau = +19$ days}; (iii)~$g_{\rm mod}(r;\tau,\nu;r_0)$, an asymmetric
law with the offset time following a power function \mbox{$\tau_\star =
\tau\,(r_\star/q)^\nu$}, where \mbox{$\tau = +19$ days} and \mbox{$\nu =
1.20$}; and (iv)~$g_{\rm mod}(r;\tau, \mu,\kappa;r_0)$, an asymmetric law
with the offset time following an exponential function \mbox{$\tau_\star =
\tau \exp[\,\mu\,(r_\star - q)^\kappa]$}, where \mbox{$\tau = +19$ days},
\mbox{$\mu = 0.000973$ AU$^{-3}$}, and \mbox{$\kappa = 3$}).  Note that,
within about 0.5~yr of perihelion, the curve presenting the exponential law
--- unlike that for the power law --- nearly coincides with the curve with
a constant offset time, and they both gradually approach the curve of the
nongravitational law symmetrical with respect to perihelion.{\vspace{0.65cm}}}
%
%
\end{figure}

The only purpose of $\tau_{\rm max}$ is to prevent the offset time $\tau_\star$
from reaching unphysically large values far from the Sun.  This is generally
not a problem with power laws (unless $\nu$ is very high), but is critically
important for exponential laws.  For example, for C/1995~O1 the exponential
laws reach \mbox{$\tau_\star = 100$ yr(!)} at \mbox{$r_\star = 34.4$ AU} when
\mbox{$\kappa = 1$}, at \mbox{$r_\star = 23.5$ AU} when \mbox{$\kappa = 2$},
and at \mbox{$r_\star = 20.7$ AU} when \mbox{$\kappa = 3$}, that is, at
distances only slightly to moderately larger than the scaling distance $r_0$.
This means that at distances greater than 20--30~AU the integration of the
comet's motion is carried with a constant offset time.  Of course, the
computed nongravitational effect 30~AU from the Sun reaches merely a
$\sim$10$^{-10}$th part of the effect at perihelion, so truncation of offset
times far from the Sun has no influence on the accuracy~of~the computed
orbit.  Because of a potential for improving the fitting of the motions of
comets, it is advisable that the perihelion-asymmetry option in the expression
for the nongravitational law discussed above be in the future incorporated
in the orbit-determination software.

\begin{figure}[ht]  
\vspace{-4.1cm}
\hspace{-0.15cm}
\centerline{
\scalebox{0.43}{
\includegraphics{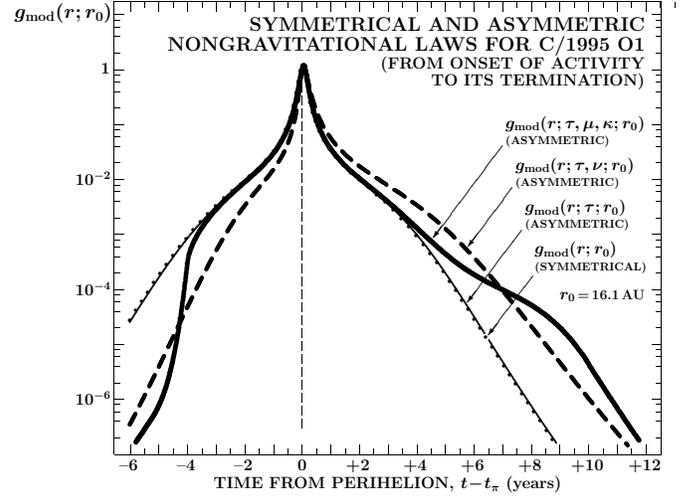}}} 
\vspace{-1.95cm}
\caption{Comparison, from 6~yr preperihelion to 12~yr post-perihelion, of
the four nongravitational laws plotted on a different scale near perihelion
in Figure~11.  Note that only the exponential and power laws fit the major
perihelion asymmetry, whereby the effects 6~yr before perihelion and 12~yr
after perihelion are essentially equal.  The offset times are limited to a
maximum, $\tau_{\rm max}$,~of~3~yr.{\vspace{0.5cm}}}
\end{figure}

%
\subsection{Loss of Mass by Outgassing Integrated\\Over the Orbit.  Trace
 Molecules}
In Sections 9.1 and 9.2 we described the major contributors to the activity of
C/1995~O1 after and before perihelion, respectively.  Our account, based on an
extensive list of referenced work, covered the production of water, carbon
monoxide, and carbon dioxide.  Before we complete the inventory of the volatile
species, we note that in-depth investigations of bright comets lead, virtually
universally, to two major conclusions: (i)~the production rate of water dominates
and (ii)~activity at large heliocentric distances is controled by carbon monoxide.
Although seldom specified, the first of the two conclusions refers to distances
near 1~AU from the Sun, which explains why the two statements are not mutually
contradictory.  As a rule, incomplete data allow to reliably determine neither
the total mass lost by outgassing per orbit, nor the relative contributions to
the total provided by water, carbon monoxide, and other species.  However,
because of the sheer amount of data collected for comet C/1995~O1, its total
outgassed mass can be derived rather accurately for water ice and carbon monoxide
and at least estimated for carbon dioxide and a number of other volatiles.

Figures 9 and 10 suggest that the $g_{\rm mod}(r;r_0)$ law, with an appropriate
choice of the scaling distance $r_0$, offers a satisfactory fit to the variations
in the production rates of carbon monoxide (linked at large heliocentric
distances to the comet's light curve) and water.  If for a particular species
{\sf X} the production rate at 1~AU from the Sun is $Q_0({\sf X})$ (in molecules
per unit time), the loss of mass by this species' outgassing integrated over the
orbit and measured from aphelion to next aphelion, is
\begin{equation}
\Delta {\cal M}({\sf X}) = Q_0({\sf X}) \: \frac{\mu_{\rm mol}({\sf X})}{A}
 \!\! \int_{(P)} \! g_{\rm mod}(r;r_0) \, dt,
\end{equation}
where $\mu_{\rm mol}({\sf X})$ is the species' molar mass, $A$ is the Avogadro
number, $P$ is the comet's orbital period, and $t$ is time.  As described by
Equation~(1) with slight modifications, the law $g_{\rm mod}(r;r_0)$ is a
dimensionless function that is symmetrical with respect to perihelion and
normalized to \mbox{$r = 1\,{\rm AU}$}.  The scaling distance $r_0$ depends
on the species' latent heat of sublimation, $L_{\rm sub}({\sf X})$, as shown
by Equation~(3).  The dimension of the integral expression in Equation~(13)
is time, defining $\Im({\sf X}) = \Im(L_{\rm sub}({\sf X}))$:
\begin{equation}
\Im({\sf X}) = \!\! \int_{(P)} \! g_{\rm mod}(r;r_0) \, dt = \frac{2}{k_{\rm
 grav}\sqrt{p}} \int_{0}^{\pi} \!\! r^2 g_{\rm mod}(r;r_0) \, d\upsilon,
\end{equation}
where \mbox{$p = q(1 \!+\! e)$},{\vspace{-0.05cm}} $q$ is the comet's
perihelion distance, $e$ its orbital eccentricity, \mbox{$k_{\rm grav} =
0.017202099$\,AU$^{\frac{3}{2}}$day$^{-1}$} the Gaussian gravitational
constant, and $\upsilon$ the true anomaly at time $t$. With $r$ in AU,
$\Im({\sf X})$ equates the production of the species {\sf X} integrated
over the orbit with the species' constant production rate, equal to that
at 1~AU from the Sun integrated over a period of $\Im$ days, which we
call an equivalent time.  If the production rate in Equation (13) is
expressed in molecules per unit time, the mass loss of the species
integrated over the orbit becomes
\begin{equation}
\Delta {\cal M}({\sf X}) = Q_0({\sf X}) \,\frac{\mu_{\rm mol}({\sf X})}{A}
 \: \Im({\sf X}).
\end{equation}
With a symmetrical law $g_{\rm mod}(r;r_0)$ one cannot account for the peak
production's offset from perihelion, but the asymmetry can be described
approximately by choosing different values of $Q_0$ and $r_0$ before and
after perihelion, allowing a minor discontinuity at perihelion.  Denoting
the respective production rates at 1~AU by $(Q_0)_{\rm pre}$ and $(Q_0)_{\rm
post}$ and the respective equivalent times by $\Im_{\rm pre}$ and $\Im_{\rm
post}$ with the scaling distances of $(r_0)_{\rm pre}$ and $(r_0)_{\rm post}$,
the total loss of mass of the species {\sf X} per orbit is
\begin{eqnarray}
\Delta {\cal M} & = & \Delta {\cal M}_{\rm pre} + \Delta {\cal M}_{\rm post}
 \nonumber \\[-0.21cm]
 & & \\[-0.21cm]
 & = & \frac{\mu_{\rm mol}}{2A}  \left[ \raisebox{0ex}[1ex][1ex]{}
 (Q_0)_{\rm pre} \Im_{\rm pre} + (Q_0)_{\rm post} \Im_{\rm post} \right] \! .
 \nonumber
\end{eqnarray}

Addressing the perihelion asymmetry in some detail, we first fit the production
rates of water, carbon monoxide, and carbon dioxide in Figures~9 and 10.  One
technique, applied to the production rates of water using~the $g_{\rm mod}(r;r_0)$ 
law, aims at determining both $r_0$ and $Q_0$.  The other technique, applied to
the production rates of carbon monoxide, derives only $Q_0$, using the scaling
distances of 20~AU and 12~AU, already marked in the two figures.  For the
production rates of carbon dioxide, based on very limited data, we use two
approaches:\ the $g_{\rm mod}(r;r_0)$ law with \mbox{$r_0 = 8.9$~AU}, which
satisfies $L_{\rm sub}$ in Table~16 and Equation~(3), and the law intrinsic
to CO$_2$, whose parameters are in Table~16.

The results in Table 19 show that the sublimation of water ice --- while
effectively confined to distances of less than 5~AU from the Sun --- still
prevails after integrating the mass loss over the orbit.  The abundance of
carbon monoxide (the component outgassed from the nucleus) makes up 14--20\%
of water in terms of the production rate at 1~AU from the Sun, but 22--32\%
of water in terms of the mass-loss rate at 1~AU, and it reaches 32\% of~the
total mass of water lost per orbit by the comet's nucleus.  A similar trend
is shown by carbon dioxide.  The summary mass of CO and CO$_2$ lost per orbit
by outgassing from the nucleus amounts to some $\frac{5}{9}$ the mass of
lost water,~so CO and CO$_2$ are hardly minor constituents of the comet's
volatile reservoir.

\begin{table}[t] 
\vspace{-4.07cm}
\hspace{3.91cm}
\centerline{
\scalebox{0.965}{
\includegraphics{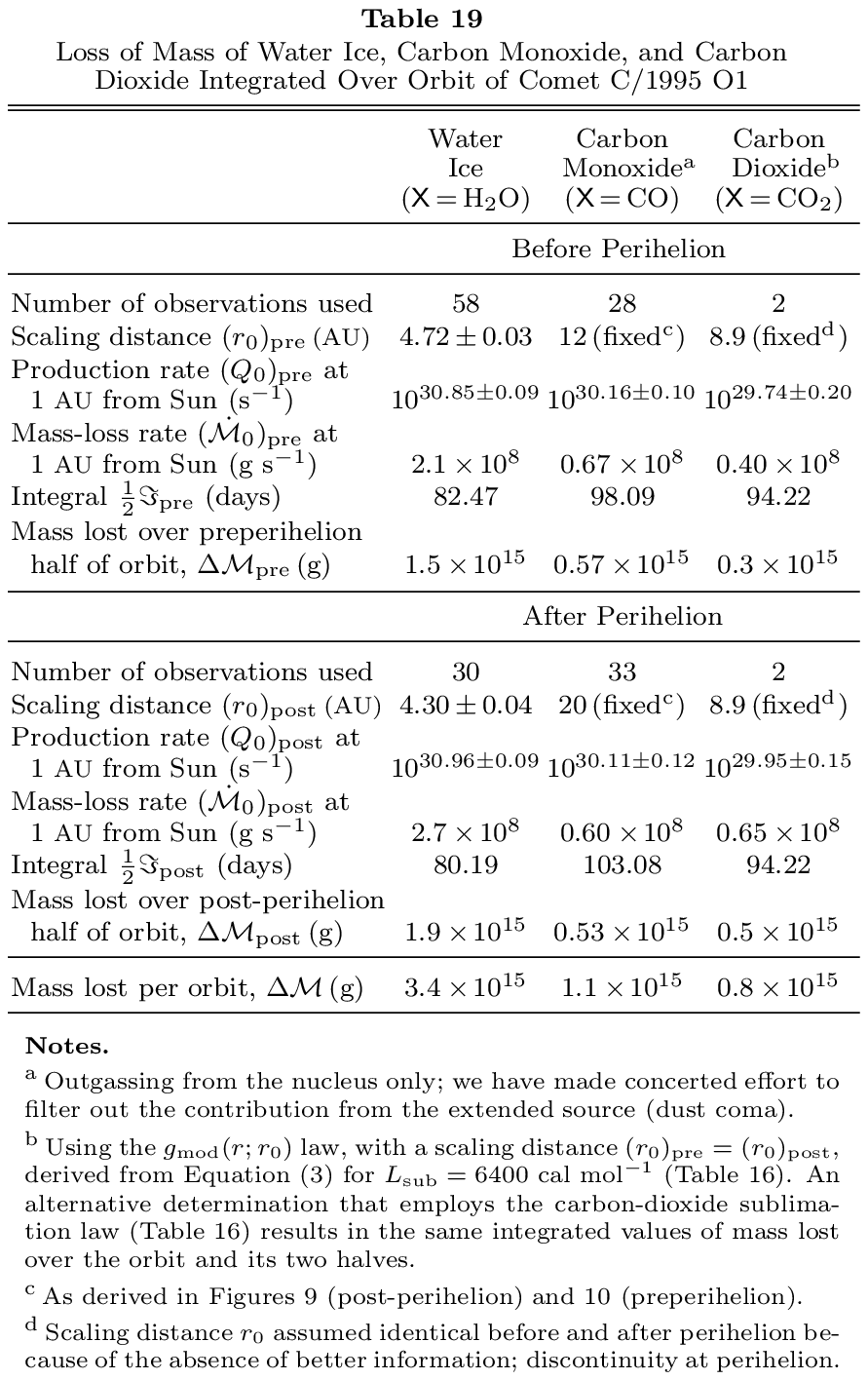}}} 
\vspace{-10.8cm}
\end{table}

Some\,asymmetry\,relative to perihelion is apparent~from the numbers in
Table~19, but not beyond errors.  More water and carbon dioxide was released
after perihelion than before perihelion, but nearly equal parts of carbon
monoxide.  However, the fit in Table~19 ignores the anomalously high rate
measured after perihelion by the ISO.  We already remarked in Section~9.2
on the inconsistencies in the measurements of the preperihelion CO production
rates between 3~AU and perihelion, an effect that complicates the mass-loss
determination.

The level of activity measured by the integrated production of the three
species is impressive, the total output per orbit amounting to \mbox{$5.3
\times \:\!\!  10^{15}$\,g} or the mass of a comet nearly 3~km in diameter
at a bulk density of 0.4~g~cm$^{-3}$.  This mass does not include dust and
a large number of additional volatiles.

The production-rate errors in Table~19 suggest that the results have an
uncertainty of about $\pm$20\% for water, around $\pm$25--30\% for carbon
monoxide, and around $\pm$40\% for the poorly observed carbon dioxide.  The
scaling distance $r_0$ for water ice is distinctly greater than 2.8~AU, an
effect that is clearly apparent from Figure~10.  The discrepancy can be
explained as a corollary to the preferential sunward outgassing; the scaling
distance of $\sim$2.8~AU applies, as noted in the text near Equations~(1) and
(3), to the orbital data fitting isothermal model; it is known (e.g., Sekanina
1988a) that in an extreme case of outgassing from the subsolar area only, the
scaling distance for water-ice sublimation becomes twice as large, 5.6~AU, and
any intermediate value is possible depending on the degree of anisotropy.  The
scaling distance also depends weakly on the Bond albedo and emissivity of the
nucleus' surface.  For carbon monoxide the scaling distance may have to do
with limited volumes of this ice's reservoirs on and below the surface, an
issue that is beyond the scope of this paper, but was discussed at some length
by Szab\'o et al.\ (2011, 2012).  The uncertainty in the scaling distance does
not fundamentally affect the results and it does not significantly increase
the error.

We note that the earliest positive detection of water was achieved when the
comet was 4.6~AU from the Sun, or 0.97$r_0$, on its way to perihelion (Biver
et al.\ 1999), while the last one when the comet was 4.4~AU, or 1.02$r_0$,
on its way out (Biver et al.\ 2002).  It may not be completely a chance that
the most distant detection of water takes place when the comet's heliocentric
distance approximately equals the scaling distance, because at these distances
the $g_{\rm mod}(r;r_0)$ is very steep, always varying as $r^{-13.9}$, so that
a very minor change in $r$ implies a dramatic change in the production rate.
If this rule should apply generally, we predict that the least volatile
species that could be detected in a comet are those for which $r_0 = q\:\!$;
in the orbit of C/1995~O1, {\vspace{-0.03cm}}an upper limit on the sublimation
heat is nearly exactly 20\,000~cal~mol$ ^{-1}$ for the isothermal model, but
slightly more than $\sim$28\,000~cal~mol$ ^{-1}$ for the model with an extreme
outgassing anisotropy (subsolar outgassing).

Although the mass lost per orbit is the prime characteristic of each species'
outgassing, it is the observed production rate near perihelion, or near 1~AU
from the Sun, that is routinely tabulated by the observers.  For species other
than water, it is fairly customary to present their production rates in units
of the water production rate.  Since an output ratio of a species relative
to water depends on whether it is given in terms of the production rate or
the mass-loss rate of the mass loss per orbit, we present the relationships
among these three measures that the isothermal model provides.  Let
$\Re_{\rm Q}({\sf X})$ be the ratio of the production rate of a species
{\sf X} to the water production rate (sometimes also called a relative
abundance) at 1~AU from the Sun,
\begin{equation}
\Re_{\rm Q}({\sf X}) = \frac{Q_0({\sf X})}{Q_0({\rm H}_2{\rm O})} \, ;
\end{equation}
let $\Re_\mu({\sf X})$ be the ratio of the respective molar masses,
\begin{equation}
\Re_\mu({\sf X}) = \frac{\mu_{\rm mol}({\sf X})}{\mu_{\rm mol}({\rm H}_2{\rm
 O})} \, ;
\end{equation}

\noindent
let $\Re_\Im({\sf X})$ be the nominal\footnote{Nominal here meaning for
\mbox{$\Re_{\rm Q} = 1$}.{\vspace{0.05cm}}} orbit-integrated production
(i.e., equivalent time) ratio,
\begin{equation}
\Re_\Im({\sf X}) = \frac{\Im({\sf X})}{\Im({\rm H}_2{\rm O})} \, ;
\end{equation}
and let $\Re({\sf X})$ be the nominal orbit-integrated mass-loss ratio,
\begin{equation}
\Re({\sf X}) = \Re_\mu({\sf X}) \, \Re_\Im({\sf X}) \, ;
\end{equation}
we define $\Re_{\cal M}({\sf X})$, the mass loss of the species {\sf X}
per orbit in units of the mass loss of water ice per orbit,

\begin{equation}
\Re_{\cal M}({\sf X}) = \frac{\Delta {\cal M}({\sf X})}{\Delta {\cal M}({\rm
 H}_2{\rm O})}  ,
\end{equation}
as a product
\begin{equation}
\Re_{\cal M}({\sf X}) = \Re_{\rm Q}({\sf X}) \, \Re({\sf X}) =
 \Re_{\rm Q}({\sf X}) \, \Re_\mu({\sf X}) \, \Re_\Im({\sf X}).
\end{equation}
%

The observers publish $\Re_{\rm Q}({\sf X})$, and it is $\Re({\sf X})$ --- and
therefore $\Re_\Im({\sf X})$ --- that we need to know in order to determine
the mass loss of {\sf X} per orbit relative to water.  Even though the actual
function is unknown, an example with carbon monoxide below illustrates that
application of Equation~(14) with the scaling distance $r_0$ tied to the
sublimation heat $L_{\rm sub}$ by Equation~(3) is robust enough to provide us
with a reasonably informative estimate for the factor that allows us to convert,
according to Equation~(22), the reported production-rate ratio $\Re_{\rm Q}({\sf
X})$ into the physically more meaningful ratio of the mass loss per orbit,
$\Re_{\cal M}({\sf X})$.

\begin{figure*}[t] 
\vspace{-5.47cm}
\hspace{-0.73cm}
\centerline{
\scalebox{0.63}{ 
\includegraphics{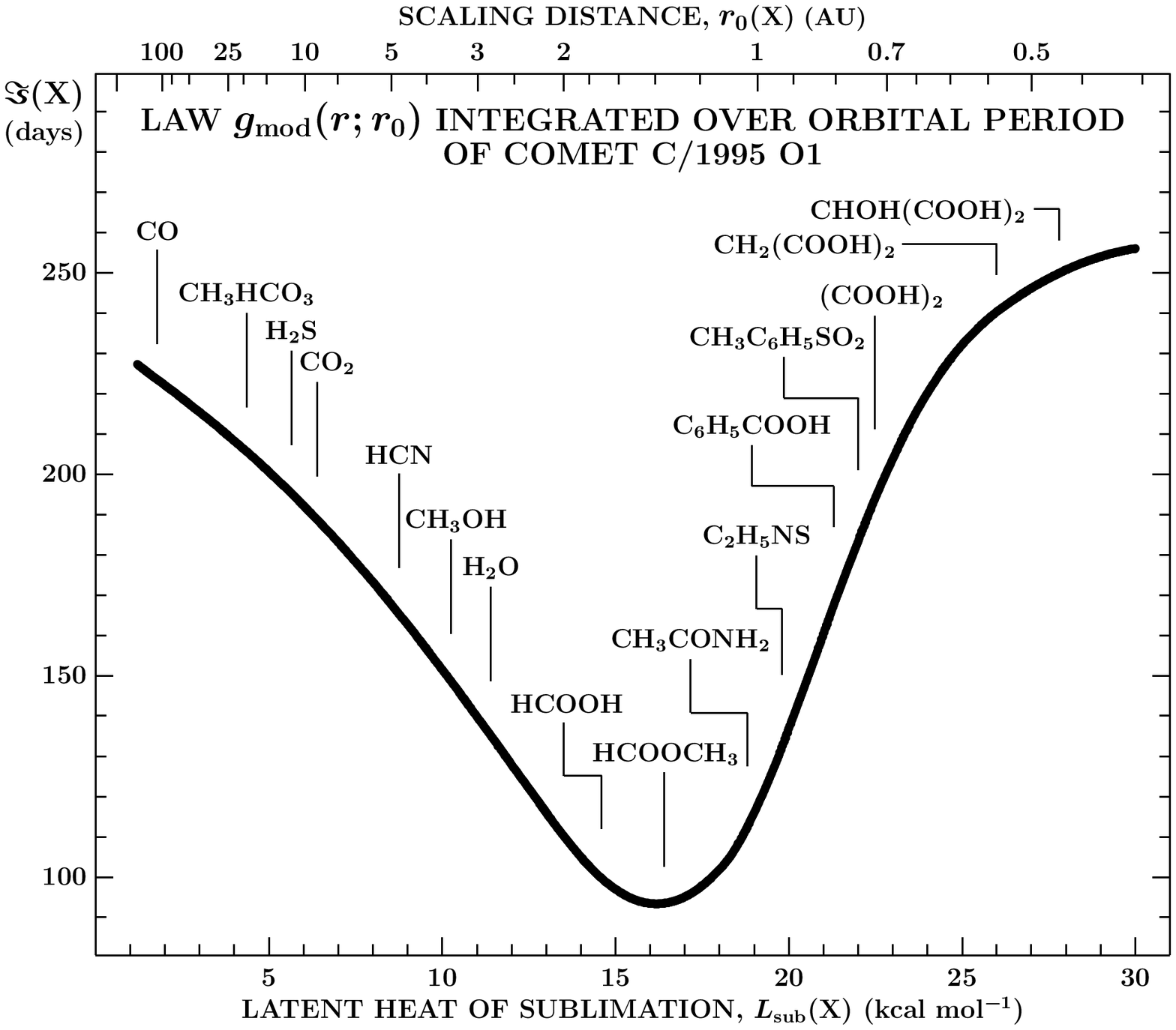}}} 
\vspace{-2.95cm}
\caption{Law $g_{\rm mod}(r;r_0)$, integrated over the orbit of C/1995 O1 as
a function of the latent heat of sublimation $L_{\rm sub}$ that determines
the scaling distance $r_0$ according to Equation~(3), and expressed as an
equivalent time $\Im$,{\vspace{-0.03cm}} defined by Equation~(14).  Contrary
to expectation, the equivalent time $\Im$ reaches a minimum near \mbox{$L_{\rm
sub} = 16\,300$ cal mol$^{-1}$}.  The locations in the graph of 16 selected
volatile species are identified by their chemical formulas.  All molecules
to the left of the curve's minimum and inclusive have been identified in
C/1995~O1, with the exception of methyl hydrogen carbonate, CH$_3$HCO$_3$.
None of the seven molecules to the right of the minimum has been
detected.{\vspace{0.55cm}}}
\end{figure*}

The dependence of the equivalent time $\Im$ on the sublimation heat $L_{\rm
sub}$ in the range of up to 30\,000~cal~mol$^{-1}$ is for an isothermal model
of the nucleus and a perihelion symmetry in outgassing displayed in Figure~13.
The gradual drop in $\Im$ on the left of the figure is understood:\ as $L_{\rm
sub}$ increases, the sublimation is effectively limited to an ever shorter
arc of the orbit around perihelion, so that \mbox{$d\Im/dL_{\rm sub} < 0$}.
However, somewhat {\vspace{-0.035cm}}surprisingly, the curve attains a minimum
at \mbox{$L_{\rm sub} = 16\,300$ cal mol$^{-1}$} and then continues to climb,
reaching the equivalent time of water [\mbox{$\Im({\rm H}_2{\rm O}) \simeq
135$ days}] at \mbox{$L_{\rm sub} \approx 20\,000$ cal mol$^{-1}$}, exceeding
the equivalent time of hyper-volatile species, \mbox{$\lim_{L_{\rm sub}
\rightarrow 0}\Im \simeq 231$ days}, at \mbox{$L_{\rm sub}\!\!$
{\gapeq}25\,000 cal mol$^{-1}$}, and converging to \mbox{$\lim_{L_{\rm sub}
\rightarrow \infty} \Im \simeq 261$ days}.  This behavior can be explained
by an increasing steepness of the sublimation curve at close proximity to
perihelion:\ most mass of such volatiles is not lost until the comet gets
within 1~AU of the Sun and their abundances{\vspace{-0.04cm}} $\Re_{\rm Q}$
diminish progressively with increasing $L_{\rm sub}$.\footnote{Note that
\mbox{$r_0 < 1.37$ AU} for \mbox{$L_{\rm sub} > 16\,300$ cal mol$^{-1}$} and
that \mbox{$r_0 < 0.58$ AU} for \mbox{$L_{\rm sub} > 25\,000$ cal mol$^{-1}$};
thus, even perihelion of C/1995~O1 is in a thermal regime dominated by
reradiation, with only a minor fraction of the solar energy spent on
sublimation of species with such a high sublimation heat.  Of course, the
$g_{\rm mod}(r;r_0)$ law is then only an approximation to the genuine
sublimation law, which is exponential.{\vspace{-0.05cm}}}  This trend is
consistent with our independent conclusion above that outgassing of species
with a sublimation heat exceeding 20\,000~cal~mol$^{-1}$ should be
increasingly rare in C/1995~O1 and that species {\vspace{-0.03cm}}with the
sublimation heat near or greater than $\sim$28\,000~cal~mol$^{-1}$ should
not be detected at all.  The shape of the curve $\Im(L_{\rm sub})$ depends
very strongly on the perihelion distance (Appendix B).

In an effort to assess a cumulative contribution~to~the orbit-integrated mass
loss by various species, we list in Table 20 the parent molecules observed
in C/1995~O1, as compiled by Bockel\'ee-Morvan et al.'s~(2004), and compare
them with a set of selected entries~from Acree \& Chickos' (2016) compendium
of organic and organo\-metallic compounds, whose latent heat of sublimation
(sublimation enthalpy) does not, in concert{\vspace{-0.03cm}} with the
constraints above, exceed 30\,000~cal~mol$^{-1}$.

The table is ordered by the latent heat~of~sublimation, $L_{\rm sub}$, which is
known to~be only a weak function of temperature.  For most entries the listed
values of $L_{\rm sub}$ were compiled by Acree \& Chickos (2016) in their
compendium from an extensive set of sources.  In the majority of cases the
results from different sources for different temperature ranges agreed with
each other quite well, often within a few percent or so.  For some molecules,
for which the sublimation heat was not listed by Acree \& Chickos (2016), we
were able to learn the information from the chemistry webbook of the National
Institute of Standards and Technology~(NIST).\footnote{For access, see {\tt
http://webbook.nist.gov/chemistry/.}} For several tabulated species, mostly
those observed in C/1995~O1, the sublimation heat apparently has not been
determined; in the majority of these instances its value could be approximated
by the sum of values of the heat of vaporization and the heat of fusion, if
both known (though usually referring to different temperatures) and is
then parenthesized.  For only three tabulated molecules we were able to find
only the heat of vaporization, which provides a lower limit to the heat of
sublimation; the heat of fusion typically amounts to only a minor fraction of
the heat of sublimation.

Highly relevant to the determination of the total mass loss by outgassing are
columns 5--7 of Table~20, in which we list, respectively, the water-normalized
ratios $\Re_\mu$, $\Re_\Im$, and $\Re$, the last one being the factor that
converts the production rate to the mass loss integrated over the orbit.  For
example, Bockel\'ee-Morvan et al.\ (2004) state that the production rate of
carbon monoxide from the nucleus~was 12\% of the water rate.  Table~20 lists
\mbox{$\Re = 2.56$} for carbon monoxide, which suggests that its mass loss
per orbit was 12\% times 2.56 = 31\%, in excellent agreement with the result
in Table~19, which gives 32\% of the water~mass~loss.

\begin{table*}[t] 
\vspace{-4.2cm}
\hspace{-0.4cm}
\centerline{
\scalebox{1}{
\includegraphics{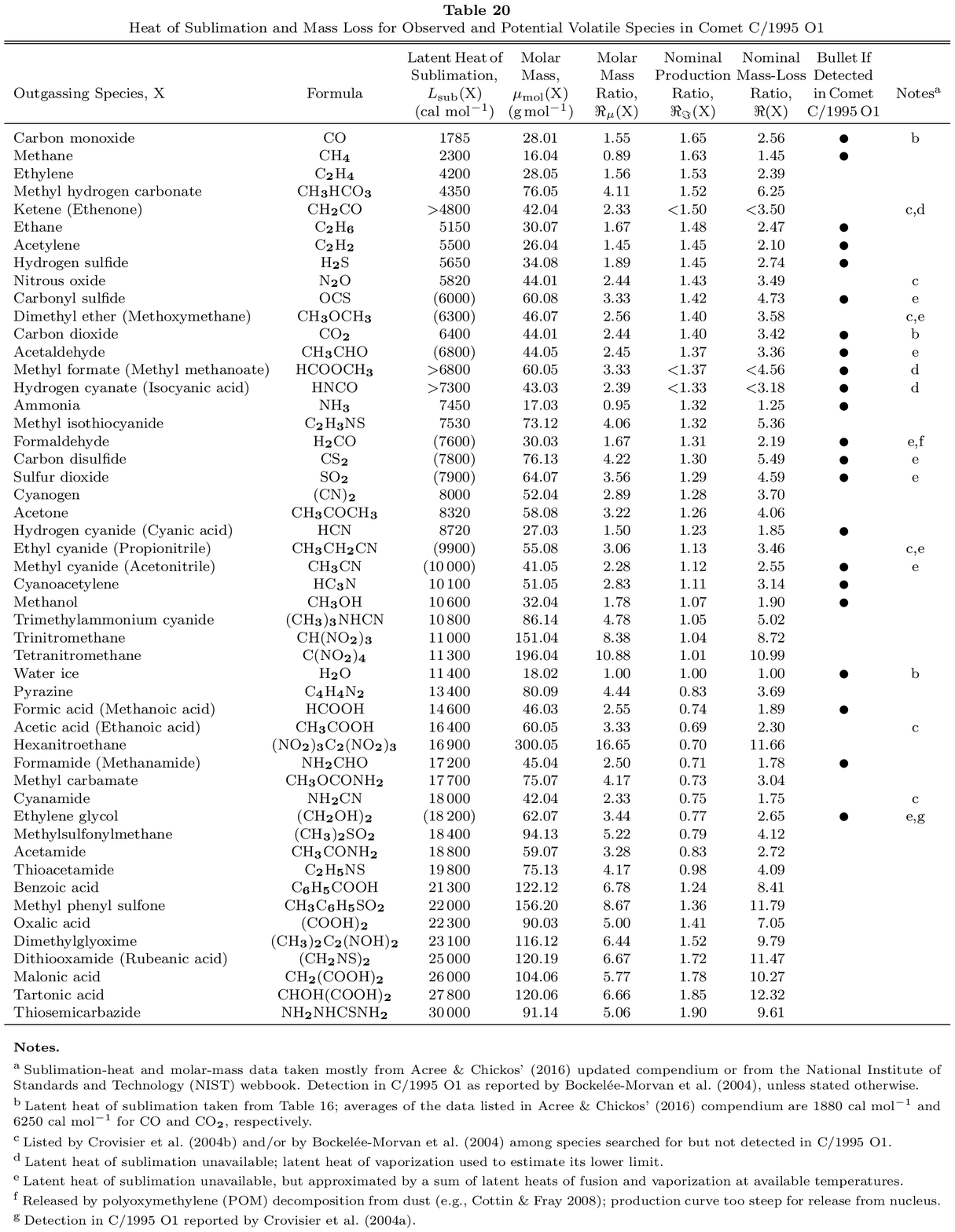}}} 
\end{table*}

Also apparent from Table~20 is that most species detected in C/1995~O1 were
more volatile than water, including carbon monoxide (the most volatile one)
and carbon dioxide.  However, three of the detected molecules were more
refractory than water; the most refractory one, ethylene glycol,\footnote{Its
detection, not included in Bockel\'ee-Morvan et al.'s (2004) compilation, was
reported by Crovisier et al.\ (2004a).{\vspace{0.13cm}}} has an estimated
sublimation heat of $\sim$18\,000~cal~mol$^{-1}$, somewhat below an upper
limit on the sublimation heat for an isothermal model.

We note from Table~20 that the nominal mass-loss ratio $\Re$ is always greater
than unity.\footnote{The three upper limits are meant to indicate that the
probable values should be some 10--20\% lower, still much greater than
unity.{\vspace{-0.15cm}}} Of the two factors that contribute to $\Re$,
$\Re_\Im$ has a generally minor effect, varying between about 0.7 and 1.9.
By contrast, $\Re_\mu$ ranges from 0.9 to nearly 17.  The heaviest molecule
detected in C/1995~O1 was carbon disulfide with \mbox{$\Re_\mu = 4.22$},
followed by sulfur dioxide (\mbox{$\Re_\mu = 3.56$}) and ethylene glycol
(\mbox{$\Re_\mu = 3.55$}).  Since the first two are more volatile than the
last one, their values of $\Re$ are proportionally higher.  Table~20 lists
eight species whose molar mass is more than six times as high as that of
water (\mbox{$\Re_\mu > 6$}), of which three are more volatile than the
two least volatile molecules detected in C/1995~O1.  Relative to their
contribution to the total production rate, such species are the most
effective contributors to the comet's total mass lost by outgassing.

\begin{table}[t] 
\vspace{-4.19cm}
\hspace{4.17cm}
\centerline{
\scalebox{0.992}{
\includegraphics{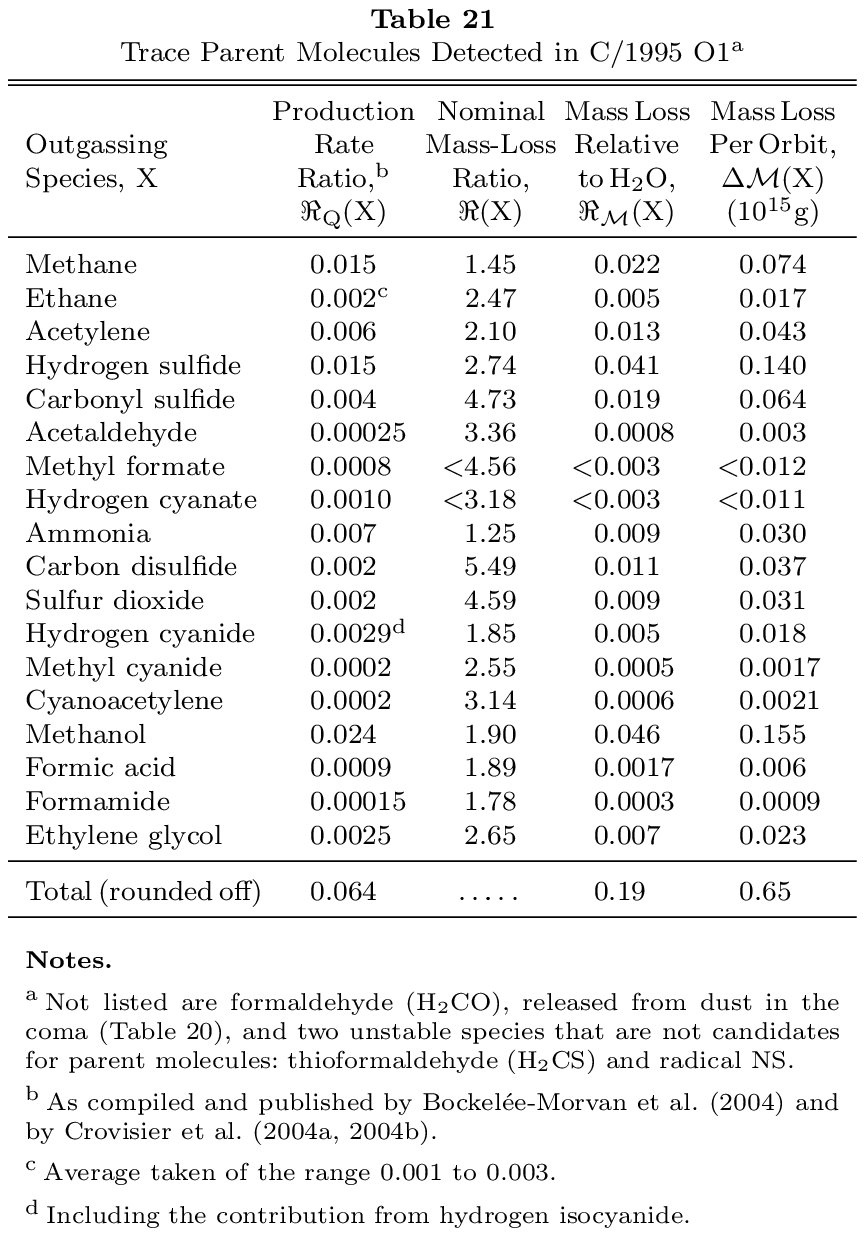}}} 
\vspace{-12.75cm}
\end{table}

The fact that the mass loss per orbit of any of the listed molecules is,
relative to water, greater than the production rate near 1~AU is consistent
with what we concluded about carbon monoxide and carbon dioxide in connection
with Table~19.  If the production rates for the trace molecules, presented
by Bockel\'ee-Morvan et al.\ (2004), refer {\vspace{-0.035cm}}essentially to
1~AU from the Sun, the mass of each of the 18 parent species\footnote{Excluding
formaldehyde, probably released from dust grains in the comet's atmosphere (e.g.,
Cottin \& Fray 2008), as well as thioformaldehyde and a radical NS (unstable
compounds that are not candidates for a parent molecule), deuterated species,
and ions.} lost over the orbit of C/1995~O1 is listed in Table~21.  By summing
up the data, we show that while their total production rate equaled merely 6.4\%
of the water production rate, their total mass lost per orbit amounted to 19\%
of the water mass loss, thus averaging $\sim$1\% per trace species.  Together
with carbon monoxide and carbon dioxide (Table~19), this makes up 75\% of
the water mass loss.  

\begin{table*}[t] 
\vspace{-4.19cm}
\hspace{-0.52cm}
\centerline{
\scalebox{1}{
\includegraphics{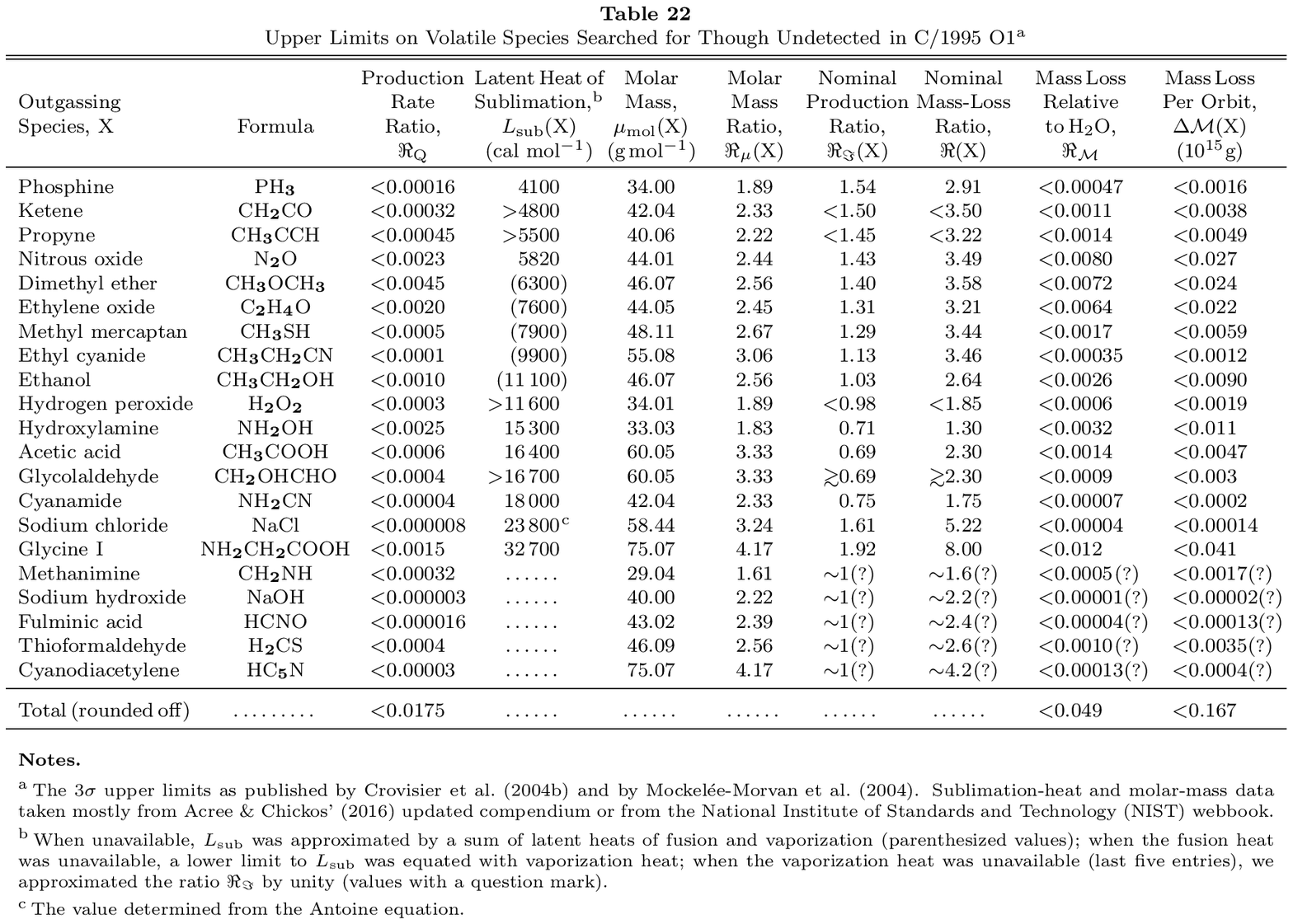}}}  
\vspace{-11.91cm}
\end{table*}

Additional molecules that were searched for, but not detected, in C/1995~O1
(Crovisier et al.\ 2004b, Bockel\'ee-Morvan et al.\ 2004) are listed in
Table~22, which again shows that an upper limit on the production rate is
much tighter than the limit on the orbit-integrated loss of mass by outgassing.
The latter is 3 times higher than the former in Table~21 and 2.8 times higher
in Table~22.

Since the parent molecules detected in C/1995~O1 (as well as in other comets)
are primarily hydrocarbons, often oxygen, nitrogen, and/or sulfur-bearing,
one wonders --- given the enormous variety of compounds into which these
elements can combine --- how many more similar, fairly volatile species
are there still waiting to be discovered in comets and, in particular, what
is their total mass relative to the mass of the water-ice reservoir.  This
question prompted Crovisier et al.\ (2004b) to inspect the distribution of
the relative abundances of more than 20 detected species, that is, their
cumulative number, ${\cal N}$, as a function of their relative production
rate, $\Re_{\rm Q}$.  Their plot shows the{\vspace{-0.04cm}} dataset to
have a tendency to vary as a power law, \mbox{${\cal N} \:
\mbox{\raisebox{0.3ex}{\scriptsize $\propto$}} \: \Re_{\rm Q}^{-0.4}$}, and
{\vspace{-0.1cm}}suggests that the dataset is increasingly incomplete at
\mbox{$\Re_{\rm Q} < 10^{-3}$}.  Crovisier et al.\ (2004b) pointed out that
this effect appeared to tie in with the well-known existence of numerous
unidentified features in diverse parts of cometary spectra.  

\begin{figure}[b] 
\vspace{-11.4cm}
\hspace{1.82cm}
\centerline{
\scalebox{0.79}{
\includegraphics{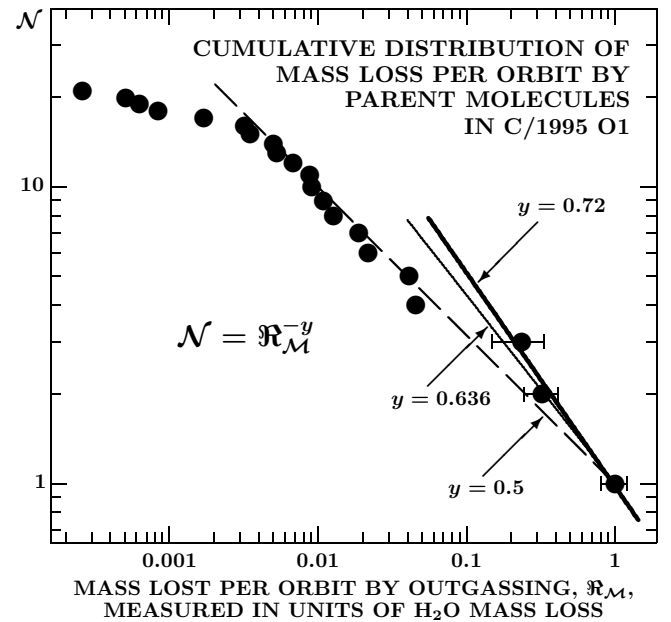}}} 
\vspace{-3.78cm}
\caption{Cumulative distribution of the orbit-integrated mass loss by the
outgassing parent molecules detected in C/1995~O1.  The number of species,
${\cal N}$, is plotted against the mass lost~by~them over the orbit,
$\Re_{\cal M}$, measured in units of the mass lost by the sublimation of
water ice, \mbox{$(0.34 \pm 0.07) \times \! 10^{16}$g}.  The total
mass{\vspace{-0.03cm}} lost per orbit by all volatile species is
\mbox{$\Re_\infty = y \, (1\!-\!y)^{-1}$}, where $-y$ is the slope of the
cumulative distribution.  In order that $\Re_\infty$ be finite, it must
be \mbox{$y < 1$}.  The thick line, \mbox{$y = 0.72$}, is a fit through the
three most abundant species (water, carbon monoxide, and carbon dioxide);
the thin line, \mbox{$y = 0.636$}, is a lower limit to the power law, based
on the number of parent species that are currently known to outgass from
the comet's nucleus; and the dashed line, \mbox{$y = 0.5$}, is a constraint
required by an extreme case, when water outgassing accounts for all the lost
mass, \mbox{$\Re_\infty = 1$}.{\vspace{-0.09cm}}}
\end{figure}

As we are primarily interested in estimating a total mass lost by outgassing
per orbit of C/1995~O1, we prepared in Figure~14 a plot of the distribution
of the orbit-integrated mass loss, $\Delta {\cal M}$, of the species presented
in Tables~19 and 21.  The dependence of a cumulative number ${\cal N}$ of the
species on the ratio $\Re_{\cal M}$ expresses the mass loss per orbit in
units of the total mass loss of water ice.  Whereas $\Re_{\cal M}$ has up
to now been a function of the species {\sf X}, we now treat the cumulative
distribution as a continuous function of $\Re_{\cal M}$ and note that the mass
loss in an interval between ${\cal N}$ and \mbox{${\cal N} \!+\! d{\cal N}$}
is \mbox{$\Re_{\cal M}({\cal N}) \, d{\cal N}$}.  We assume the function
${\cal N}$ to vary as a power of $\Re_{\cal M}$ (with \mbox{${\cal N} \!=\!
1$} for \mbox{$\Re_{\cal M} \!=\! 1$}),\\[-0.2cm]
\begin{equation}
{\cal N} = \Re_{\cal M}^{-y},\\[-0.1cm]
\end{equation}
so that \mbox{$d{\cal N} = -y \Re_{\cal M}^{-(y+1)}d \Re_{\cal M}$}, and
for the total mass loss per orbit experienced by all outgassing species,
$\Re_\infty$, we get
\begin{equation}
\Re_\infty \!=\!\! \int_{1}^{\infty} \!\!\! \Re_{\cal M} \,d{\cal N} =
 -y \!\! \int_{1}^{0} \!\! \Re_{\cal M}^{-y} \, d\Re_{\cal M} = \frac{y}{1
 \!-\!y},
\end{equation}
where \mbox{$0.5 < y < 1$}.  The limits on $y$ are dictated by
requiring that $\Re_\infty$ be finite, positive, and greater than unity.
The infinite upper limit in the first integral implies merely a very large
set under consideration that includes species with the abundances many orders
of magnitude lower than the abundance of water, so that \mbox{${\cal N}
\rightarrow \infty$} when \mbox{$\Re_{\cal M} \rightarrow 0$}, in compliance
with Equation~(23).

A tighter condition for $y$ follows from our above finding that the mass lost
per orbit by all known outgassing species, including water, was 175\% of the
mass loss of water, which implies \mbox{$y = 0.636$} for the lower limit to
$y$, as depicted in Figure~14.  Since most data points in the figure
congregate along \mbox{$y \simeq 0.5$}, their distribution is inconsistent
with the required law and suggests that the set is substantially incomplete.
A fit through the three most abundant species --- water, carbon monoxide, and
carbon dioxide --- predicts that \mbox{$y = 0.72 \pm 0.12$}.  Combining the
constraints, a probable range for the relative loss of mass by outgasssing per
orbit of C/1995~O1 amounts to \mbox{$2 \mbox{\lapeq} \Re_\infty \mbox{\lapeq}
5$}, {\vspace{-0.04cm}}so that water ice may account for as~\mbox{little} as
$\sim$20\% of the orbit-integrated outgassed mass.  If the cumulative
distribution of sublimating species does follow a power law, their number
with mass loss per orbit greater than 10\% of the water mass loss is
predicted to be between four and seven (compared to three known); those
with $\Re_{\cal M}$ greater than 1\%, between 21 and 46 (compared to 9
known, an incompleteness of more than 50\%); those with $\Re_{\cal M}$
greater than 0.1\%, between 100 and $\sim$320 (compared to 17 known, an
incompleteness of 83--95\% or a factor of about 6 to 19); and those with
$\Re_{\cal M}$ greater than 0.01\%, between $\sim$460 and $\sim$2150
(compared to 21 known, an incompletness by a factor of 22 to 100).  For
this last $\Re_{\cal M}$ limit, the number of species to outgass from the
nucleus equals $\sim$760, if \mbox{$y = 0.72$}.

These findings suggest that in terms of the total loss of mass from
C/1995~O1 by outgassing, water ice contributes at best about 50\% and may
contribute as little as 20\%, although this last estimate is only a
crude limit.  A broad variety of compounds, mostly organic, account for
the rest; many of them may have a molecular mass substantially greater
(\mbox{$\Re_\mu \!\gg\! 1$}), and their equivalent sublimation time longer
(\mbox{$\Re_\Im \!>\! 1$}), than water, so that summarily they could
contribute to the comet's total mass loss per orbit by outgassing much
more than implied by their production-rate ratio $\Re_{\rm Q}$.  In
addition, one cannot rule out that C/1995~O1 carried{\vspace{-0.04cm}}
many hundreds of parent molecules with a mass-loss ratio \mbox{$\Re_{\cal
M} \!>\! 10^{-3}$} and perhaps more than a thousand with \mbox{$\Re_{\cal
M} \!>\! 10^{-4}$}.

We next compare this predicted variety of species in C/1995~O1 with the
number of diverse molecular formulas generated as a sum of all possible
combinations~of specific elements and subject to prescribed assumptions.
Disregarding for a moment the chemical issues and excluding the homonuclear
molecules, the number of mathematically possible compounds{\vspace{-0.06cm}}
is a function of (i)~a number $N$ of the elements, \mbox{${\cal E}^{(i)}
\,(i \!=\! 1, 2,\ldots, N$)}, that make~up the molecules and (ii)~maximum
numbers{\vspace{-0.09cm}} $K_i$ of atoms, \mbox{${\cal E}_{k_i}^{(i)} \,
(k_i \!=\!  0, 1,\ldots, K_i)$}, {\vspace{-0.04cm}}that each of the elements
contrib\-utes to the molecules.  A general chemical formula of such compounds
is
\begin{equation}
{\cal E}_{k_1}^{(1)} {\cal E}_{k_2}^{(2)}\ldots\,{\cal E}_{k_N}^{(N)} \!; \:\;(i
 \!=\! 1, 2, \ldots, N; \; k_i \!=\! 0, 1, \ldots, K_i),
\end{equation}
where \mbox{$k_i \!=\! 0$} means that an $i$-th element does not contribute to
the particular molecule.  However, the exclusion of the homonuclear molecules
requires that there always be at least two elements (\mbox{$N \!>\! 1$}), $i$
and \mbox{$j \!\neq\! i$}, in any molecule for which \mbox{$k_i \!>\! 0$}
and \mbox{$k_j \!>\!  0$}.  The complete set of compounds is symbolically
denoted as
\begin{equation}
\left[{\cal E}_{K_1}^{(1)} {\cal E}_{K_2}^{(2)}\ldots\,
 {\cal E}_{K_N}^{(N)}\right]
\end{equation}
and the number ${\cal N}$ of mathematically possible molecular compounds is
\begin{equation}
{\cal N}=\!\prod_{i=1}^{N}(K_i\!+\!1)-\!\left(\!1 \!+\!\sum_{i=1}^{N}
 K_i\!\right).
\end{equation}
The molecules consisting of $N\!-\!1$, $N\!-\!2$, \ldots , 2 elements that
make up subsets of the set (26) are included~in~${\cal N}$.\footnote{If
homonuclear molecules should be included among the mathematically possible
compounds, then \mbox{$N \!=\! 1$} is allowed and Equation~(27) changes to
\mbox{${\cal N} = \Pi_{\scriptscriptstyle i=1}^{\scriptscriptstyle N} (K_i
\!+\! 1) \!-\! (N \!+\! 1)$}.}  Note that for \mbox{$K_1\!=\!K_2\!= \ldots
=\!K_N\!=\!K$} Equation~(27) {\vspace{-0.04cm}}is simplified to \mbox{${\cal
N}\!=\!(K\!+\!1)^N\!-\!(1\!+\! K\:\!\!N)$}.  We next illustrate this exercise,
including the chemical properties of the resulting compounds, on an example
of a molecular set consisting of three elements, each of which contributes
up to two atoms per molecule, [C$_2$H$_2$O$_2$] (\mbox{$N \!=\! 3$};
\mbox{$K_{\rm C} \!=\! K_{\rm H} \!=\! K_{\rm O} \!=\! 2$}), in Table~23;
{\vspace{-0.04cm}}we show that, of the total number of compounds,
\mbox{${\cal N} = (2 \!+\! 1)^3 \!-\! (1 \!+\! 2 \!\times\!3) = 20$},
fully 16 (80\%) have been detected outside the Solar System, whereas 8
(40\%) have been observed in C/1995~O1, five (25\%) of them being parent
molecules and the rest dissociation products.  Two molecules seen in
interstellar or circumstellar space have not been detected in the comet
in spite of attempts to do so. Among the four~compounds not observed in
either environment is glyoxal;\footnote{From our standpoint, glyoxal is
the most interesting among three C$_2$H$_2$O$_2$ isomers; the other two
are acetylenediol HOCCOH and acetolactone H$_2$COCO.} two of the three
remaining ones --- ethylenedione and ethynediolide (together with its isomer
glyoxalide) --- are derivatives of glyoxal (Dixon et al.\ 2016).  The spectrum
of solid glyoxal was found to have fundamental bands between 3.5~$\mu$m and
18.2~$\mu$m (Durig \& Hannum 1971).  The last undetected molecule is the
hydrocarboxyl functional group COOH, of fleeting existence as a separate
compound; it is contained in all carboxylic acids, one of which (formic
acid) is listed in Table~23, while another (glyoxylic acid) is related to
glyoxal; the oxalic acid has the formula of the group's dimer (Table~20).

\begin{table}[t] 
\vspace{-4.2cm}
\hspace{4.21cm}
\centerline{
\scalebox{1}{
\includegraphics{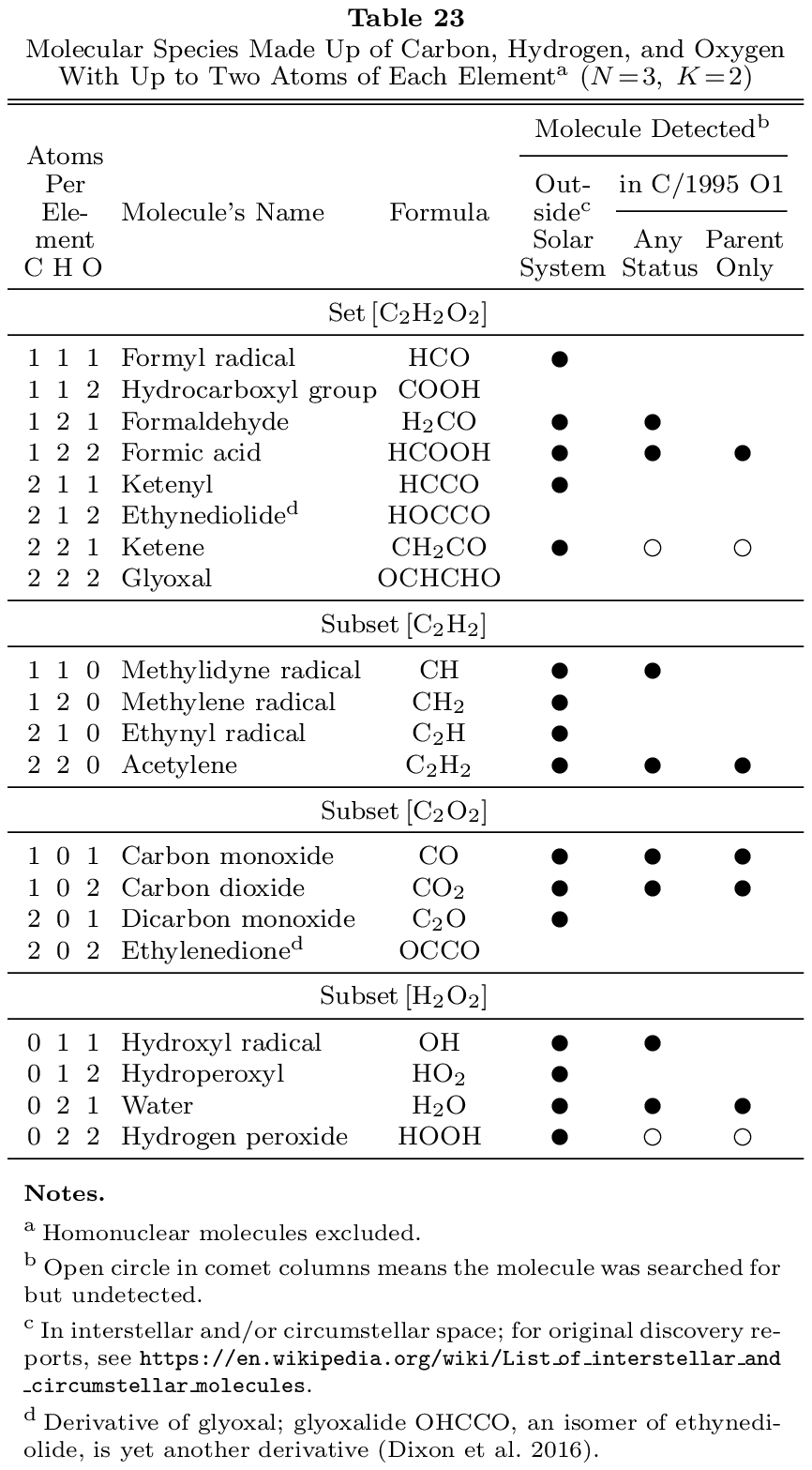}}} 
\vspace{-9.3cm}
\end{table}

A total number of molecules detected in the interstellar medium up to now
is about 200 --- nearly 10 times the number of parent molecules observed
{\vspace{-0.04cm}}in C/1995~O1~--- with a peak molar mass well over
100~g~mol$^{-1}$, either as gas or as ice.  Some are radicals that could
not exist in a cometary nucleus without being bound chemically to other,
more stable ices.  A majority of interstellar molecules are made up of up
to five elements:\ carbon, hydrogen, oxygen, nitrogen, and sulfur.  The
number of hydrogen and carbon atoms seldom exceeds six each per molecule;
oxygen rarely contributes more than two to three atoms per molecule, while
the other elements mostly not more than one atom per molecule (see footnote
c to Table~23).  Under these conditions, a mathematically complete set,
[C$_6$H$_6$O$_{2-3}$NS], consists of \mbox{${\cal N}\!\approx\! 570$--770}
diverse molecules.  On the one hand, some of these formulas might not
correspond to any~\mbox{chemical} at all, others may not have the necessary
thermophysical properties.  On the other hand, this range of numbers does not
account for isomers (and polymers),~whose~tally should increase rapidly with
both $N$ and $K_i$.  In fact, isomers exist even among the compounds presented
in Table~23.  In addition to the isomers of glyoxal, referred to above, there
is hydroxymethylene (HCOH), an isomer of formaldehyde; ethynol (HC$_2$OH), an
isomer of ketene; and dioxyrane (H$_2$CO$_2$), an isomer of the formic acid.

If one expects a degree of analogy between the presence of complex molecules
in the interstellar medium and in comets, and detection incompleteness by
a factor of three to four in the interstellar medium on the assumption
that a great majority of compounds complies with a symbolical formula
[C$_6$H$_6$O$_{2-3}$NS] in Equation~(26), then a degree of detection
incompleteness of parent molecules in a comet like C/1995~O1 is estimated
at a factor of $\sim$20--30 or even more.  It is worth remarking that the
estimate for interstellar molecules is in line with our nominal estimate
of $\sim$760~volatile species in C/1995~O1 whose mass loss is greater than
0.01\% of the water mass loss, based on the data in Figure~14.  In any
case, the present inventory of parent molecules in C/1995~O1 appears to be
extremely incomplete, by a factor much greater than 10, and the mass loss
of water is quite probably less than 50\% of the total outgassed mass lost
per orbit.  The conclusion on the existence of hundreds of very complex
molecules that contribute to the process of outgassing from comets is also
broadly consistent with very recent results from the Rosetta Mission, which
suggest that carbon in dust particles from comet 67P/Churyumov-Gerasimenko
is bound in organic matter of high molecular weight (Fray et al.\ 2016)
and which support the presence of polyoxymethylene (POM) and other polymers
in the dust grains as a plausible source of organic matter in gaseous phase,
such as formaldehyde (Wright et al.\ 2015).  A potential presence of
polycyclic aromatic hydrocarbons (PAHs), such as phenanthrene, anthracene,
etc., as discussed by Bockel\'ee-Morvan et al.\ (2004) and others, would
only strengthen this general conclusion.

As a caveat, the degree of approximation to the law of outgassing by the
isothermal model is open to question, as the nominal orbit-integrated production
ratio $\Re_\Im$ depends on the distribution of the sublimation rate over~the
nucleus.~For a distribution other than that conforming~to the isothermal model,
the constant in Equation~(3){\vspace{-0.07cm}} exceeds 19\,100~AU$^{\frac{1}{2}}$
cal mol$^{-1}$, a maximum of 27\,000~AU$^{\frac{1}{2}}$ cal mol$^{-1}$ being
reached when the sublimation proceeds from a subsolar area only.  The scaling
distances for the water sublimation curves in Table~19 would be fitted best
by{\vspace{-0.05cm}} choosing for the constant the~values~of~24\,800~AU$^{
\frac{1}{2}}$ cal mol$^{-1}$ before and 23\,600~AU$^{\frac{1}{2}}$ cal
mol$^{-1}$ after perihelion.  Nonetheless, use of the isothermal model is
warranted by its~successful application in orbital studies and by expectation
that the scaling distances for other molecules are likely to be affected
similarly, resulting in Table~21 in a net $\Re_\Im$ effect close to nil except
for the species whose $L_{\rm sub}$ exceeds $\sim$16\,000 cal mol$^{-1}$, i.e.,
formamide and ethylene glycol, whose contributions would moderately drop.

\subsection{Production of Dust and Its Mass Loading of\\the Gas Flow}
Further insight into the issue of the mass lost over the orbit by outgassing is
provided by investigating the mass-loss rates of dust and by examining the
terminal velocities of dust particles.  Since it is well known that, owing to a
fairly flat particle-size distribution, the large-sized dust (submillimeter and
larger grains) provides a dominant contribution to the total particulate mass
lost, we focus on the investigations of C/1995~O1 based on~imaging observations
at submillimeter and longer wavelengths.

Jewitt \& Matthews (1999) observed the comet~at~850 $\mu$m on 16 dates in 1997
between February~9~(1.27~AU~pre\-perihelion) and October 24{\vspace{-0.03cm}}
(3.17~AU post-perihelion), deriving {\vspace{-0.05cm}}a~dust~mass-loss~rate of
\mbox{$(1.6 \pm 0.5) \times\!10^9$}\,g~s$^{-1}$ at~1~AU before perihelion and
\mbox{$(1.8 \pm 0.4) \times\!10^9$}\,g~s$^{-1}$ at 1~AU after perihelion, and
an orbit-integrated mass loss of \mbox{$3 \times\! 10^{16}$}\,g, equivalent
to the mass of a comet more than 5~km in diameter, at an assumed density of
0.4~g~cm$^{-3}$.  They employed a blackbody approximation and a model of
an optically thin, spherically symmetrical, steady-state coma, by-passing
the particle-size distribution function by adopting a dust opacity of
0.55~cm$^2$~g$^{-1}$ at 850~$\mu$m.  They equated the mass production rate
with a ratio of the particles' mass to their residence time in the beam.
However, the residence time depends not only on the ejection speed of the
particles (and therefore on their size), but also on the direction of their
motion relative to the observer.  In particular, particles moving in directions
close to the line of sight have a residence time much longer than particles
moving perpendicular to the line of sight, a scenario considered by Jewitt \&
Matthews (1999).  Sekanina \& Kracht (2014) developed a technique that properly
accounts for the effect of residence time in a spherically symmetrical,
steady-state coma environment, also employing a more realistic approximation
for the particle velocity.  This technique relates a total mass production
rate of dust at time $t$,
\begin{equation}
\dot{\cal M}_{\rm d}(t) \equiv \dot{\cal M}_{\rm d}(s_0,s_\infty) = \!
 \dot{N}_{\rm d}(t) \!\! \int_{s_0}^{s_\infty} \!\!\! {\textstyle
 \frac{4}{3}} \pi \rho_{\rm d} s^3 f_{\rm d}(s) \, ds,
\end{equation}
and a geometric cross-sectional area of all dust particles (of assumed
sphericity) ejected from the nucleus that at time $t$ reside in a beam whose
radius at the comet is $r_{\rm b}$,
\begin{equation}
X_{\rm d,r_{\rm b}}\!\equiv\!X_{\rm d,r_{\rm b}}\:\!\!(s_0,\:\!\!s_\infty)\!
 =\!\!\! \int_{s_0}^{s_\infty} \!\!\!\!\!\! \pi s^2 f_{\rm d}(s)\,ds\!\!
 \int_{-\infty}^{t}\!\!\!\!\!\dot{N}_{\rm d}(t) \:\! \Phi(s,\:\!\!t;\:\!\!
 r_{\rm b}) \, dt,
\end{equation}
where $f_{\rm d}(s) \, ds$ is a normalized distribution function of particle
radii $s$ whose lower and upper limits are,{\vspace{-0.055cm}} respectively,
$s_0$ and $s_\infty$, $\rho_{\rm d}$ is a particle bulk density, $\dot{N}_{\rm
d}(t)$ is the number of particles released at $t$ per unit time,~and
\mbox{$0 \!\le\! \Phi(s,\:\!\!t;\:\!\!r_{\rm b}) \!\le\! 1$} is a
function extracting the beam grains.

The cross-sectional area of the dust in the beam is measured either
from the scattered light (in the visible spectrum) or, as done by Jewitt
\& Matthews (1999), from the thermal emission.  In the former case, it
is~the particles' scattering properties (such as an albedo and phase
function) that complicate the problem, in the latter case it is the
particles' emissivity \mbox{$\varepsilon_{\rm d} \!\leq \! 1$}.

In the case of far infrared or submillimeter observations at wavelength
$\lambda$, the information provided by the measured thermal flux density
${\cal F}_\lambda$ in the beam is
\begin{equation}
{\cal F}_\lambda \!\sim \!\!\! \int_{s_0}^{s_\infty} \:\!\!\!\!\!\!
 \varepsilon_{\rm d}(s)\, \pi s^2 f_{\rm d}(s) \, ds \!\!
 \int_{-\infty}^{t} \!\!\!\!\!  \dot{N}_{\rm d}(t) \:\!
 \Phi(s,t;r_{\rm b})\,dt,
\end{equation}
but since a variation with particle size of the emissivity at $\lambda$
is unknown, the flux density is interpreted in terms of a blackbody's
cross-sectional area, that is, it is assumed that \mbox{$\varepsilon_{\rm
d}(s) \!=\! 1$} for any $s$ between $s_0$ and $s_\infty$.  And since
particles whose radii $s$ satisfy a condition
\begin{equation}
 x = \frac{2 \pi s}{\lambda} \ll 1
\end{equation}
do not radiate efficiently and \mbox{$\lim_{s \rightarrow 0} \,
\varepsilon_{\rm d}(s) \!=\!  0$}, the true geometric cross-sectional area
of the particles in the beam should in fact be much greater than offered by
the flux-density measurement.  To account for the effect of emissivity, we
note that there must exist particle sizes \mbox{$s_{\rm min} \!>\! s_0$} and
\mbox{$s_{\rm max} \!\leq\! s_\infty$} such that
\begin{equation}
\int_{s_0}^{s_\infty} \:\!\!\!\!\!\! \varepsilon_{\rm d}(s) \,\pi s^2 \phi_{\rm
 d}(s)\,ds = \!\!\! \int_{s_{\rm min}}^{s_{\rm max}} \:\!\!\!\!\!\! \pi s^2
 \phi_{\rm d}(s) \, ds \!=\!\! X_{\rm d}(s_{\rm min},s_{\rm max}),
\end{equation}
{\vspace{-0.04cm}}where $\phi_{\rm d}(s)$ equals \mbox{$f_{\rm d}(s)
\int_{-\infty}^{t} \,\ldots\, dt$} from Equation (30) and $X_{\rm d}$
is the measured blackbody cross-sectional~area. \mbox{This formalism
approximates the unknown variations in} $\varepsilon_{\rm d}(s)$
by introducing~\mbox{discontinuities}~at~$s_{\rm min}$~and~$s_{\rm
max}$, with\,\mbox{$\varepsilon_{\rm d}(s) = 0$} at\,\mbox{$s_0
\!\leq\! s \!<\! s_{\rm min}$\,and\,$s_{\rm max} \!  \leq\!  s
\!\leq\! s_\infty$},~but~with \mbox{$\varepsilon_{\rm d}(s) =
1$}~at~\mbox{$s_{\rm min} \!\leq\! s \!\leq\!  s_{\rm max}$}.  In
the case examined here one can safely adopt \mbox{$s_{\rm max} \!=\!
s_\infty$}.  If the variations~in~the emissivity can, for example,
plausibly be \mbox{approximated} \mbox{by \mbox{$\varepsilon_{\rm d}\!
sim x$} for \mbox{$x \!<\! x_0$} and by \mbox{$\varepsilon_{\rm d}
\simeq 1$} for \mbox{$x \!\geq\! x_0$}, their total} effect is accounted
for by{\vspace{-0.055cm}} taking \mbox{$2 \pi s_{\rm min}/ \lambda \simeq
{\textstyle \frac{1}{2}}x_0$}, that is, \mbox{$s_{\rm min} \simeq \lambda
x_0/4 \pi$}.  As $x_0$ is near unity, for submillimeter wavelengths this
condition implies that \mbox{$s_{\rm min} \!\gg\! s_0$}, which greatly
affects the results, as seen below.

Based on the work by Sekanina \& Kracht (2014), the relationship between the
total mass production rate of dust and the cross-sectional area $X_{\rm
d}(s_{\rm min},s_{\rm max})$ is then
\begin{equation}
\dot{\cal M}_{\rm d} = \frac{8\rho_{\rm d}}{3\pi r_{\rm b}} X_{\rm d}(s_{\rm
 min},s_{\rm max}) \, \frac{{\displaystyle \int_{s_0}^{s_\infty} \!\!\! s^3}
 f_{\rm d}(s) \,ds}{{\displaystyle \int_{s_{\rm min}}^{s_{\rm max}}} \!\!
 {\displaystyle \frac{1}{v_{\rm d}(s)}} \,s^2 f_{\rm d}(s) \, ds},
\end{equation}
where $v_{\rm d}(s)$ is a size-dependent ejection velocity of the dust
particles that generally is a function of heliocentric distance.  We adopt
a power law for the size distribution,
\begin{equation}
f_{\rm d}(s) \, ds = C \! \left( \! \frac{s_0}{s} \!\right)^{\!\alpha}\!\!,
 \;\;\;\;\int_{s_0}^{s_\infty} \!\!\!\! f_{\rm d}(s) \, ds = 1,
\end{equation}
where $\alpha$ is the distribution's index or power and $C$ is a
normalization constant; and make use of an expression previously employed for
the ejection velocity (Sekanina~\& Kracht 2014),
\begin{equation}
v_{\rm d}(s) = \frac{v_0}{1 \!+\! \chi \sqrt{s}}\,
\end{equation}
with $v_0$ and $\chi$ being the parameters, $v_0$ potentially varying with
heliocentric distance.  The total mass production rate is then as follows
\begin{eqnarray}
\dot{\cal M}_{\rm d} & = & \frac{8 v_0 \rho_{\rm d} s_\infty}{3 \pi r_{\rm b}}
 \, \frac{(3 \!-\! \alpha)(\frac{7}{2} \!-\! \alpha)}{4 \!-\! \alpha} \left(
 \! \frac{s_\infty}{s_{\rm max}} \! \right)^{\!3-\alpha} \nonumber \\[0.14cm]
 & & \times \, (1 \!-\! \epsilon_{\infty}^{4-\alpha}) \,
 X_{\rm d}(s_{\rm min},s_{\rm max}) \\[0.04cm]
 & & \times \! \left[ ({\textstyle \frac{7}{2}} \!-\! \alpha) (1 \!-\!
 \epsilon^{3-\alpha}) \!+\! (3 \!-\! \alpha)(1 \!-\!
 \epsilon^{\frac{7}{2}-\alpha}) \, \chi \sqrt{s_{\rm max}}
 \right]^{\!-1} \!\! , \nonumber
\end{eqnarray}
where \mbox{$\epsilon_\infty \!=\! s_0/s_\infty$}, \mbox{$\epsilon = s_{\rm
min}/s_{\rm max}$},{\vspace{-0.03cm}} and \mbox{$\alpha \neq 3$, $\frac{7}{2}$,
and 4}; otherwise the powers of $s_\infty$, $s_{\rm max}$, $\epsilon_\infty$,
and/or $\epsilon$ should~be replaced with the respective logarithms.

Before we apply Equation (36) to Jewitt \& Matthews' (1999) set of the
cross-sectional data, we carefully select the model parameters.  We first focus
on the constants that define the size distribution function, to which the
results are most sensitive.  The index $\alpha$ was for C/1995~O1 determined
numerous times and in different ways; a useful compilation, published by Lasue
et al.\ (2009), shows that \mbox{$3.0 \leq \alpha \leq 3.7$}.  The entries
most relevant to the important submillimeter and larger particles were provided
by the investigations based, at least in part, on the ISO observations, whose
spectral reach extended to nearly 200~$\mu$m.  Min et al.\ (2005) derived
\mbox{$\alpha = 3.48$} from the ISO observations made in September 1996, about
six months before perihelion, while Harker et al.\ (2002) deduced from the
observations covering nine months that \mbox{$\alpha = 3.4$} in 1996 when the
comet was nearly 3~AU from the Sun on its way to perihelion, but \mbox{$\alpha
= 3.7$} when the comet was within three months of perihelion.  These numbers
are in good agreement with an earlier result by Gr\"{u}n et al.\ (2001), who
derived \mbox{$\alpha = 3.5$}.\footnote{Both Gr\"{u}n et al.\ (2001) and Harker
et al.\ (2002) used a~size distribution law that was introduced by Sekanina \&
Farrell (1982) and subsequently employed extensively by Hanner (e.g., Hanner
1983, 1984).  The Sekanina-Farrell law differs from a power law for submicron-
and micron-sized grains, but in terms of $\alpha$ both laws are practically
identical for submillimeter-sized and larger particles.}  We adopt
\mbox{$\alpha = 3.55$} as an optimum mean value; we estimate its error at
$\pm$0.15.

A lower boundary to the particle size distribution, $s_0$, has a negligible
effect on the mass production rate; we employ \mbox{$s_0 = 0.01 \mu$m}, since
Min et al.\ (2005) showed that incorporating grains of amorphous olivine
and pyroxene of this minute size improves a fit to the comet's observed
spectral energy distribution.

An upper boundary to the size distribution, $s_\infty$, influences the total
mass production rate much more significantly than the lower boundary.  We
assume that the largest particles must reach an escape velocity from the
nucleus at the time they decouple from the gas flow, which according to
Probstein (1969) is at a distance of $\sim$20~nucleus' radii.  The size of
the nucleus is among the subjects of Part~II of this study; here we only
mention that Szab\'o et al.'s (2012) results suggest{\vspace{-0.04cm}} that
at the relevant distance the escape velocity amounts to \mbox{$v_{\rm esc}
\simeq 4$ m s$^{-1}$}.  From Equation~(35) the particle radius at the upper
boundary of the size distribution function is then equal to
\begin{equation}
s_\infty = \left( \! \frac{v_0}{v_{\rm esc}}\!-\!1\!\right)^{\!\!2}\!\chi^{-2},
\end{equation}
so that this task is reduced to finding the particle velocity parameters $v_0$
and $\chi$.  

Jewitt \& Matthews (1999) calculated that the velocity of particles 1~mm in
{\vspace{-0.04cm}}radius ejected from C/1995~O1 was 80~m~s$^{-1}$ at 1~AU from
the Sun (although they eventually used a value three times {\it lower\/}),
assuming that the diameter of the nucleus equasled 40~km.  With its
dimensions nearly twice as large (Szab\'o et al.'s 2012), the ejection
velocity of the millimeter-sized particles was more likely to amount to
at least 110~m~s$^{-1}$.  In fact, Vasundhara \& Chakraborty's (1999)
direct fit to dust-coma features in the comet's images taken between 1997
February 18 and May 2 imply for these particles ejection velocities that
are still higher, 123--145~m~s$^{-1}$.

\begin{table}[ht]  
\vspace{-4.23cm}
\hspace{4.21cm}
\centerline{
\scalebox{1}{
\includegraphics{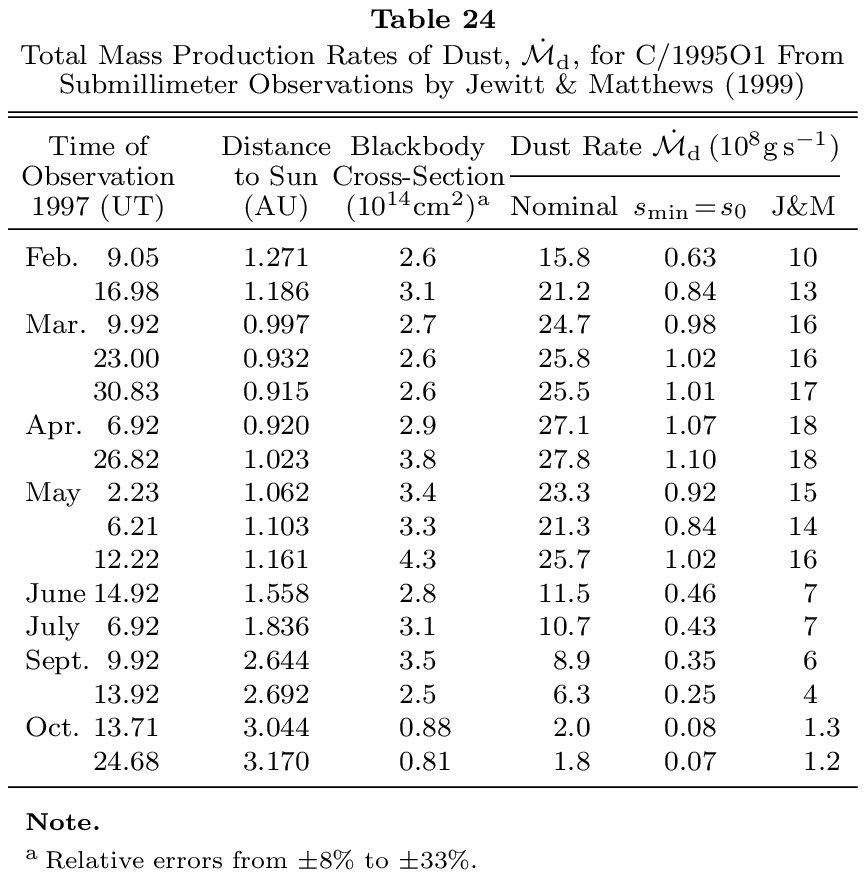}}} 
\vspace{-16.6cm}
\end{table}

At the other end of the~\mbox{dust-particle}~size~\mbox{spectrum},\,the
ejection velocities~of~\mbox{submicron-sized}~grains~that~made up the leading
boundaries of recurring expanding halos were determined by Braunstein et
al.\ (1999) from images taken over a period of 61~days and approximately
centered~on~\mbox{perihelion}.  Corrected for the projection effects, the
ejection veloci\-ties averaged \mbox{$670 \pm 70$ m s$^{-1}$}, ranging from
603~m~s$^{-1}$\,to 775~m~s$^{-1}$, with no systematic trends between 0.91~AU
and 1.10~AU from~the~Sun.~\mbox{Vasundhara} \& Chakraborty (1999) derived the
{\vspace{-0.04cm}}ejection velocities~of \mbox{up to 650 m s$^{-1}$\,for
submicron-sized particles near\,1\,AU,} with a rate of decrease with
{\vspace{-0.075cm}}heliocentric distance $r$ only slightly steeper than
$r^{-\frac{1}{2}}$ (to be adopted here).~To~reconcile these high ejection
velocities of microscopic grains with the velocities of \mbox{130--140 m
s$^{-1}$} for millimeter-sized particles, the parameters in Equation~(35)
{\vspace{-0.03cm}}should equal, after rounding off, \mbox{$v_0 \!=\! 700$ m
s$^{-1}$} at{\vspace{-0.06cm}} 1~AU from the Sun and \mbox{$\chi \!=\! 13$
cm$^{-\frac{1}{2}}$}.

With these values and an escape velocity of $\sim$4~m~s$^{-1}$ inserted
into Equation~(37), the upper boundary of the particle size distribution is at
\mbox{$s_\infty \!=\! 180$ cm}.  Referring to our previous arguments, we adopt
this same number for $s_{\rm max}$, while with a rather conservative value of
\mbox{$x_0 \!\approx\! 1.2$} ($x_0$ is expected to be near unity), we find
\mbox{$s_{\rm min} \!= 80\;\mu$m}.

The remaining physical parameter in Equation~(36) --- a particle bulk density
--- is rather uncertain,~\mbox{because} available information is indirect,
based on research of comets other than C/1995~O1.  Exposure of Stardust's
sample collector to impacts of dust grains ejected from 81P/Wild~2 points
to a very broad range {\vspace{-0.03cm}}of bulk densities, from compact
particles with \mbox{$\rho_{\rm d} \!\sim\! 3$ g cm$^{-3}$} to highly porous
aggregates for which $\rho_{\rm d}$ is as low as $\sim$0.3~g~cm$^{-3}$ (e.g.,
H\"{o}rz et al.\ 2006).  A recent investigation of the dust detected by the
{\it Grain Impact Analyzer and Dust Accumulator\/} (GIADA) on board the
Rosetta spacecraft led to a conclusion {\vspace{-0.05cm}}that for
67P/Churyumov-Gerasimenko a mean dust-particle bulk {\vspace{-0.06cm}}density
equals $0.80_{-0.07}^{+0.84}$\,g~cm$^{-3}$ (Fulle et al.\ 2016).  We
henceforth adopt \mbox{$\rho_{\rm d} \!=\! 0.8$ g cm$^{-3}$}.

With the known angular beam radius of 7$^{\prime\prime\!}$.65, the linear
radius becomes \mbox{$r_{\rm b} = 5550 \Delta$ km}, where $\Delta$ is the
comet's geocentric distance in AU.  This completes the prerequisites for
deriving the total mass production rates of dust, which are in Table~24
listed in a column marked ``Nominal'' and compared with the numbers by
Jewitt \& Matthews (1999), in the ``J\&M'' column, and with the rates
derived in a case of the neglected emissivity effect (in a column marked
\mbox{``$s_{\rm min} \!= s_0$''}). 

Table 24 suggests an unexpectedly good correspondence between our nominal mass
production rates and the rates determined by Jewitt \& Matthews (1999);~our
solution offers rates that, on the average, are only 1.55 times higher
than are theirs.  We are convinced that this agreement is fortuitous, in part
because the enormous differences between both approaches are demonstrated by
the employed ejection veloc\-ity:\ our value of 137~m~s$^{-1}$ for grains
1~mm in radius, is a factor of 5.5 times higher than the value used by Jewitt
\& Matthews.  According to their Equation~(6), our nominal rates should have
been higher by the same factor.  In addition, the maximum radius of the
particles escaping from the nucleus in their model was $\sim$3~cm, while our
estimate is 60 times larger, a factor that also increases the production rate
as seen from our Equation~(36).  This means that some of the other quantities
that impact the resulting rates must have been strongly underestimated by
Jewitt \& Matthews (1999), one of them being the residence times of grains
moving in directions other than perpendicular to the line of sight, as
pointed out by us earlier.

Comparison of the columns 4 and 5 in Table 24 illus\-trates that ignoring the
effect of low emissivity for particles smaller than about 0.1 the wavelength
yields unacceptable results.  In this case, assigning improperly a unit
emissivity to grains as small as 10$^{-5}$ the wavelength implies the dust
production rates that are too low by a factor of about 25.

As for the uncertainties involved in our nominal mass production rates in
Table~24, we note that they are most sensitive to the size distribution index
$\alpha$.  The error of $\pm$0.15 brings about a scatter from 0.57 the nominal
rate for \mbox{$\alpha \!=\! 3.7$} to 1.65 the nominal rate for \mbox{$\alpha
\!=\! 3.4$}.  The uncertainties in the other parameters are of lesser impact.
A change by a factor of two in a particle size at the upper boundary of the
distribution, $s_\infty$, results in a range from 0.77 to 1.31 the nominal
rate.  The same change in a particle size at the distribution's lower boundary,
$s_0$, has an entirely negligible effect on the production rates, but a change
{\vspace{-0.01cm}}in the lower limit of the size of the particles that
contribute to the measured thermal flux density, $s_{\rm min}$, does affect
the rates moderately; a change by a factor of two causes the rates to vary
from 0.87 to 1.15 the nominal rate.  The column \mbox{$s_{\rm min} = s_0$}
{\vspace{-0.01cm}}in Table~24 shows a change in the rates after the nominal
value of $s_{\rm min}$ was reduced by a factor of 8000.  Equation~(36)
implies that changes in the particle density and ejection velocity project
linearly as changes in the production rate (disregarding a minor effect from
the parameter $\chi$).

\begin{table}[t]  
\vspace{-4.21cm}
\hspace{4.23cm}
\centerline{
\scalebox{1}{
\includegraphics{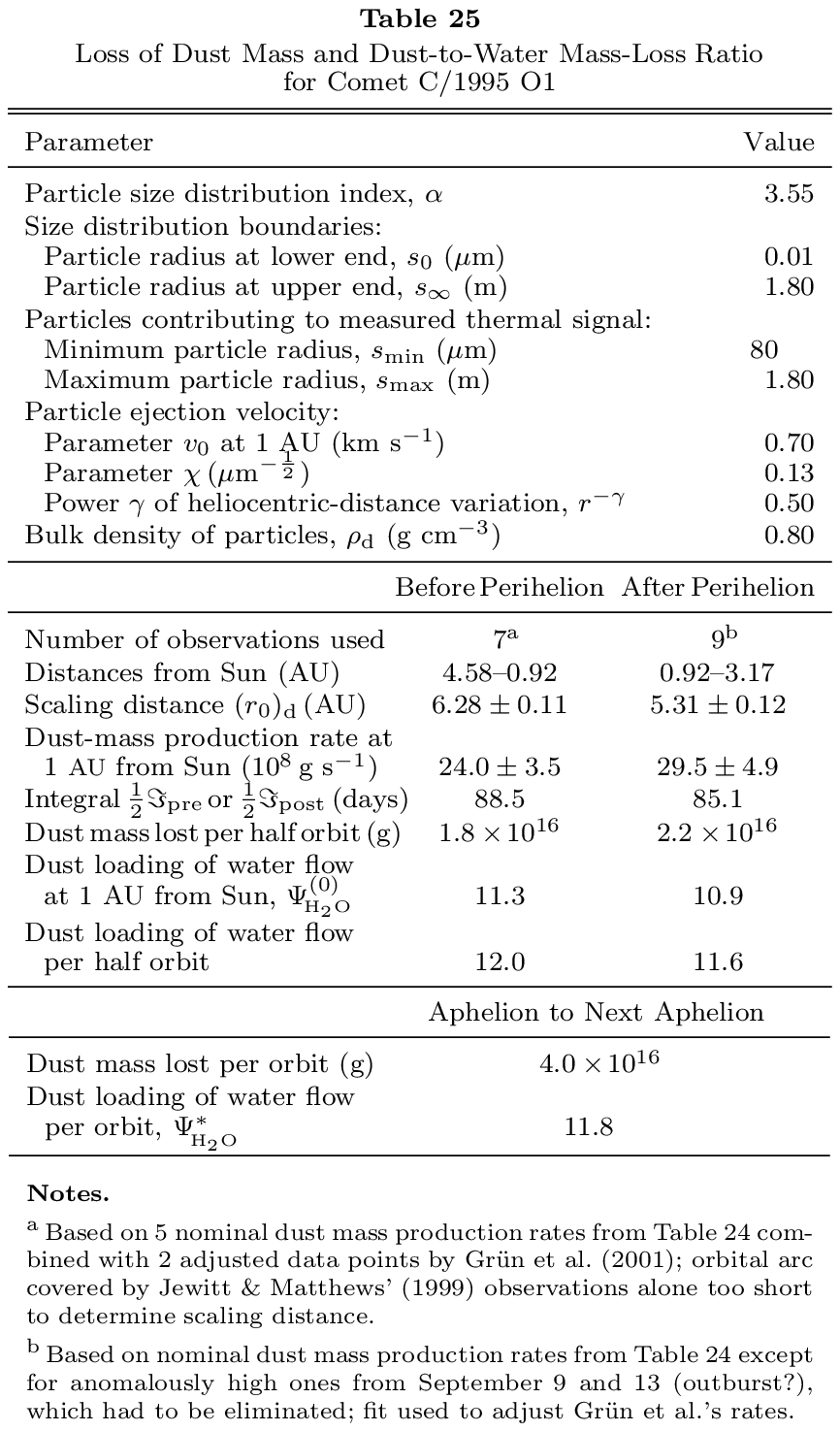}}}  
\vspace{-10.6cm}
\end{table}

The next tasks were to test whether the mass production variations of dust
with time could be fitted by the same $g_{\rm mod}$-type law as the gas
production variations; to estimate a total loss of dust mass{\vspace{-0.08cm}}
from C/1995~O1, derived by integrating this production rate $\dot{\cal M}_{\rm
d}$ over the orbit from aphelion to next aphelion; and to compare it to the
mass loss of water presented in Table~19.  We began by ascertaining that
the more extensive set of the post-perihelion nominal data points from
Table~24 is satisfied by a law $g_{\rm mod}[r;(r_0)_{\rm d,post}]$ and by
determining the tabulated scaling distance $(r_0)_{\rm d,post}$.  We then
turned to the preperihelion data in Table~24, which cover a shorter range
of heliocentric distances.  We noticed that, fortunately, Gr\"{u}n et al.\
(2001) reported the dust production rates at three larger heliocentric
distances in 1996 and 1997, derived from the ISO observations.  The
chronologi\-cally last of these data points, from 1997 December~30, when
the comet was 3.90~AU from the Sun, was used by us to adjust Gr\"{u}n et
al.'s scale of dust production to agree with ours.  It turned out that
when their production rate at 3.9~AU was multiplied by a factor of 1.3,
it fitted the value predicted for this time by the post-perihelion modified
law.  We then adjusted Gr\"{u}n et al.'s (2001) preperihelion data points,
referring to 1996 April~27 (the comet at 4.58~AU from the Sun) and to 1996
October~7 (2.82~AU from the Sun) by the same factor, and linked them with the
five preperihelion nominal production rates from Table~24; we obtained a rather
satisfactory fit by applying another modified law, $g_{\rm mod}[r;(r_0)_{\rm
d,pre}]$.\footnote{Additional observations of the comet in a spectral range
(near or beyond 1~mm) that should warrant a proper account of the contributions
from massive grains were reported by Senay et al.\ (1997) at three wavelengths
between 1.1~mm and 2.1~mm in late February 1997 and by de~Pater et al.\ (1998)
at wavelengths \mbox{2.6--3.5 mm} and 7.0--13.3~mm during March and April
1997.  We do not use these observations in our computations because they cover
the same period of time as the nominal data points in Table~24, but note that
Senay et al.\ derived a fairly low production rate of 3.2$\,\times\:\!$10$^8
$\,g~s$^{-1}$ five weeks before perihelion, while de Pater et al.\ deduced a
production rate on the order of 10$^9$\,g~s$^{-1}$ within four weeks of
perihelion, in conformity with the data we employ.} Together with a summary
of the adopted parameters, the dust-production results are listed in Table~25,
separately for either orbital branch as well as for the whole orbit.

We found that the orbit-integrated mass loss of dust by C/1995~O1 equaled
\mbox{$4 \times \:\!\! 10^{16}$\,g}, close to --- but slightly higher than ---
the total by Jewitt \& Matthews (1999).  The dust production appears to be
only marginally, and within 1$\sigma$ errors, higher after perihelion than
before perihelion.  Figure~15 illustrates that the data points are fitted
by the two $g_{\rm mod}$-type laws quite satisfactorily.  However, we would
not expect the laws to be applicable to heliocentric distances larger than
approximately 4.5~AU.  The production rate of carbon monoxide catches up
with the water production rate at \mbox{3.6--3.7}~AU, thus gradually taking
control over the process of dust release, including non-sublimating water-ice
grains, at larger distances from the Sun.

The mass ratio of the production of dust to the production of water is called
in Table~25 and hereafter a {\it mass loading of the water flow by dust\/} or
shortly{\vspace{-0.035cm}} a {\it dust loading of the water flow\/}.  We
{\vspace{-0.04cm}}distinguish a {\it normalized dust loading\/},
$\Psi_{\!_{{\rm H}_2{\rm O}}}^{(0)}$, given as{\vspace{-0.05cm}}
\begin{equation}
\Psi_{\!_{{\rm H}_2{\rm O}}}^{(0)}\!\!=\!\frac{\mbox{mass production
 rate of dust
 at$\:$1\,{\small AU}$\:$from$\:$Sun}}{\mbox{mass$\:$production$\:$rate$\:$of
 water$\:$at\,1\,{\small AU}\,from$\:$Sun}},
\end{equation}
whose values are different before and after perihelion; and a dust loading
integrated over the orbit, or simply a {\it dust loading per orbit\/} or
{\vspace{-0.08cm}}{\it integrated dust loading\/}, $\Psi_{\!_{{\rm H}_2{\rm
O}}}^\ast$, defined as
\begin{equation}
\Psi_{\!_{{\rm H}_2{\rm O}}}^\ast \!\!=\! \frac{\mbox{total mass of dust lost
 by ejection per orbit}}{\mbox{total mass of sublimated water ice per orbit}},
\end{equation}
which consists of the preperihelion and post-perihelion contributions to the
total.  When data are limited to a heliocentric distance $r$ different from
1~AU, one may only be able to determine{\vspace{-0.045cm}} a {\it nominal
dust loading\/}, $\Psi_{\!_{{\rm H}_2{\rm O}}}\!(r)$, at that particular
distance, which --- assuming the validity of a $g_{\rm mod}$-type law for
both water and dust --- is related to the normalized dust loading by
\begin{equation}
\Psi_{\!_{{\rm H}_2{\rm O}}}\!(r) = \Psi_{\!_{{\rm H}_2{\rm O}}}^{(0)} \!\!
 \left[ \frac{(r_0)_{\rm d}^n +\! 1}{(r_0)_{_{{\rm H}_2{\rm O}}}^n \!+\!
 1} \, \frac{(r_0)_{_{{\rm H}_2{\rm O}}}^n \!+ r^n}{(r_0)_{\rm d}^n +
 r^n} \right]^{\!k} \!\! ,
\end{equation}
where $n$ and $k$ are the exponents of the modified law, as defined below
{\vspace{-0.06cm}}Equation~(1), while $(r_0)_{\rm d}$ and $(r_0)_{_{{\rm
H}_2{\rm O}}}$ are, respectively, the scaling distances for the dust (from
Table~25) and water (from Table~19), which both apply to either the
preperihelion or post-perihelion branch of the orbit, as do the values
of the nominal and normalized dust loading of the water flow.

\begin{figure*}  
\vspace{-4.82cm}
\hspace{-0.2cm}
\centerline{
\scalebox{0.73}{
\includegraphics{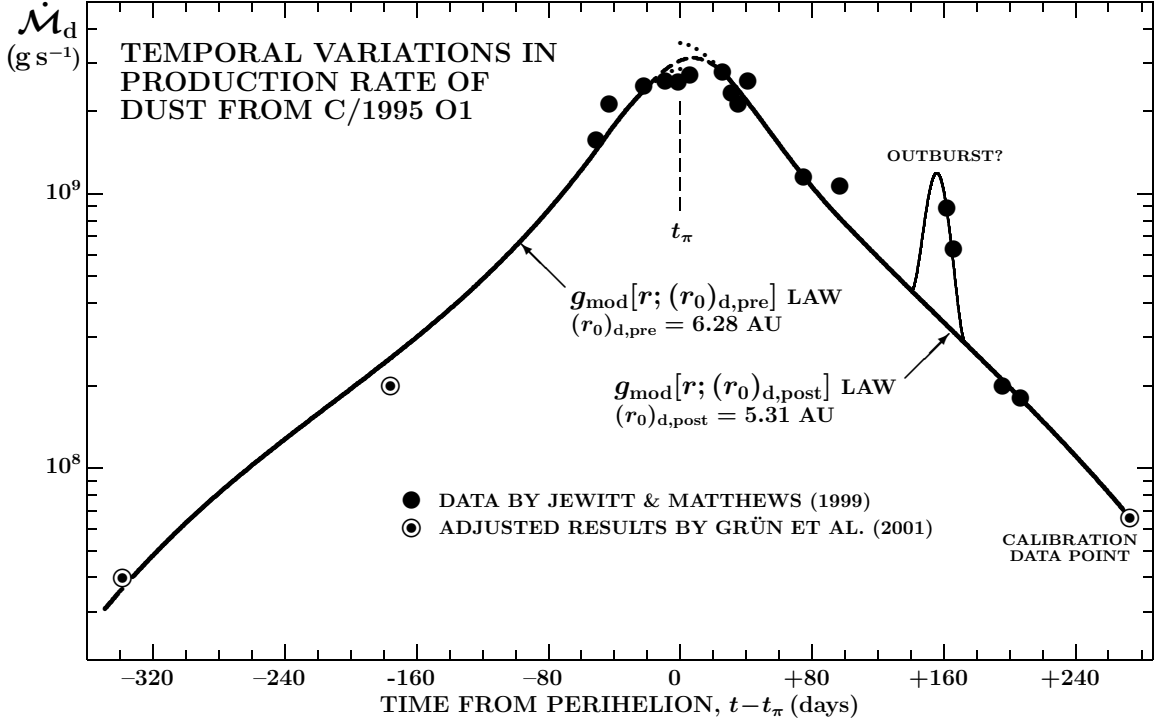}}}  
\vspace{-7.28cm}
\caption{Curve of the dust production rate of C/1995~O1 as a function of time.  
It was derived using the technique described in the text, applied to the dust
cross-sectional data at a submillimeter wavelength (Jewitt \& Matthews 1999),
which were linked to the adjusted ISO-based production rates by Gr\"{u}n et
al.\ (2001), and fitted by a modified law, $g_{\rm mod}\!\left[r;(r_0)_{\rm
d}\right]$, separately before and after perihelion (thick curve).  Of the 11
Jewitt-Matthews post-perihelion data points, 9 were successfully matched by
a $g_{\rm mod}$ law, the two discarded ones possibly suggesting the presence
of a minor outburst some 160~days after perihelion.  The data point by Gr\"{u}n
et al.\ (2001) at 3.90~AU from the Sun (+273~days from perihelion) was employed
to calibrate their other observations; it required a multiplication factor of
1.3 to fit the $g_{\rm mod}$ curve.  The preperihelion data by Gr\"{u}n et
al.\ at 4.58~AU and 2.82~AU from the Sun were then adjusted by this factor,
linked to the other five preperihelion data points, and all subsequently
fitted by a $g_{\rm mod}$ law, as depicted in the figure.  At perihelion the
two laws are disconnected, the post-perihelion production rate being nominally
23\% higher than the preperihelion rate, as illustrated by the dots.  This
discrepancy is bridged by an empirical dashed curve, peaking shortly after
perihelion.{\vspace{0.6cm}}}
\end{figure*}

A logarithmic differentiation of Equation (40),
\begin{equation}
\frac{d \ln \Psi_{\!_{{\rm H}_2{\rm O}}}(r)}{d \ln r} = \frac{nk r^n \! \left[
 (r_0)_{\rm d}^n  - (r_0)_{_{{\rm H}_2{\rm O}}}^n \right]}{\left[
 (r_0)_{\rm d}^n + \mbox{\raisebox{0ex}[1.5ex][1ex]{$r^n$}} \right]
 \!\!\!\: \left[ (r_0)_{_{{\rm H}_2{\rm O}}}^n \!+ r^n \right]} ,
\end{equation}
implies that it is the difference between the scaling~distances of the
$g_{\rm mod}$ laws for the production of dust and water, respectively, that
exclusively determines whether the dust loading increases or decreases with
heliocentric distance:\ when the scaling distance for the dust production
is greater, as is the case for C/1995~O1 both before and after perihelion,
the dust loading increases with increasing distance from the Sun.  However,
{\vspace{-0.05cm}}$\Psi_{\!_{{\rm H}_2{\rm O}}}$ does not diverge, as
\mbox{$\lim_{r \rightarrow \infty} d \Psi_{\!_{{\rm H}_2{\rm O}}}/dr = 0$}.
The limit, equaling
\begin{equation}
\lim_{r \rightarrow \infty} \Psi_{\!_{{\rm H}_2{\rm O}}}(r) = \Psi_{\!_{{\rm
 H}_2{\rm O}}}^{(0)} \!\! \left[ \frac{(r_0)_{\rm d}^n + 1}{(r_0)_{_{{\rm
 H}_2{\rm O}}}^n \!+ 1} \right]^{\!k} \!\!\! \simeq \Psi_{\!_{{\rm
 H}_2{\rm O}}}^{(0)} \!\! \left[ \! \frac{(r_0)_{\rm d}}{(r_0)_{_{{\rm H}_2{\rm
 O}}}} \! \right]^{\!nk} \!\! ,
\end{equation} 
is to be employed in practice with caution because of the aforementioned
transformation of the dust ejection process from a water dominated mode
to a carbon-monoxide dominated mode near 4~AU from the Sun.

Table 25 suggests that our results~for~both~the~normalized and integrated dust
loading~of~the~\mbox{water}~flow~range between about 11 and 12 and that they
are at most~only marginally~higher prior to~perihelion.  The~\mbox{ parameters}
from Tables~19 and 25 allow one to predict the~mass~loading by dust of the
{\vspace{-0.05cm}}water flow, $\Psi_{\!_{{\rm H}_2{\rm O}}}\:\!\!(r)$, as well
as that of the combined flow of water and carbon monoxide, $\Psi_{\!_{{\rm
H}_2{\rm O+CO}}}\:\!\!(r)$, and of water, carbon monoxide,{\vspace{-0.07cm}}
and carbon dioxide, $\Psi_{\!_{{\rm H}_2{\rm O+CO+CO}_2}\:\!\!\!}(r)$, for
a broad variety {\vspace{-0.04cm}}of heliocentric distances $r$.  The results
of these computations, in Table~26, show that the dust loading increases
systematically, both before and after perihelion, only for the flow of
water.  Once we consider a more realistic case, a combined flow of water
with other species, the dust loading peaks and then may drop at larger
heliocentric distances.  This behavior is a corollary of Equation~(41)
written for a flow or more than one species, because the scaling distance
of dust is smaller than the scaling distances of carbon monoxide and carbon
dioxide both before and after perihelion.  We note that in the case of a
combined flow of water, carbon monoxide, and carbon dioxide, which in Table~26
approximates the true conditions in the comet's atmosphere most closely, the
dust loading never exceeds 11 before perihelion and 10 after perihelion.  The
table also allows one to derive a dust loading of a flow of carbon monoxide,
$\Psi_{\!_{\rm CO}}$, and of carbon dioxide, $\Psi_{\!_{{\rm CO}_2}}$:
\\[-0.3cm]
\begin{eqnarray}
\Psi_{\!_{\rm CO}}\!(r) & = & \left[ \Psi_{\!_{{\rm H}_2{\rm O+CO}}}^{-1}\!(r)
 - \Psi_{\!_{{\rm H}_2{\rm O}}}^{-1}\!(r) \right]^{-1} \nonumber \\[-0.22cm]
 & & \\[-0.22cm]
\Psi_{\!_{{\rm CO}_2}}\!(r) & = & \left[ \Psi_{\!_{{\rm H}_2{\rm
 O+CO+CO}_2}}^{-1}\!(r) - \Psi_{\!_{{\rm H}_2{\rm O+CO}}}^{-1}\!(r)
 \right]^{-1}. \nonumber
\end{eqnarray}
\begin{table}  
\vspace{-4.2cm}
\hspace{4.22cm}
\centerline{
\scalebox{1}{
\includegraphics{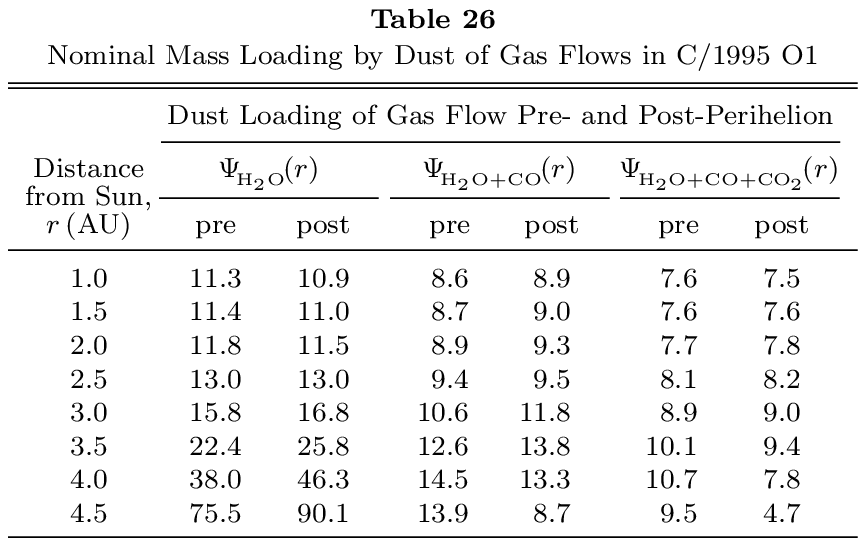}}} 
\vspace{-19.75cm}
\end{table}

Before comparing the numbers in Table~26 with the results reported elsewhere,
a caveat first.  Whereas care is usually taken, as it should, to clearly
distinguish between a dust loading of the {\it water\/} flow, $\Psi_{\!_{{\rm
H}_2{\rm O}}}$, {\vspace{-0.07cm}}on the one hand and a dust loading of the
{\it gas\/} flow (that is, of the total flow of a number{\vspace{-0.04cm}}
of volatile species), $\Psi_{\rm gas}$ (\mbox{$\Psi_{\rm gas} \!<\!
\Psi_{\!_{{\rm H}_2{\rm O}}}$}), on the other hand, the two quantities are
not always differentiated (e.g., A'Hearn et al.\ 1995; Schleicher et al.\
1997), apparently with a tacit, but --- as seen from Table 26 --- rather
questionable, premise that water dominates the sublimation process.

A high mass loading by dust in C/1995 O1~was commented on by numerous
researchers.  With one exception, the reported values of the dust loading
based on the observations at submillimeter and millimeter wavelengths are
in good to excellent agreement with the numbers in Table~26.  The exception
is an observation at wavelengths between 1.1~mm and 2.1~mm in late February
1997, five weeks before perihelion, by Senay et al.\ (1997), who reported for
{\vspace{-0.04cm}}$\Psi_{\!_{{\rm H}_2{\rm O}}}$ a fairly low value of 3.5.
On the other hand, Jewitt \& Matthews (1999), whose data we employ extensively,
estimated a lower limit to $\Psi_{\!_{{\rm H}_2{\rm O+CO}}}${\vspace{-0.04cm}}
at 5 close to perihelion; having accounted for the contribution from carbon
monoxide as a fifth of the contribution from water, they{\vspace{-0.05cm}}
effectively estimated a lower limit to $\Psi_{\!_{{\rm H}_2{\rm O}}}$~at~6.
For greater heliocentric distances, the dust loading was reported by Gr\"{u}n
et al.\ (2001) to equal 9 at 4.58~AU before perihelion, 6 at 2.82~AU before
perihelion, and 10 at 3.90~AU after perihelion.  For the gas production they
employed the combined rate of water and carbon monoxide.  At 4.58~AU before
{\vspace{-0.01cm}}perihelion we find from Table~19 a gas production rate of
\mbox{3.0$\, \times$10$^6$\,g s$^{-1}$}\,(15\% H$_2$O, 85\% CO), so that
\mbox{$\Psi_{\!  _{{\rm H}_2{\rm O+CO}}} \!\simeq 10$}; {\vspace{-0.07cm}}at
2.82~AU before perihelion, \mbox{2.4$\,\times$10$^7$\,g s$^{-1}$}\,(70\%
H$_2$O, 30\% CO), so {\vspace{-0.07cm}}that \mbox{$\Psi_{\!_{{\rm H}_2{\rm
O+CO}}}\!\simeq 6$}; and at 3.90~AU after perihelion, \mbox{4.9$\,\times
$10$^6$\,g s$^{-1}$}\,(33\% H$_2$O, 67\% CO),{\vspace{-0.06cm}} so that
\mbox{$\Psi_{\!_{{\rm H}_2{\rm O+CO}}} \!\simeq 10$}, in excellent agreement
with Gr\"{u}n et al.\ (2001).  Since we adjusted their dust production rates
by a factor of 1.3 in Figure~15, the dust loading of the H$_2$O+CO flow with
these numbers becomes 13, 8, and 13, respectively, still in good agreement,
given that the uncertainty in Gr\"{u}n et al.'s dust production rates was
estimated by the authors at a factor of two in the least.  

Three investigations employed a dynamical approach, and in this case it
was a dust loading of the total~gas~flow, $\Psi_{\rm gas}$, that was
determined.  Fitting the dust morphological features in the head of
C/1995~O1,~\mbox{Vasundhara}~\& Chakraborty (1999) determined a lower
limit of 4.8$\,\pm\:\!$1.4 to the dust loading from observations on three
days between six weeks before perihelion and one month~\mbox{after}
\mbox{perihelion.  Applying a modified Probstein-Finson meth-} od of
analysis to the comet's dust tail, Fulle et al.\ (1998) concluded that
at heliocentric distances greater than $\sim$4.4~AU preperihelion, during the
CO-driven activity, the dust loading \mbox{$\Psi_{\!_{\rm CO}}$} must have
{\vspace{-0.04cm}}been at least 5; they deduced a dust production rate of
\mbox{8$\, \times$10$^6$\,g s$^{-1}$} some 320~days before perihelion, about
a factor of 5 below the curve in Figure~15.  Table~19 suggests that the CO
production rate was in fact near \mbox{2.7$\,\times$10$^6$\,g s$^{-1}
$},~leading to a dust loading of only 3 with Fulle et al.'s (1998)~dust
production rate.  From Table~26 we find \mbox{$\Psi_{\!_{\rm CO}} \simeq 17$}
at~a distance of 4.5~AU.  Fulle et al.'s dust~rate estimate of 5$\,\times$10$
^5$g~s$^{-1}$ in 1993, at a heliocentric distance of 13~AU implies
\mbox{$\Psi_{\!_{\rm CO}} \!\!\sim \! 100$}.  This value is probably much too
high and only formally consistent~with~\mbox{Sekanina}'s~(1996)~con\-clusion
that the dust loading of the CO flow~must~have been greater than 15 over a
range of heliocentric distances centered on 6.7~AU, derived from low expansion
rates of a sequence of spiral features observed in the comet's head during
the months immediately following discovery.  These constraints suggest that
the dust~production rate varied, on the average, more steeply than $r^{-5}$
between 13~AU and 6.7~AU and as approximately $r^{-2}$ between 6.7~AU and
4.6~AU before perihelion.

Lisse et al.\ (1999) determined a dust production{\vspace{-0.04cm}} rate of
$\sim$1.4$\, \times$10$^8$\,g~s$^{-1}$ and a dust loading of the gas flow of
$\sim$5 from a 5--13~$\mu$m spectral energy distribution on 1996 October 31,
when the comet was 2.54~AU from the Sun.  With this dust rate we find a
loading of 5.9 for water and 4.3  for water plus carbon monoxide.  From
Table~25, we compute a dust production rate about 2.2 times higher.
Similarly, their estimate of 1$\,\times$10$^{16}$\,g for an orbit-integrated
loss of dust is a factor of 3--4 lower than Jewitt \& Matthews' (1999) and
ours.

Finally, several investigations determining the mass loading by dust were
based on photometry or spectro\-photometry of scattered sunlight at optical
wavelengths; with no exception, the mass of dust ejecta (and thus the loading)
was in these studies underestimated usually by a wide margin.  It is well
known that millimeter-sized particles dominate the mass distribution,
while micron-sized particles the cross-sectional distribution of cometary
dust ejecta.\footnote{This effect is nicely illustrated by McDonnell et al.'s
(1991) Figure~13 for the dust ejecta from 1P/Halley detected during the
Giotto encounter.{\vspace{0.08cm}}}  Optical observations always sample
primarily the smaller grains and fail to adequately account for the larger
ones, as remarked by Jewitt \& Matthews (1999).  This effect manifests
itself in the dust production rates reported by Weaver et al.\ (1997, 1999b)
for heliocentric distances exceeding 2.4~AU both before and after perihelion,
derived via a proxy parameter {\it Af}$\rho$ (A'Hearn et al.\ 1995) from
their HST and IUE observations.\footnote{The values of {\it Af}$\rho$, derived
from the spectra of the International Ultraviolet Explorer (IUE), were based
of course on scattered sunlight at UV rather than optical wavelengths; one
{\vspace{-0.038cm}}of two Echelle spectrographs worked in a spectral range
of up to 3300~{\AA} at most.}  The dust production rates are much too low by
about one order of magnitude, even though the computed dust loading of the
water flow exceeds unity for all entries in Table I of Weaver et al.\ (1999b).

Schleicher et al.\ (1997) reported the dust production data between 25 July
1995 and 15 February 1997 only in terms of {\it Af}$\rho$, which they did
not convert to the dust production rates.  They did, however, point out that
the gas production (including OH) was in C/1995~O1 about 20 times higher than
in 1P/Halley, while the dust production was more than 100-fold greater at
comparable heliocentric distances.  Since the best estimate of the dust
production rate of 1P is 5$\,\times$10$^7 $\,g~s$^{-1}$ (McDonnell et al.\
1991)~at~the encounter time of the Giotto spacecraft (0.90~AU from the Sun
{\vspace{-0.06cm}}post-perihelion), one should expect a rate of at least
4$\,\times$10$^9$\,g~s$^{-1}$ for C/1995~O1 at 1~AU from the Sun after
perihelion, which is by more than 2$\sigma$ higher than predicted by the
model in Table~25 --- still a fair agreement, especially considering that
this comparison involves a premise of comparable perihelion asymmetry for
the two objects.  However, when one employs a conversion factor between {\it
Af}$\rho$ and the dust production rate recommended by A'Hearn et al.\ (1995),
{\vspace{-0.03cm}}one obtains \mbox{$\sim$0.9--1.0$\,\times$10$^9$\,g s$^{-1}$}
at 1~AU before perihelion, a factor of $\sim$2.5 lower than from Table~25.  The
source of this discrepancy is again the fact that {\it Af}$\rho$ systematically
underrates the dust production; the data that Schleicher et al.\ (1999)
{\vspace{-0.03cm}}referred to for 1P are represented by \mbox{{\it Af}$\rho
\!=\! 10^{3.9}$\,cm} at 1~AU preperihelion {\vspace{-0.04cm}}(A'Hearn et al.\
1995) and therefore by a dust rate of only 8$\,\times$10$^6 $\,g~s$^{-1}$.
On the one hand, this rate indeed is more than 100 times lower than the rate
for C/1995~O1, but on the other hand it is also more than 6 times lower than
McDonnell et al.'s (1991) post-perihelion rate for 1P normalized to 1~AU from
the Sun.  The lesson appears to be that comparison of two comets in terms of
{\it Af}$\rho$ provides a correct dust production rate for the comet of
interest, if a correct dust production rate is available for the other comet.

Monitoring the dust distribution between 4.6~AU and 2.9~AU before perihelion
in a red passband, Rauer et al.\ (1997) derived production rates much too low
by a factor of 2 to 4 relative to the curve in Figure~15; their values of the
dust loading of the flow consisting of water and carbon monoxide, 2.3 through
6.0, then become 9 through 13, in good agreement with the numbers in Table~26.

A more extensive investigation by Weiler et al.\ (2003), covering a range of
heliocentric distances from 4.6~AU to 2.9~AU preperihelion (using in part
the observations examined by Rauer et al.\ 1997) and from 2.8~AU to 12.8~AU
post-perihelion, also began by determining {\it Af}$\rho$, which was converted
to a dust production rate on certain assumptions concerning the dimensions and
activity of the nucleus and regarding the size distribution, terminal velocity,
bulk density, geometric albedo, and phase function of dust particles.  The fit
was relatively insensitive to variations in some of the tested parameters and
parameteric functions with an exception of the active fraction of the surface:\
19\% of a nucleus 30~km in radius provided the best match; this indicated an
active area of about 2150~km$^2$.  Unfortunately, Weiler et al.'s resulting
dust production rates within 4.6~AU of the Sun were up to almost a factor of 10
lower than the rates by Gr\"{u}n et al.\ (2001) or those implied by the curve
in Figure~15.  At 7~AU post-perihelion, Weiler et al.'s dust rate was close to
1$\,\times$10$^6$\,g~s$^{-1}$, again at least one order of magnitude lower than
needed to satisfy Sekanina's (1996) dust-loading condition at $\sim$6.7~AU
preperihelion.  And at 12.8~AU post-perihelion, their dust rate of $\sim$2$
\,\times$10$^5$\,g~s$^{-1}$ was a factor of 2--3 lower than Fulle et al.'s
(1998) dust rate at 13~AU preperihelion.  Weiler et al.'s major result of a
nearly constant dust{\vspace{-0.04cm}} loading of \mbox{$\Psi_{\!_{{\rm
H}_2{\rm O+CO}}} \! \simeq 1.1 \pm 0.3$}, was lower than the dust loading
derived in other studies, a problem admitted by the authors.  But a
discrepancy of even one order of magnitude in the dust loading could not
account for Weiler et al.'s conclusion that merely 1.1$\,\times$10$^{11}$\,g
of dust was lost by the comet over its orbit about the Sun, an amount of
mass that was by more than five orders of magnitude (sic!) lower than our,
Jewitt \& Matthews' (1999), and Lisse et al.'s (1999) estimates.

To summarize, we are convinced that only investigations based on thermal flux
observations at wavelengths greater than about 100~$\mu$m and/or on carefully
executed dynamical analysis of the dust ejecta provide reliable information
on the total losses of dust from C/1995~O1.  These results suggest that the
comet's production of dust exceeded a rate of 10$^9$\,g~s$^{-1}$ near
perihelion and that the total loss of dust per orbit was on the order of
10$^{16}$\,g.  The dust loading of the gas flow was more than a factor of
10 for water (even in the proximity of perihelion) and close to 10 for
water plus carbon monoxide.  Such a high dust loading is bound to affect
downward the dust-particle velocities, an issue that is addressed next.

\subsection{Effects of Dust Loading on Terminal Velocities of Microscopic
 Grains, and Expansion of Dust Halos}
A mass loading by dust of the gas flow has a significant effect on the drag
acceleration that the gas imparts to the dust particles during their liftoff
from the nucleus.  The terminal velocity $v_{\rm d}(s)$ that a particle of
radius $s$ attains at the time its motion decouples from the gas flow is a
complicated function of the interaction (Probstein 1969), but the relationship
between $v_{\rm d}$ and $\Psi_{\rm gas}$ is simplified and can be written in
closed form for a particle of trivial dimensions; for its terminal velocity,
$(v_{\rm d})_{\rm lim}$, Probstein provides an expression:
\begin{eqnarray}
(v_{\rm d})_{\rm lim} & = &  \lim_{s \rightarrow 0} v_{\rm d}(s) = \left(
 \! \frac{2\,T_{\rm gas}}{1 \!+\! \Psi_{\rm gas}} \! \right)^{\!\frac{1}{2}}
 \!\! \left\{ {\mbox{\raisebox{0cm}[2.5ex][-2.5ex]{}}} (c_{\rm p})_{\rm gas}
 \right. \nonumber \\[-0.33cm]
 & & \left. \right. \\[-0.28cm]
 & & \left. \!\!\times\! \left[ 1 \!+\! {\textstyle \frac{1}{2}} (\gamma_{\rm
 gas} \!-\! 1) \, {\mbox{\bf M}}_{\rm gas}^2 \right] \!+ c_{\rm dust}
 \Psi_{\rm gas} {\mbox{\raisebox{0cm}[2.5ex][-2.5ex]{}}} \right\}^{\!
 \frac{1}{2}} \nonumber
\end{eqnarray}
where $T_{\rm gas}$, {\bf M}$_{\rm gas}$, $(c_{\rm p})_{\rm gas}$, and
$\gamma_{\rm gas}$ are, respectively, the temperature, the initial Mach
number, the specific heat capacity at constant pressure, and the heat
capacity ratio (Poisson constant) for the gas that drives the dust away
from the nucleus, and $c_{\rm dust}$ is the specific heat capacity of the
dust particle.  We note that $(v_{\rm d})_{\rm lim}$ is identical with the
parameter $v_0$ from Equation~(35); by its substitution into Equation~(44),
we obtain for a dust loading of the water flow
\begin{equation}
\Psi_{\!_{{\rm H}_2{\rm O}}} \!=\! \frac{2(c_{\rm p})_{_{{\rm H}_2{\rm O}}}
 T_{_{{\rm H}_2{\rm O}}} \!\! \left[ 1 \!+\! {\textstyle \frac{1}{2}}(
 \gamma_{_{{\rm H}_2{\rm O}}} \!\!-\! 1) {\mbox{\bf M}}_{_{{\rm H}_2{\rm
 O}}}^2 \! \right] \!-\! v_0^2}{v_0^2 \!-\! 2 \:\! c_{\rm dust} T_{_{{\rm
 H}_2{\rm O}}}} .
\end{equation}
Since \mbox{$\gamma_{_{{\rm H}_2{\rm O}}} \!=\! 1.33$} and{\vspace{-0.04cm}}
\mbox{{\bf M}$_{_{{\rm H}_2{\rm O}}} \!\!<\! 1$} for \mbox{$\Psi_{\!_{{\rm
H}_2{\rm O}}} \!>\! 0$}, the second term in the square brackets is much smaller
than unity.  This equation is solved by computing a first-guess value of
$\Psi_{\!_{{\rm H}_2{\rm O}}}\!\!$ {\vspace{-0.06cm}}with this term ignored;
by finding an initial Mach number from an approximate formula,
%
%
\begin{equation}
{\mbox{\bf M}}_{_{{\rm H}_2{\rm O}}} \! = 1 - \frac{\Psi_{\!_{{\rm H}_2{\rm
 O}}}^{\frac{3}{5}}}{0.65 \!+\! \Psi_{\!_{{\rm H}_2{\rm O}}}^{\frac{3}{5}}};
\end{equation}
and by iterating Equations (45) and (46) until they converge; Equation~(46)
provides a reasonable approximation to \mbox{\bf M}$_{_{{\rm H}_2{\rm O}}}\!$
for all values of \mbox{$\Psi_{\!_{{\rm H}_2{\rm O}}}$}.

Our objective is now to apply Equation (45) to appropriate observations in
order to test whether the very high dust loading of the water flow that
resulted from comparison of the production rates of water and dust is
independently confirmed in this fashion.  The best approximation to a
particle of trivial dimensions, which --- according to Equation~(44) ---
is accelerated to the highest terminal velocity, is obviously provided by
the smallest ejected grains.  We already remarked that Min et al.\ (2005)
advocated the presence of grains as small as 0.01~$\mu$m in radius in order
to fit the comet's spectral energy distribution.  An overabundance in
C/1995~O1 of unusually small submicron-sized dust grains --- equal to or
smaller than 0.1~$\mu$m in radius, especially near perihelion --- was
reported by numerous researchers (e.g., Williams et al.\,1997, Harker et
al.\,1999, Lisse et al.\,1999, Hayward et al.\ 2000).  These grains,
{\vspace{-0.02cm}}with the highest terminal velocities near 700~m~s$^{-1}$,
populating the leading boundaries of a succession of expanding dust halos,
were observed extensively over a period of about two months around perihelion,
as already noted in Section~9.5.  Besides their velocities, application of
Equation~(45) requires the knowledge of the mineralogical composition of the
submicron-sized grains, because the heat capacity $c_{\rm dust}$ is strongly
temperature dependent.

The composition of microscopic dust from C/1995~O1 was examined many times
with use of a variety of techniques (e.g., Hanner et al.\ 1999, Wooden et al.\
1999, 2000, Hayward et al.\ 2000, Gr\"{u}n et al.\ 2001, Harker et al.\ 2002,
Moreno et al.\ 2003, Min et{\vspace{-0.02cm}} al.\ 2005).~Below~we~compare
the results of two very different investigations, one by Harker et al.\
(2002)\footnote{Harker et al.'s (2002) results were amended in an important
erratum that was published two years later; see the reference.} and the other
by Hayward et al.\ (2000); both dealt with the grain populations over the
critical time near perihelion.  Harker et al.\ considered five categories
of dust:\ amorphous carbon, amorphous and crystalline olivine, and amorphous
and crystalline pyroxene (orthopyroxene).  Their synthetic spectral energy
distribution over a range from 1~$\mu$m to 50~$\mu$m, designed to model
the comet at 0.93~AU from the Sun (10 days after perihelion), was fairly
consistent with the contributions, by mass, of fully 64\% of crystalline
olivine with a radiative equilibrium temperature of 220\,K; 9\% of
crystalline orthopyroxene with a temperature of 320\,K; 12\% of amorphous
pyroxene (closest to Mg$_{0.5}$Fe$_{0.5}$SiO$_3$) with 435\,K (for grains
0.1~$\mu$m in radius); and even smaller contributions from amorphous olivine
and carbon.
%
%

Hayward et al.\ (2000) combined their investigation of the thermophysical
properties of the microscopic dust in C/1995~O1 with analysis of particle
dynamics and the morphology of the dust halos.  They concluded that~the
thermal emission of the halos arose from submicron-sized particles subjected
to radiation pressure accelerations smaller than the Sun's gravitational
acceleration, which explains a fairly uniform spacing of the halos, controled
by a constant ejection velocity of the dust that populated their leading
boundaries.  These relatively low accelerations are consistent with silicate,
but not carbonaceous, material.  Whereas carbon grains accounted largely for
the continuum near 8~$\mu$m and almost exclusively for the \mbox{3--5}~$\mu$m
continuum, silicates dominated the 10~$\mu$m region of the thermal spectrum.
Unlike Harker et al.\ (2002), Hayward et al.\ concluded that amorphous
pyroxene was the most abundant silicate (at least 40\% by mass) and that the
pyroxenes contributed almost two-thirds of the silicate grain population.
Hayward et al.\ also pointed out that even though cometary silicates tend
to be magnesium rich (whose temperatures are typically below the blackbody
temperature), a strong contamination by absorbing material raises their
temperature above the blackbody temperature regardless of their intrinsic
composition.  Hayward et al.\ determined that the continuum color temperature
was elevated in the halos of C/1995~O1; measured from the spectra taken over
a period of 1997 March \mbox{24--28} it averaged 395\,K at 0.92~AU from the
Sun, implying a superheat of 1.36.

Even though carbon grains with temperatures in excess of 500\,K were present
in C/1995~O1, we dismiss their role in populating the leading boundaries of
the expanding halos because they were subjected to much higher radiation
pressure accelerations (greatly exceeding the Sun's gravitational
acceleration) than were silicate grains, thus falling increasingly behind in
the course of the halos' expansion.  We compute the specific heat capacity
$c_{\rm dust}$ for two effective dust-particle temperatures; after
normalization to 1~AU from the Sun these are 210\,K [corresponding to the
dominant particles in Harker et al.'s~(2002) model] and 380\,K [consistent
with Hayward et al.'s (2000) model].  Although radiative{\vspace{-0.08cm}}
equilibrium temperatures do not necessarily vary as $r^{-\frac{1}{2}}$,
we will use this power law~as an admissible approximation in a narrow
range of heliocentric distances near 1~AU.

The variations in the specific heat capacity of solids with the temperature
$T$ are known to folow the Debye law at very low $T$ and to converge to a
constant at very high $T$.  For our application of Equation~(45) it will
suffice to approximate $c_{\rm dust}(T)$ by
\begin{equation}
c_{\rm dust}(T) = \frac{\xi_0 T^3}{1 \!+\! \xi_1 T \!+\! \xi_2 T^2 \!+\!
 \xi_3 T^3},
\end{equation}
which is readily seen to satisfy either of the two conditions, as
\mbox{$c_{\rm dust} \simeq \xi_0 T^3$} when \mbox{$T \!\rightarrow\!
0$} and \mbox{$c_{\rm dust} \!\rightarrow\! \xi_0/\xi_3$} when \mbox{$T
\!\rightarrow\! \infty$}.  The constants $\xi_0, \ldots, \xi_3$ are
determined by fitting appropriate data.  Combining the Debye law for
low temperatures with the Dulong-Petit law for high temperatures, we
obtain from Equation~(47) an estimate for the Debye temperature, $T_D$,
\begin{equation}
T_D = 2\pi \, {^{^{\scriptstyle 3}}} \!\!\!\! \sqrt{\frac{3\pi}{10}\:\!\xi_3}\,.
\end{equation}

In practice, we employed the heat capacities that were computed by Yomogida
\& Matsui (1983) for a number of meteorites --- as comet-dust analogs --- in
a temperature range from 100\,K to 500\,K (at a 50\,K step) from their mineral
compositions and specific-heat data compiled by Touloukian (1970a, 1970b).
The numbers are known to be rather compatible with the more recent results
by Consolmagno et al.\ (2013).  To streamline{\vspace{-0.03cm}} the
cubic fit, the averaged expressions of $T^3\!/c_{\rm dust}$ for the nine
standard temperatures between 100\,K and 500\,K were linked with a synthetic
data point at \mbox{$T \!=\! 20$\,K} that was varied until a condition was
satisfied that required, in accordance with a Debye-temperature normalized
heat-capacity curve, that at $T_D$ the heat capacity reach about 95\% of its
limiting value, which in the notation of Equation~(47) equals $\xi_0/\xi_3$.
This constraint was equivalent to assigning \mbox{$c_{\rm dust}(T) =
0.0085$\,J\,g$^{-1}$K$^{-1}$} at \mbox{$T = 20$\,K}, which resulted in the
following formula for a~representative~heat capacity of the meteorites:
\begin{equation}
c_{\rm dust}(T) = \frac{2.60 \, T^3}{1 \!+\! 7.03 \, T \!+\! 0.28 \, T^2
 \!+\! 2.52 \, T^3},
\end{equation}
where $T$ is expressed in units of 100\,K and{\vspace{-0.04cm}} $c_{\rm
dust}(T)$~comes out in J\,g$^{-1}$K$^{-1}$.  For~the~Debye temperature this
relation offers \mbox{$T_D = 838$\,K} and{\vspace{-0.05cm}} \mbox{$c_{\rm
dust}(T_D) = 0.98$ J\,g$^{-1}$K$^{-1}$}, while \mbox{$(c_{\rm dust})_{\rm
lim} \!=\!  \lim_{T \rightarrow \infty} c_{\rm dust}(T) \!=\! 1.03$ J\,g$^
{-1}$K$ ^{-1}$},~so~that,~indeed, \mbox{$c_{\rm dust}(T_D)/(c_{\rm
dust})_{\rm lim} = 0.95$}.  For olivine~and~pyrox\-ene the Dulong-Petit
law indicates that $(c_{\rm dust})_{\rm lim}$ equals,{\vspace{-0.03cm}}
respectively,~1.01~and 1.07~J\,g$^{-1}$K$^{-1}$, in good agreement with
the result from Equation~(49), thus suggesting that in terms of the specific
heat capacity the meteorites are acceptable analogs for the dust in C/1995~O1.
For the critical temperatures between 210\,K and 380\,K at 1~AU from the Sun,
$c_{\rm dust}$ varies from 0.60 to 0.84~J\,g$^{-1}$K$^{-1}$, in a range that
we employ below in the applications of Equation~(45).

The two quantities for water vapor in Equation~(45) that still need to be
{\vspace{-0.06cm}}addressed are its temperature, $T_{\!_{{\rm H}_2{\rm O}}}$,
and specific heat capacity{\vspace{-0.03cm}} at constant pressure, $(c_{\rm
p})_{_{{\rm H}_2{\rm O}}}$. The tem\-perature of water vapor is lower than
the temperature of water ice on the nucleus, $T_{\rm ice}$, from which it
sublimates.  The relationship between the two is a function of the initial
Mach number, {\bf M}$_{_{{\rm H}_2{\rm O}}}$, and the heat{\vspace{-0.04cm}}
capacity ratio, $\gamma_{_{{\rm H}_2{\rm O}}}$; the gas dynamics approach
{\vspace{-0.05cm}}provides the following expression (e.g., Cercignani 1981):
\begin{eqnarray}
T_{_{{\rm H}_2{\rm O}}} & \!= & T_{\rm ice} \! \left[1 + (\pi \!+\! 1)\:\!
 \Omega^2 - 2 \pi^{\frac{1}{2}} \Omega \:\! (1\!+\!\Omega^2)^{\frac{1}{2}}
 \:\!\!\right] \nonumber \\[-0.08cm]
 & = & T_{\rm ice} \!\left[ (1 \!+\! \Omega^2)^{\frac{1}{2}} -
 \pi^{\frac{1}{2}} \Omega \right]^2 \\[-0.08cm]
 & \simeq & T_{\rm ice} \! \left( 1 - 2 \pi^{\frac{1}{2}} \Omega \right)
 \;\;\; {\rm for} \;\, \Omega \ll 1, \nonumber
\end{eqnarray}
where
\begin{equation}
\Omega = \mbox{\bf M}_{_{{\rm H}_2{\rm O}}} \:\!
 \frac{\gamma_{_{{\rm H}_2{\rm O}}} \!-\! 1}{\gamma_{_{{\rm H}_2{\rm O}}}
 \!+\! 1}
 \left( \frac{\gamma_{_{{\rm H}_2{\rm O}}}}{8} \! \right)^{\!\frac{1}{2}} \!\!
 = 0.058 \, \mbox{\bf M}_{_{{\rm H}_2{\rm O}}}.
\end{equation}
Since in the presence of dust \mbox{{\bf M}$_{_{{\rm H}_2{\rm O}}} \!<\! 1$},
$\Omega$ is{\vspace{-0.05cm}} much smaller than unity, making the approximation
in the last line of Equation~(50) reasonably accurate.

The temperature of water ice was, as a function of heliocentric distance,
determined by solving the energy balance on the comet's nucleus, using the
isothermal model (Section~4).  Within a few tenths of AU of a unit
heliocentric distance, it can be approximated with high accuracy (to
better than $\pm$0.1\,K) by an interpolation formula
\begin{equation}
T_{\rm ice} = 194.7 - 16.2 \, (r \!-\! 1) + 5.64 \, (r \!-\! 1)^2,
\end{equation}
where $r$ is in AU.  At heliocentric distances 0.9~AU to 1.1~AU the
temperature of ice varies only by 3.2\,K; even the extreme temperature at
the subsolar point is merely $\sim$10\,K higher.

The specific heat capacity of water vapor at~\mbox{constant} pressure,
$(c_{\rm p})_{_{{\rm H}_2{\rm O}}}$, is nearly independent{\vspace{-0.04cm}}
of temperature between at least 50\,K and 250\,K; to two decimal places, it
is approximated by 1.85~J\,g$^{-1}$K$^{-1}$ in this entire range (Freedman
\& Haar 1954, Wagner \& Pruss 2002, Murphy \& Koop 2005).
%
%
%
%
%

We are now ready to calculate the mass loading by dust of the water flow near
perihelion to check whether the high loading rates, exceeding 10, established
from comparison of the production rates in Table~26, are corroborated.  As
mentioned in Section~9.5, Braunstein et al.\ (1999) systematically
investigated the expansion rate of the concentric dust halos of C/1995~O1
on 13~days between 25 February 1997 (33~days before perihelion) and 27 April
1997 (26~days after perihelion).  Derived from the spacing of the leading
boundaries of successive halos and an accurately determined rotation period,
the expansion velocity was an average of the terminal velocities of the
smallest ejected grains, whose radii were estimated at 0.1~$\mu$m at the
most.  As such, this halo expansion velocity, $v_{\rm exp}$, should be by
at least several tens of meters per second lower than the limiting velocity
$v_0$.  Nonetheless, substitution of the halo expansion velocity $v_{\rm
exp}$ for the limiting velocity $v_0$ in Equation~(45) still is a very good
approximation that should provide us with a fairly tight {\it upper\/} limit
on the mass loading by dust of the water vapor flow near perihelion of
C/1995~O1.\footnote{The effect of projection onto the plane of the sky could
have slightly been overcompensated by Braunstein et al.\ (1999) in some
cases in which the halos emanated from sources that were rather far from
the subsolar latitude; still, the deprojected velocities are a much better
measure of $v_0$ than the uncorrected velocities.}
\begin{table}[t]
\vspace{-4.2cm} 
\hspace{4.22cm}
\centerline{
\scalebox{1}{
\includegraphics{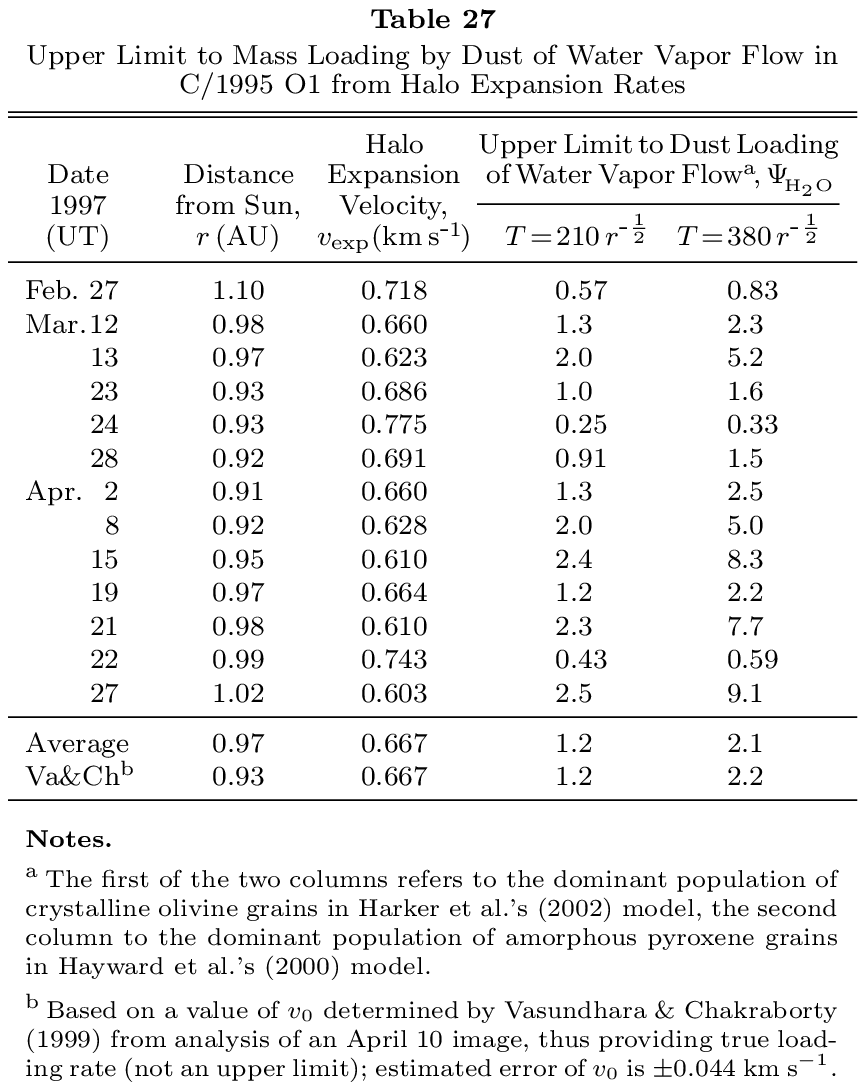}}}  
\vspace{-14.3cm}
\end{table}

Surprisingly, Table 27 demonstrates that when applied to Braunstein et al.'s
(1999) halo expansion velocities, Probstein's (1969) model provides us with
the dust loading rates $\Psi_{\!_{{\rm H}_2{\rm O}}}$ that are, on the
{\vspace{-0.04cm}}average, at least~one~order of magnitude lower than are the
expected numbers (Table~26).  A dust loading of $\sim$11 would require at
1~AU from the Sun that the expansion velocities be lower than 0.53~km~s$^{-1}$
for the dust temperature of 210\,K and lower than 0.60~km~s$^{-1}$ for 380\,K.
This effect could not possibly be caused by systematic errors in Braunstein
et al.'s (1999) expansion-velocity measurements, because a similar result
was independently obtained by Vasundhara \& Chakraborty's (1999), who
investigated the dynamics of the dust features.  Unfortunately, the April~10
image was the only image that Vasundhara \& Chakraborty analyzed from the
period of time between late February and late April of 1997, when the
straightforwardly interpretable concentric halos were observed (Braunstein
et al.\ 1999).  The result by Vasundhara \& Chakraborty is listed in Table~27
for comparison; in this case the projection issue was mute, because one of
their parameters was directly related to the tabulated value of $v_0$.  Of
interest is the statement by these authors that a {\it lower\/} limit on the
dust loading of the {\it gas\/} flow was equal to 3.4$\,\pm\,$1.0.  Given that
water was only part of the gas mix, this result contradicts a dust loading
of the water flow of \mbox{$<$1.2--2.2}, but not of $\sim$11.  Indeed, a
nominal {\it lower\/} limit on the sublimation rate that Vasundhara \&
Chakraborty {\vspace{-0.03cm}}(1999) offer for the April 10 image is
2.6$\,\times$10$^{-5}$\,g~cm$^{-2}$~s$^{-1}$, while an {\it average\/}
sublimation rate of water ice at the same heliocentric distance is, according
to the isothermal model, 1.1$\,\times$10$^{-5}$\,g~cm$^{-2}$~s$^{-1}$, or
0.4 the nominal lower limit.  Unfortunaely, the derived sublimation rate
is burdened by a very high error, so that this sublimation-rate (but not the
dust-loading) argument is rather weak.  The net result of this discussion is
that the {\small \bf sublimation of water~ice alone could not explain the high
\mbox{expansion}~\mbox{velocities}\/} of the prominent dust halos and that
{\small \bf outgassing by other parent molecules should have contributed to the
effect\/}.  This conclusion is reminiscent of, and supports, the conclusion
we made in Section 9.4 that water ice did not account for{\vspace{-0.05cm}}
more than $\frac{5}{9}$ the total mass outgassed from{\vspace{-0.05cm}}
C/1995~O1 and may have --- in an extreme case --- accounted for as little
as $\frac{1}{5}$.

We are now in a position to infer, in general terms, possible properties
of the {\small \bf missing parent molecules\/}; we search for a type of
volatiles, each of which is, on the one hand, very highly loaded with dust,
\mbox{$\Psi_{\rm gas} \rightarrow \infty$}, yet, on the other hand, can ---
near 1~AU from the Sun --- accelerate submicron-sized silicate grains to
terminal velocities of nearly 0.7~km~s$^{-1}$.  Returning to Equation~(44),
we find a limit for infinitely high dust loading,
\begin{equation}
\lim_{\Psi_{\rm gas} \rightarrow \infty} \!\!\! (v_{\rm d})_{\rm lim} = v_0
 = (2\,c_{\rm dust}T_{\rm gas})^{\frac{1}{2}}.
\end{equation}
This is a condition for the {\it temperature of a heavily loaded gas\/} as a
function of the velocity and specific heat~\mbox{capacity} of the smallest
(silicate) grains in the expanding halos, which implicitly involves the
dust-particle temperature.  However, the gas temperature is essentially
identical with the temperature of the sublimating solid (i.e., a nonwater
ice), \mbox{$T_{\rm gas} \!\rightarrow\! T_{\rm solid}$} because for a very
high mass loading by dust the initial Mach number of the gas flow approaches
zero \mbox{({\bf M}$_{\rm gas} \!\rightarrow\! 0$)} and \mbox{$\Omega
\!\rightarrow\! 0$} from Equations~(51) and (50), so that
\begin{equation}
T_{\rm solid} = \frac{v_0^2}{2\,c_{\rm dust}}.
\end{equation}

Next we find a solution to Equation (54) for a heliocentric distance of
Hayward et al.'s (2000) thermal infrared spectra, \mbox{$r = 0.92$ AU}.  
{\vspace{-0.05cm}}From Equation~(49) it follows that the specific heat
capacity \mbox{$c_{\rm dust} \!= 0.62$\,J\,g$^{-1}$\,K$^{-1}$} for the
crystalline {\vspace{-0.04cm}olivine grains that Harker et al.\ (2002)
advocated (\mbox{$T \!=\! 220$\,K}), but \mbox{$c_{\rm dust} \!=
0.85$\,J\,g$^{-1}$\,K$^{-1}$} for{\vspace{-0.01cm}} the amorphous
pyroxene grains preferred by Hayward~et~al.\ (\mbox{$T \!=\! 395$\,K}).~With
\mbox{$v_0 \!=\!  0.667$ km s$^{-1}$}\,Equation~(54) gives:
\begin{equation}
T_{\rm solid} \!= 359\,{\rm K}\;{\rm (Harker)} \;\;{\it or} \;\;
 262\,{\rm K}\;{\rm (Hayward)}.
\end{equation}

The relatively cool grains of Harker et al.\ can be ruled out, because no
ice can have a temperature of $\sim$360\,K on the surface of a comet at
a heliocentric distance of 0.92~AU, at which the blackbody temperature
is 290\,K.  However, the hot grains suggested by Hayward et al.\ do
fit the proposed hypothesis, which thus implies that {\small \bf
low-volatility$\:$ices$\:$that$\:$sublimate$\:$just$\:$below$\:$the$\:$blackbody temperature could upon release accelerate hot grains to terminal
velocities near 0.70~km~s{\boldmath $^{-1}\!$}}.  This conclusion is not
meant to contest the participation of water in the process of dust ejection,
but to emphasize that the low-volatility ices (owing to their higher
temperature upon sublimation) are critical for imparting the grains
their high velocites, otherwise unattainable.  Still hotter grains, up to
440\,K (e.g., Hanner et al.\ 1999), imply even lower $T_{\rm solid}$,
down to 253\,K.  And since $\Psi_{\rm gas}$, though high, is finite,
these values of $T_{\rm solid}$ tend to be upper limits.{\hspace{0.25cm}}

\begin{figure}[b]
\vspace{-9.7cm}
\hspace{0.67cm}
\centerline{
\scalebox{0.65}{
\includegraphics{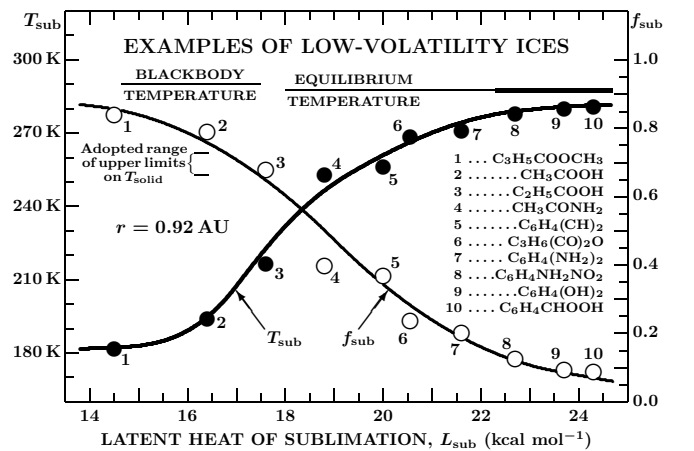}}} 
\vspace{-3.42cm}
\caption{Sublimation temperature, $T_{\rm sub}$ (solid circles), and a
fraction of the impinging solar radiation that is spent on sublimation,
$f_{\rm sub}$ (open circles), for ten organic low-volatility ices at a
heliocentric distance of 0.92~AU.  In the order of increasing heat of
sublimation, $L_{\rm sub}$, they are:\ methyl methacrylate (1), acetic
acid (2), propanoic acid (3), acetamide (4), pyrocatechol (5), glutaric
anhydride (6), m-phenylenediamine (7), nitranilin or 3-nitroaniline (8),
hydroquinone (9), and parahydroxybenzaldehyde (10).  The blackbody
temperature at this distance from the Sun is 290\,K, while the equilibrium
temperature of a nonsublimating body with a Bond albedo of 4\% and a unit
emissivity is 287\,K, a limit that no ice at that albedo and emissivity
can exceed.  Seven of the species have $T_{\rm sub}$ near or exceeding
the critical temparature $T_{\rm solid}$.  The correlation between the
sublimation heat and sublimation temperature is high, with scatter not
exceeding a few K at most.  In the displayed range of the heat of
sublimation (\mbox{14,000--25,000 cal mol$^{-1}$}), the fraction of the
solar energy that is spent on the sublimation (as opposed on the thermal
reradiation) spans about one order of magnitude.  A range of upper limits
on the ice temperature $T_{\rm solid}$ that supports a dust-grain velocity
of 0.667~km s$^{-1}$ is also depicted.{\vspace{-0.35cm}}}
\end{figure}
\begin{table*}
\vspace{-4.21cm}
\hspace{-0.52cm}
\centerline{
\scalebox{1}{
\includegraphics{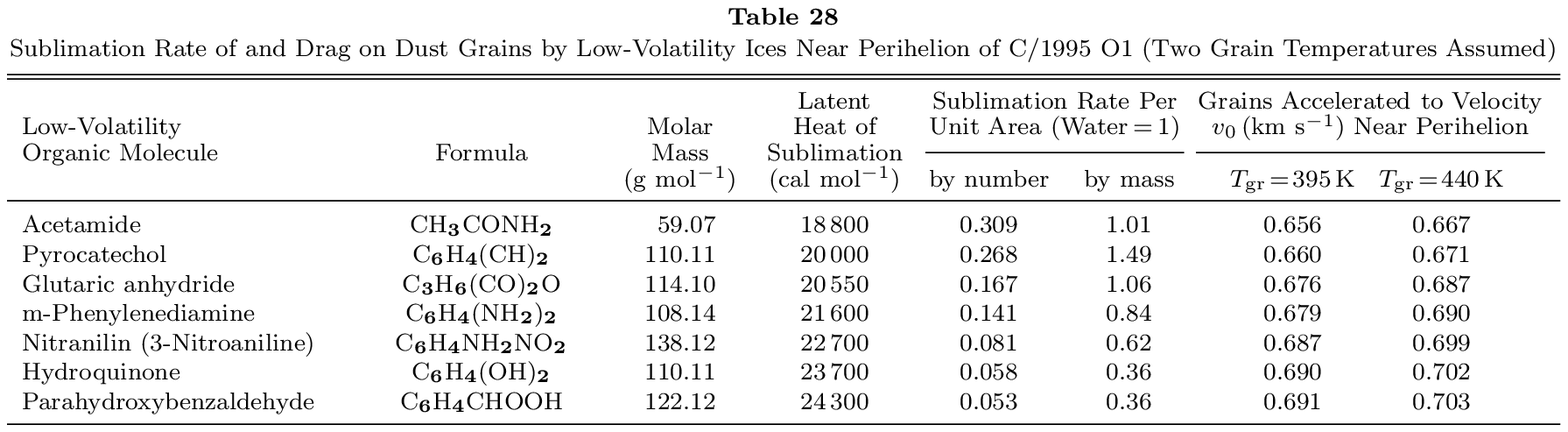}}} 
\vspace{-19.8cm}
\end{table*}

In an effort to further learn about these low-volatility ices, in Figure~16
we present a plot of sublimation temperatures, $T_{\rm sub}$, at a
heliocentric distance of 0.92~AU for ten organic molecules whose sublimation
heat, $L_{\rm sub}$, exceeds 14,000 cal mol$^{-1}$, which is about 20\%
higher than the sublimation heat of water ice.  The molecules were selected
out of a very limited pool of organic compounds for which --- in addition
to $L_{\rm sub}$ --- the saturated pressure was available as a function of
temperature from the NIST webbook (see footnote~10) in terms of the three
constants of the Antoine equation, even though sometimes not, unfortunately,
covering the needed temperature range, thus requiring an extrapolation.
Figure~16 shows a high degree of correlation between the two quantities.
Of particular interest are the critical values of $T_{\rm sub}$ and the
corresponding values of $L_{\rm sub}$ that are close to and higher than
$T_{\rm solid}$.  For our suggested range of \mbox{253\,K$\:\leq \:\!\!
T_{\rm solid} \:\!\!  \leq\:$262\,K} one finds~that \mbox{$T_{\rm sub} \!=\!
T_{\rm solid}$} for the ices with a sublimation heat between about 19\,000
and 20\,000~cal~mol$^{-1}$.  All ices with a higher sublimation heat
accelerate minuscule dust particles to terminal velocities that are higher
than 0.667~km~s$^{-1}$, while the ices with only a moderately lower
sublimation heat still should support velocities of about this magnitude.

Not all species in the range of Figure~16 should necessarily fit the curve of
$T_{\rm sub}(L_{\rm sub})$ as closely as do the ten molecules plotted, and the
relationship should not be extrapolated to \mbox{$L_{\rm sub}\!< 14\,000$ cal
mol$^{-1}$}; indeed, water ice has on the given assumptions (isothermal model
and the adopted albedo and emissivity) at 0.92~AU a temperature of 196\,K
and would leave an offset of more than 15\,K from the curve.  In fact, the
sublimation temperature of water is a bit higher than those of the two most
volatile molecules plotted in Figure~16.

At the other end of the sublimation-heat range, the curve can confidently be
extrapolated (with an error of a few K), because of the blackbody temperature
limit.~A stronger constraint is provided by the equilibrium temperature of
a nonsublimating body, if the albedo is higher than the emissivity drop from
unity.  For 0.92~AU from the Sun, the equilibrium temperature of our
isothermal object is 287\,K.  The temperature of the least volatile ice
among the ten is only 6\,K below this limit.

Also plotted in Figure 16 are the variations with $L_{\rm sub}$ in the
fraction of the incident solar radiation that is being spent on the ice's
sublimation, $f_{\rm sub}$.  When \mbox{$f_{\rm sub} \!\rightarrow 0$}, the
solar energy is spent in its entirety of reradiation (heat transport beneath
the nucleus' surface being neglected), the surface temperature of the ice
varies~as~an inverse square root of heliocentric distance, and the sublimation
rate varies exponentially.  When \mbox{$f_{\rm sub} \!\rightarrow 1$}, the
Sun's radiation is spent entirely on sublimation,~the~surface temperature is
nearly constant, and the sublimation rate varies as an inverse square of
heliocentric distance.  The first three most volatile species in Figure~16
use at 0.92~AU a greater fraction of the energy on sublimation, whereas the
opposite is the case with the seven least volatile molecules.  The fraction
$f_{\rm sub}$ spans about one order of magnitude between the most and the
least volatile ices in the figure and the sublimation and reradiation break
even for a sublimation heat of $\sim$18\,800~cal~mol$^{-1}$, when
\mbox{$f_{\rm sub} \!=\! \frac{1}{2}$}.{\vspace{0.05cm}}

Sublimation temperatures close to or above the~\mbox{critical} temperature
$T_{\rm solid}$ that are necessary to accelerate the hot submicron-sized
silicate grains to terminal velocities near 0.7~km~s$^{-1}$ are reached by
only seven among the molecules plotted in Figure~16, one of them marginally.
These are merely examples among a sheer number~of~ex\-isting organic compounds,
for most of which needed data are unavailable.  Listed in Table~28, these
seven molecules illustrate what types of species, whose existence was implied
in Section~9.4, are believed to have amply contrib\-uted to the activity of
C/1995~O1.  They are, first~of~all, {\small \bf low-volatility ices, with
a~\mbox{minimum}~\mbox{sublimation}~heat near 19\,000 cal mol{\boldmath
$^{-1}$}}; in addition, they are mostly~{\small \bf very heavy molecules,}
several times heavier than water, with the molar mass typically exceeding 100;
as a corollary, their {\small \bf sublimation rates per unit surface~area are,
by mass,~comparable to that of water ice}, even though they are much lower
in terms of number of molecules because of their lower volatility.  Given the
large number of more simple organic compounds already detected in C/1995~O1,
their {\small \bf enormous~overall~\mbox{number}~and~\mbox{variety} make their
summary major~\mbox{contribution} to~\mbox{activity}} compelling, whereas a
{\small \bf sublimation heat from 19\,000 to 25\,000~cal~mol{\boldmath
$^{-1}$}} for many of them makes, near 1~AU from the Sun, this population
of organic~\mbox{compounds} to be both {\small \bf sufficiently volatile to
outgas vigorously}~and~to {\small \bf sublimate at temperatures high enough to
\mbox{accelerate} hot submicron-sized grains of dust to high~\mbox{terminal}
\mbox{velocities}} to explain observational evidence that is inconsistent with
water ice-based activity.  While many may never be detected individually, we
propose that, in their sum, these ices are the {\small \bf molecular compounds
that were responsible for the incompleteness of the integrated mass-loss
distribution} in Section~9.4 (Figure~14).

Large organic molecules whose sublimation heat exceeds
$\sim$25\,000~cal~mol$^{-1}$ sublimate too insignificantly at heliocentric
distances near 1~AU, with their role becoming progressively marginalized.
We do not attend to this category of material, although it becomes
increasingly important for comets with perihelion distances~$\ll$1\,AU.

Even though none of the seven molecules in Table~28 has been detected in
C/1995~O1 (or any other comet), one molecule --- ethylene glycol, with a
sublimation heat estimated at $\sim$18\,200~cal~mol$^{-1}$, near the lower
boundary of the category of organic ices that is proposed to drive hot
microscopic dust into the atmosphere --- is known (Crovisier et al.\ 2004a),
as already pointed out.  

\section{Ramifications, and Future Work}
The proposed population of heavy organic molecules of low volatility has
implications for the dynamical behavior of the nucleus of C/1995~O1.
Although individually the abundances of these compounds represent only a
minor fraction of the water abundance, their sheer number and a variety
of their composition may make them to summarily contribute a mass of gas 
that is comparable to --- if not greater than --- the mass of water vapor.
Accordingly, the heavy organic low-volatility ices could upon their release
add substantially to the total momentum that the nucleus is exposed to while
revolving about the Sun and thus augment the nongravitational effect that is
measured as part of the orbit determination.

Part II of this investigation will focus on the issues related to the nucleus
of C/1995~O1.  Because of an enormous size of the nucleus (see Szab\'o et
al.\ 2011, 2012 for updates), the sublimation-driven nongravitational effects
were expected to be trivial and their eventual incorpora\-tion into the
equations of motion was accompanied by much reluctance (e.g.,\,Marsden 1999).
Nonetheless,~their introduction turned out to be essential to improving the
orbital solution; although our preferred Weight System~II (Table~10)
offers a total{\vspace{-0.07cm}} nongravitational acceleration of \mbox{$A
= \!\sqrt{A_1^2 \!+\! A_2^2 \!+\!  A_3^2} = (0.707 \!\pm\! 0.039) \! \times
\! 10^{-8}$AU day$^{-2}$}{\vspace{-0.07cm}}~at 1\,AU from{\vspace{-0.05cm}}
{\vspace{-0.04cm}}the Sun, only $\sim\!\!\frac{2}{3}$ the acceleration derived
by {\vspace{-0.04cm}}Marsden (1999) and $\frac{1}{2}$ that determined by
Szutowicz et al.\ (2002) and by Kr\'olikowska (2004) from much shorter orbital
arcs, the effect on the orbital velocity integrated over one revolution about
the Sun still reaches as much as \mbox{$2.46 \pm 0.14$ m s$^{-1}$}.  In
fact, Sosa \& Fern\'andez (2011) pointed out that while they employed the
nongravitational effects to successfully predict masses and dimensions for a
number of long-period comets, the application to C/1995~O1 failed utterly, an
exception they admitted was a puzzle:\ the predicted diameter of the nucleus
was a factor of seven much too small and the fraction of the surface that
was active came out to be more than seven times the predicted surface area!
Given that the nongravitational acceleration is unquestionably genuine and
well determined, the {\it qualitative\/} explanation offered by Sosa \&
Fern\'andez is the comet's {\it extreme hyperactivity\/} and, parenthetically,
a smaller nucleus.  If so, its enormous cross-sectional area of 4300~km$^2$
(Szab\'o et al.\ 2011,~2012) is still unexplained.  The nature of the orbital
motion of C/1995~O1 thus remains an enigma to this day.

With all available evidence to be scrutinized in Part~II, there appears to
be no escape from a conclusion that the nucleus was an unresolved and rather
compact cluster of fragments (some possibly still in contact) with dimensions
of up to at least 10~km, the detected nongravitational acceleration
being that imparted to the most~massive object.  Even though the
nongravitational accelerations on the active fragments that were less
massive were accordingly higher, they triggered perturbations relative to
the central mass that remained undetected.
%
%

\section{Conclusions}
The prime conclusions from this first part of a comprehensive investigation
of C/1995~O1 are as follows:

(1) With thousands of astrometric observations covering an orbital arc
of 17.6~yr, C/1995~O1 offered~us~an~opportunity to derive a set of orbital
elements of exception\-ally high quality.  As a result of an extensive
examination of the data and in-depth dynamical analysis, we present two
nominal nongravitational sets of orbital elements, which are distinguished
from each other by different systems of weighting the critical observations
at both ends of the orbital arc.  The superior solution, based on Weight
System~II (the single pre-discovery position of weight 20, the last three
positions of weight 15 each), fits 1950~obser\-vations from 1993--2010 with
an rms of $\pm$0$^{\prime\prime\!}$.64; 1631 observations with a residual
exceeding 1$^{\prime\prime\!}$.5 in either coordinate were eliminated in
the process of orbit improvement.

(2) Our orbit-determination method fully accommo\-dates Marsden's (1999)
proposal that the comet under\-went a preperihelion close encounter with
Jupiter in the previous return to the Sun, in the 23rd century BCE, and
that it was then captured from an Oort-Cloud-type orbit.  Although we
establish unequivocally that the encounter and capture occurred in $-$2251,
three Jovian revolutions earlier than originally determined by Marsden
(1999), we find his proposal entirely plausible, legitimate, and~very
compelling. The perijove distance equaled 10.73~Jupi\-ter's equatorial
radii (= 767\,100~km = 0.005128~AU).

(3) The nongravitational parameters of the solution based on Weight
System~II imply that an orbit-integrated sublimation-driven effect on
the comet's orbital velocity amounted to 2.46~m~s$^{-1}$, with the radial
component positive and dominant, the normal component negative and equal
to about $\frac{3}{4}$ the radial component, and the transverse component
two orders of magnitude smaller and poorly determined.  The scaling
distance of the modified nongravitational law applied equals 15.4~AU,
$\sim$5.5 times the scaling distance for water ice.

(4) We computed the ephemeris for the comet's~\mbox{return} to perihelion in
$-$2250 and found it to have~been~a~fairly favorable apparition for the
northern-hemisphere observ\-er, with a peak apparent brightness estimated
at $-$2.~We painstakingly searched for a historical record that might
refer to C/1995~O1, but the time-frame uncertainties in the 23rd century
BCE prevented us from making any definite identification.

(5) The orbital motion of C/1995 O1 was extrapolated forward in time, with
a prediction for the next return to perihelion in the year 4393 and
afterwards in 5451.  It is noted not only that the actual orbital period is
getting shorter with time, from 4246~yr between the years $-$2250 and 1997,
to 2396~yr between the years 1997 and 4393, and to 1058~yr between the years
4393 and 5451; but that the comet's orbital evolution is and will for some
time be driven by orbital-cascade resonance with Jupiter, from 1:358 to
1:202 and then to approximately 1:89.  We studied this dynamical
process recently in connection with a pair of long-period comets and
singled it out as momentous for a rapid inward drifting of aphelia of
comets in nearly-parabolic orbits (Sekanina \& Kracht 2016).

(6) Voluminous data on the production rates of water and other parent
molecules sublimating from the nucleus of C/1995~O1 over a wide range of
heliocentric distances as well as data on the brightness (as a proxy) at~very
large distances from the Sun allowed us to show that the variations in
activity were expressible by the same~type~of law as the nongravitational
acceleration in the comet' orbital motion, but with different values of
the scaling distance that is a function of the latent heat of sublimation.

(7) The comet's activity was asymmetric relative to per\-ihelion, especially
at very large heliocentric distances.  The carbon-monoxide-driven activity
is described by a scaling distance of 12~AU before perihelion, but 20~AU~after
perihelion.  The scaling distance of the orbital solution with Weight System
II is almost exactly halfway~between the two.  The ways of incorporating the
perihelion asymmetry into the nongravitational law are addressed.

(8) Employing appropriate scaling distances that fit the gas-abundance
observations, the total mass lost per orbit of C/1995~O1 was determined for
three{\vspace{-0.04cm}} parent mole\-cules, equaling 3.4$\,\times$10$^{15}$\,g
for water, 1.1$\times$10$^{15}$\,g for carbon monoxide, and
0.8$\,\times$10$^{15}$\,g for carbon dioxide.  In their sum, they are
equivalent to the mass of a sphere 1.7~km in diameter at a density of
0.4~g~cm$^{-3}$, with water accounting for less than $\frac{2}{3}$ of
the total mass.

(9) Outgassing by a number of additional, mostly~organic, parent molecules
identified in C/1995~O1 (as com\-piled by Bockel\'ee-Morvan et al.\ 2004)
was examined~in the same fashion.  For each of 18 trace~\mbox{compounds},~the
mass lost per orbit was computed from the nongravitational law with a
scaling distance determined from~the latent heat of sublimation, and their
sum found to~add another $\sim$0.7$\,\times$10$^{15}$g to the total
production of gas;~the trace molecules searched for but not detected
were not counted.  Because of their high molar mass, the organic
molecules contribute disproportionately more masswise, compared to their
{\vspace{-0.01cm}}abundances.~Their mass brings the total mass lost per
orbit by outgassing to 6$\,\times$10$^{15}$g in the least and the share
of water to not more than 57\%.

(10) On account of a sheer number and variety of compounds into which the
elements C, H, O, N, S (possibly others) could combine, and because of a
large tally of as yet unidentified bands in the spectrum, particularly in
the near-infrared, submillimeter, and microwave regions, one can expect the
presence in the nucleus of numerous additional, mostly organic, ices currently
unknown.~We first examined the number of observed species as a function
of their orbit-integrated mass loss relative to water.  The data for C/1995~O1
(by far the most extensive for any comet) suggested that the cumulative
number of species followed a power law of the normalized mass loss per~orbit
with a slope near $-$0.7 and that, indeed, this sample appeared to be
substantially incomplete, by a factor of up to five for the molecules
whose total mass loss per orbit was $\geq$1\% of the mass loss of water
ice, by a factor of up to 19 for the molecules whose total mass loss was
$\geq$0.1\% of that of water ice, and by a factor of as much as 100 for the
molecules whose total mass loss was $\geq$0.01\% of that of water ice.
One could expect that in C/1995~O1 the number of parent molecules of the
last category may be as high as several hundred.  For comparison, the
number of molecular compounds detected in interstellar and/or
circumstellar space is about 200.

(11) The described exercise also allows one to estimate the comet's
total mass loss per orbit by outgassing by summing up the contributions
from all species.  Because of fairly large uncertainties involved, it
is possible to provide only probable limits, which are about 2 to 5
times the mass loss of water per orbit, with the upper limit
representing merely an educated guess.

(12) To check a maximum possible number of compounds produced under certain
constraints, we developed a simple mathematical model for generating
molecular formulas by combining specific elements with an arbitrary but
limited number of atoms per element.  We ascertained that a mathematically
complete set of compounds consisting of up to six atoms of carbon and
hydrogen, up to two to three atoms of oxygen, and up to one atom of
nitrogen and sulfur --- mimicking the limits for most compounds
detected in the interstellar and/or circumstellar medium --- numbered
570--770 molecules (not counting isomers but including formulas that refer
to no genuine chemicals), on the same order of magnitude as the number
of parent molecules derived from the outgassing mass-loss distribution of
species.

(13) Examining the production rates of dust, based on submillimeter and
microwave observations, we found that they too can be fitted by the same
type of law~as the nongravitational acceleration in the comet's orbital
motion.  Establishing that the comet's production of dust integrated over
the orbit about the Sun amounted to some 4.0$\times$10$^{16}$\,g (in fair
agreement with the estimate by Jewitt \& Matthews 1999) and comparing it
with the orbit-integrated production of parent mole\-cules, we concluded
that the mass loading by dust of the water flow (the dust-to-water mass-loss
ratio) integrated over the orbit about the Sun reached 11.8, while the dust
loading of the combined flow of water, carbon monoxide, and carbon dioxide
equaled 7.5.  The dust loading of the total gas flow from C/1995~O1 may have
been as low as 3.

(14) Another piece of evidence for the existence of a large number of parent
molecular compounds in the nucleus of C/1995~O1 is provided by the terminal
veloc\-ities of submicron-sized silicate grains that populated the
comet's expanding halos, observed for about two months near perihelion.
The halos expanded with velocities close to 0.7~km~s$^{-1}$, which --- as
implied by a hydrodynamic model --- is too high to be accounted for by a drag
acceleration exerted by sublimating water molecules, given the extremely high
mass loading of the water flow by the dust.  Probstein's (1969) theory was
used to compute an upper limit on the expected mass loading by dust from the
observed halo-expansion velocities that was typically one order of magnitude
lower than the dust loading derived directly from the production data.  This
contradiction provides strong evidence that the terminal velocities of
submicton-sized silicate particles were supported not only by the expanding
flow of water vapor but by additional sublimating matter as well.

(15) Further application of the hydrodynamic model suggests that a likely
critical driver of the hot microscopic silicate dust in the halos of
C/1995~O1 was~a~mixture~of heavy organic molecules of relatively low
{\vspace{-0.03cm}}volatility (with a range of sublimation heat of
\mbox{$\sim$19\,000--25\,000~cal~mol$^{-1}$)} with
sublimation temperatures just below the~black\-body temperature at
heliocentric distances near 1~AU.

(16) These heavy organic molecules of low volatility appear to be
identical with the missing compounds that we found to be responsible for
a major incompleteness of the observed distribution of sublimating ices
sorted by the relative orbit-integrated mass loss.  In Part~II we will
argue that the presence of these as yet undetected species should also
slightly alleviate the problem of the sizable nongravitational effects
detected in the orbital motion of C/1995~O1, although the primary source
of the disproportion of the momentum transferred to the nucleus by the
mass lost by outgassing appears to be the nature of the nucleus as a
compact cluster of massive fragments in near-contact with one another,
a bold hy\-pothesis --- to be addessed in Part~II --- that, given the
strong evidence, is extremely difficult to avoid.\\[-0.1cm]

This project, whose orbital part was inspired by the insightful paper
written by the late B.\ G.\ Marsden~in 1998, was initiated by the first
author (ZS) years ago in collaboration with P.\ W.\ Chodas, whose
contribution is hereby acknowledged and greatly appreciated.  The
orbital results presented in this paper were achieved~thanks primarily
to the second author (RK). This~research was carried out in part at the
Jet Propulsion Labo\-ratory, California Institute of Technology, under
contract with the National Aeronautics~and~Space~Ad\-min\-istration.\\[-0.2cm]

\begin{center}
APPENDIX A\\[0.15cm]
MOST RECENT EFFORTS TO OBSERVE\\COMET C/1995 O1\\
\end{center}
\vspace{-0.2cm}

\indent
The 2012 attempt by D.\ Herald to observe this comet (Section~2) was not
the only effort of this kind, following the 2010 astrometric observations
made with the 220-cm f/8.0 Ritchey-Chr\'etien telescope at the European
Southern Observatory (ESO) (S\'arneczky et al.\ 2011; Szab\'o et al.\ 2011).
The comet was in fact observed with the 820-cm Antu unit of the ESO's~Very
Large Telescope (VLT) as a stellar object of an average $V$ magnitude of
24.20 and $R$ magnitude 23.72 on 2011 Oct 5, 23, and 25 (Szab\'o et al.\
2012), but no astrometry is available.  Further information is provided
in a blog by N.\ Howes, G.\ Sostero, and E.\
Guido,\footnote{\tt http://remanzacco.blogspot.it/2012/10/deep-south.html.}
who employed remotely the 200-cm f/10.0 Siding Spring-Faulkes South telescope
to search for the comet on five occasions between 2012 Sept 25 and Oct 9;
it was detected on none of 45 to 75-minute images with a limiting magnitude
from 21.5 to 23.5.  All these efforts point consistently to a conclusion
that the object of magnitude 21.8, exposed by Herald with his 40-cm
reflector on 2012 Aug 7, could not be the comet, in agreement with the
observer's own final statement.\\[-0.2cm]

\begin{center}
APPENDIX B\\[0.15cm]
SHAPE OF THE CURVE $\Im({\sf X})$ AS FUNCTION OF\\A COMET'S
 PERIHELION DISTANCE\\
\end{center}
\vspace{-0.2cm}

\indent
While, in general, the $g_{\rm mod}(r;r_0)$ function is integrated numerically,
approximate values of $\Im({\sf X})$ can be found in closed form in the
hypothetical cases of an infinitely volatile species, when \mbox{$L_{\rm sub}
\! \rightarrow \! 0$} (\mbox{$r_0 \! \rightarrow \! \infty$}), and an
infinite\-ly refractory species, when \mbox{$L_{\rm sub} \! \rightarrow \!
\infty$} (\mbox{$r_0 \! \rightarrow \! 0$}).  Expressing the normalization
coefficient $\psi$ in terms of $r_0$ and the exponents $m$, $n$, and $k$,
\begin{equation}
g_{\rm mod}(r;r_0) = r^{-(m + nk)} \! \left( \frac{r_0^{-n} \!+\! 1}{r_0^{-n}
 \!+\! r^{-n}} \! \right)^{\!\!k} \! \! = r^{-m} \! \left(\! \frac{r_0^n + 1}
 {r_0^n  +  r^n} \! \right)^{\!\!k} \!\!,
\end{equation}
we find
\begin{eqnarray}
\lim_{r_0 \rightarrow \infty} g_{\rm mod}(r;r_0) & = & r^{-m} \;\;\;\;\;\;\;
 \;\;\; {\rm for} \; L_{\rm sub} \! \rightarrow \! 0, \nonumber \\[-0.33cm]
 & & \\[-0.33cm]
\lim_{r_0 \rightarrow 0} g_{\rm mod}(r;r_0) & = & r^{-(m + nk)} \;\;
 {\rm for} \; L_{\rm sub} \! \rightarrow \! \infty. \nonumber
\end{eqnarray}
Thus, the integral $\Im(L_{\rm sub})$ from Equation~(14) may in either case be
written as
\begin{eqnarray}
\Im & = & \frac{2}{k \sqrt{p}} \int_{0}^{\pi} \!\!\left( \! \frac{1 \!+\! e
 \cos v}{p} \!\right)^{\!\!\zeta} \! dv  \nonumber \\[-0.2cm]
 & & \\[-0.2cm]
 & = & \frac{2^{\zeta + 2}}{k} \,p^{-(\zeta + \frac{1}{2})} \!\!\!
 \int_{0}^{\frac{1}{2}\pi} \!\!\! \cos^{2 \zeta} \! u \! \left[ 1 \!-\!
 \Delta e (1 \!-\! {\textstyle \frac{1}{2}} \sec^2 \! u) \right]^\zeta \! du ,
 \nonumber
\end{eqnarray}
where \mbox{$u = \frac{1}{2}v$}, \mbox{$\Delta e = 1 \!-\! e$} and \mbox{$\zeta
= m \!-\! 2$} for \mbox{$L_{\rm sub} \!\rightarrow \! 0$} and \mbox{$\zeta =
m \!+\! nk \!-\! 2$} for \mbox{$L_{\rm sub} \!\rightarrow \! \infty$}.  For
C/1995~O1 as a nearly parabolic comet, \mbox{$\Delta e \!\rightarrow \! 0$}
and the term with $\Delta e$ has the nature of a minor correction to the
$\cos^{2 \zeta}\!u$ term.  Let us re\-place the variable $u$ with an
appropriate constant value of $\langle u \rangle$, so that \mbox{$\sec^2
\langle u \rangle = \eta$} and the integration is readily executable.  We
obtain
\begin{eqnarray}
\Im & = & \frac{\sqrt{\pi}}{k} \, 2^{\zeta+1} p^{-(\zeta+\frac{1}{2})} \,
 \frac{\Gamma(\zeta \!+\! \frac{1}{2})} {\Gamma(\zeta \!+\! 1)}
 \left[ \raisebox{0cm}[0.75ex][0.75ex]{}1 - \zeta (1 \!-\! {\textstyle
 \frac{1}{2}} \eta) \Delta e \right. \nonumber \\[-0.25cm]
    &   &  \\[-0.25cm]
    &   & \left. +  {\textstyle \frac{1}{2}} \zeta (\zeta \!-\! 1) (1 \!-\!
 {\textstyle \frac{1}{2}} \eta)^2 (\Delta e)^2 - \, \ldots \,\, \right],
 \nonumber
\end{eqnarray}
where $\Gamma$ is the Gamma function.

Let us now insert the exponents $m$, $n$, and $k$ for $\zeta$ and calculate
the ratio of the two limiting equivalent times; we obtain
\begin{eqnarray}
\frac{\Im_\infty}{\Im_0} & = & \frac{\lim_{L_{\rm sub} \rightarrow \infty}
 \Im(L_{\rm sub})}{\lim_{L_{\rm sub} \rightarrow 0} \Im(L_{\rm sub})}
 \nonumber \\[0.1cm]
 & = & \left( \frac{2}{p} \right)^{\!\! nk} \frac{\Gamma(m \!+\! nk \!-\!
 \frac{3}{2})}{\Gamma(m \!+\! nk \!-\! 1)} \, \frac{\Gamma(m \!-\!
 1)}{\Gamma(m \!-\! \frac{3}{2})} \nonumber \\[-0.22cm]
 & & \\[-0.24cm]
 & & \times \left\{ \raisebox{0cm}[1ex][1ex]{} 1 \!-\! \left[ {\textstyle
 \frac{1}{2}} (m \!-\! 2)(\eta_0 \!-\! \eta_\infty) \right. \right.
 \nonumber \\[-0.05cm]
 & & \!\!\! \left. \left.  \raisebox{0cm}[0.75ex][0.75ex]{}
 + nk(1 \!-\!  {\textstyle \frac{1}{2}} \eta_\infty) \right] \Delta e +
 \raisebox{0cm}[1ex][1ex]{} \ldots \, \right\}, \nonumber
\end{eqnarray}
where $\eta_0$ is the value of $\eta$ for \mbox{$L_{\rm sub} \rightarrow 0$}
and $\eta_\infty$ its value for \mbox{$L_{\rm sub} \rightarrow \infty$}.  By
numerical integration we determine that \mbox{$\Im_0 = 231.07$ days},
requiring \mbox{$\eta_0 = 19.36$} and therefore \mbox{$\langle u_0 \rangle =
76^\circ\!.9$}.  On the other hand, \mbox{$\Im_\infty = 260.98$ days}, so
that \mbox{$\eta_\infty = 1.02$} and \mbox{$\langle u_\infty \rangle =
8^\circ\!.0$}.

Equation (60) shows that the ratio $\Im_\infty/\Im_0$ is a very steep function
{\vspace{-0.043cm}}of the orbit parameter \mbox{$p = q (1\!+\!e)$}, varying as
$p^{-23.5}$.  For parabolic orbits the $\Delta e$ correction term disappears
and \mbox{$p = 2q$}, so that \mbox{$\Im_\infty/\Im_0 = 0.138\,q^{-23.5}$}.
This formula implies that \mbox{$\Im_\infty = \Im_0$} for a parabolic orbit
with \mbox{$q = 0.919$ AU}, only very slightly larger than the perihelion
distance of C/1995~O1.  On the other hand, \mbox{$\Im_\infty = 10 \, \Im_0$}
for \mbox{$q = 0.833$ AU} and \mbox{$\Im_\infty = 100 \, \Im_0$} for \mbox{$q
= 0.756$ AU}.  Thus the shorter the perihelion distance is, the more dominant
becomes the orbit-integrated mass loss of more refractory species in
comparison with more volatile species, much more so than their reported
production-rate ratio $\Re_{\rm Q}$ indicates.

\begin{center}
{\footnotesize REFERENCES}
\end{center}
\vspace*{-0.4cm}
\begin{description}
{\footnotesize
\item[\hspace{-0.3cm}]
Acree, W., \& Chickos, J. S. 2016, J. Phys. Chem. Ref. Data, 45,{\linebreak}
 {\hspace*{-0.6cm}}033101
\\[-0.57cm]
\item[\hspace{-0.3cm}]
A'Hearn, M. F., Millis, R. L., Schleicher, D. G., et al.\ 1995,
 Icarus,{\linebreak}
 {\hspace*{-0.6cm}}118, 223
\\[-0.57cm]
\item[\hspace{-0.3cm}]
Azreg-A\"{\i}nou, M. 2005, Monatshefte f\"{u}r Chemie, 136, 2017;
 also,{\linebreak}
 {\hspace*{-0.6cm}}2014, eprint arXiv:1403.4403
\\[-0.57cm]
\item[\hspace{-0.3cm}]
Baldet, F. 1950, Annuaire pour l'an 1950, Bureau des Longitudes,{\linebreak}
 {\hspace*{-0.6cm}}B.1
\\[-0.57cm]
\item[\hspace{-0.3cm}]
Baldet, F., \& de Obaldia, G. 1952, Catalogue G\'en\'eral des
 Orbites{\linebreak}
 {\hspace*{-0.6cm}}de Com\`etes de l'an $-$466 \`a 1952. (Paris:\ Centre
 National de la{\linebreak}
 {\hspace*{-0.6cm}}Recherche Scientifique)
\\[-0.57cm]
\item[\hspace{-0.3cm}]
Biver, N., Bockel\'ee-Morvan, D., Colom, P., et al.\ 1997, Science,{\linebreak}
 {\hspace*{-0.6cm}}275, 1915
\\[-0.57cm]
\item[\hspace{-0.3cm}]
Biver, N., Bockel\'ee-Morvan, D., Colom, P., et al.\ 1999, Earth Moon{\linebreak}
 {\hspace*{-0.6cm}}Plan., 78, 5
\\[-0.57cm] 
\item[\hspace{-0.3cm}]
Biver, N., Bockel\'ee-Morvan, D., Crovisier, J., et al.\ 2002, Earth{\linebreak}
 {\hspace*{-0.6cm}}Moon Plan., 90, 5
\\[-0.57cm] 
\item[\hspace{-0.3cm}]
Bockel\'ee-Morvan, D., \& Rickman, H. 1999, Earth Moon Plan., 79,{\linebreak}
 {\hspace*{-0.6cm}}55
\\[-0.57cm] 
\item[\hspace{-0.3cm}]
Bockel\'ee-Morvan,\,D., Crovisier,\,J., Mumma,\,M.\,J., \& Weaver,\,H.\,A.{\linebreak}
 {\hspace*{-0.6cm}}2004, in Comets II, ed.\,M.\,C.\,Festou,\,H.\,U.\,Keller,\,\&\,H.\,A.\,Weaver{\linebreak}
 {\hspace*{-0.6cm}}(Tucson, AZ: University of Arizona), 391
\\[-0.57cm] 
\item[\hspace{-0.3cm}]
Braunstein,\,M., Womack,\,M., Deglman,\,F., et al.\,1999, Earth Moon{\linebreak}
 {\hspace*{-0.6cm}}Plan., 78, 219
\\[-0.57cm]
\item[\hspace{-0.3cm}]
Breasted, J. H. 1906, Ancient Records of Egypt (Second Series).{\linebreak}
 {\hspace*{-0.6cm}}(Chicago:\ University of Chicago Press)
\\[-0.57cm]
\item[\hspace{-0.3cm}]
Brueckner, G. E., Howard, R. A., Koomen, M. J., et al.\ 1995,{\linebreak}
 {\hspace*{-0.6cm}}Solar Phys., 162, 357
\\[-0.57cm]
\item[\hspace{-0.3cm}]
Capria, M. T., Coradini, A., \& De Sanctis, M. C. 1997, in Inter-{\linebreak}
 {\hspace*{-0.6cm}}actions Between Planets and Small Bodies, XXIII IAU General
 {\hspace*{-0.6cm}}Assembly, Joint Disc.\ 6, 39
\\[-0.57cm]
\item[\hspace{-0.3cm}]
Capria, M. T., Coradini, A., \& De Sanctis, M. C. 2002, Earth{\linebreak}
 {\hspace*{-0.6cm}}Moon Plan., 90, 217
\\[-0.57cm]
\item[\hspace{-0.3cm}]
Cercignani, C. 1981, Progr.\ Astronaut.\ Aeronaut., 74, 305
\\[-0.57cm]
\item[\hspace{-0.3cm}]
Colom, P., G\'erard, E., Crovisier, J., et al.\ 1999, Earth Moon Plan.,{\linebreak}
 {\hspace*{-0.6cm}}78, 37
\\[-0.57cm] 
\item[\hspace{-0.3cm}]
Combi, M. R., Reinard, A. A., Bertaux, J., et al.\ 2000, Icarus,~144,{\linebreak}
 {\hspace*{-0.6cm}}191
\\[-0.57cm] 
\item[\hspace{-0.3cm}]
Consolmagno, G. J., Schaefer, M. W., Schaefer, B. E., et al.\ 2013,{\linebreak}
 {\hspace*{-0.6cm}}Plan.\ Space Sci., 87, 146
\\[-0.57cm]
\item[\hspace{-0.3cm}]
Cottin, H., \& Fray, N. 2008, Space Sci.\ Rev., 138, 179
\\[-0.57cm] 
\item[\hspace{-0.3cm}]
Crovisier, J. 1999, Earth Moon Plan., 79, 125
\\[-0.57cm] 
\item[\hspace{-0.3cm}]
Crovisier, J., Brooke, T. Y., Hanner, M. S., et al.\ 1996, A\&A, 315,{\linebreak}
 {\hspace*{-0.6cm}}L385
\\[-0.57cm] 
\item[\hspace{-0.3cm}]
Crovisier, J., Leech, K., Bockel\'ee-Morvan, D., et al.\ 1997, Science,{\linebreak}
 {\hspace*{-0.6cm}}275, 1904
\\[-0.57cm] 
\item[\hspace{-0.3cm}]
Crovisier, J., Leech, K., Bockel\'ee-Morvan, D., et al.\ 1999, in The{\linebreak}
 {\hspace*{-0.6cm}}Universe As Seen by ISO, ESA-SP 427, ed.\ P.\ Cox \& M.\ F.\
 Kes-{\linebreak}
 {\hspace*{-0.6cm}}sler (Noordwijk, Netherlands:\ ESTEC), 137
\\[-0.57cm] 
\item[\hspace{-0.3cm}]
Crovisier, J., Bockel\'ee-Morvan, D., Biver, N., et al.\ 2004a, A\&A,{\linebreak}
 {\hspace*{-0.6cm}}418, L35
\\[-0.57cm]
\item[\hspace{-0.3cm}]
Crovisier, J., Bockel\'ee-Morvan, D., Colom, P., et al.\ 2004b, A\&A,{\linebreak}
 {\hspace*{-0.6cm}}418, 1141
\\[-0.57cm] 
\item[\hspace{-0.3cm}]
Dello Russo, N., Mumma, M. J., DiSanti, M. A., et al.\ 2000,~Icarus,{\linebreak}
 {\hspace*{-0.6cm}}143, 324
\\[-0.57cm] 
\item[\hspace{-0.3cm}]
de Pater, I., Forster, J., Wright, M., et al. 1998, AJ, 116, 987
\\[-0.57cm]
\item[\hspace{-0.3cm}]
Despois, D. 1999, Earth Moon Plan., 79, 103
\\[-0.57cm] 
\item[\hspace{-0.3cm}]
DiSanti, A., Mumma, M. J., Dello Russo, N., et al.\ 1999, Nature,{\linebreak}
 {\hspace*{-0.6cm}}399, 662
\\[-0.57cm] 
\item[\hspace{-0.3cm}]
Dixon, A. R., Xue, T., \& Sanov, A. 2016, J.\ Chem.\ Phys., 144,{\linebreak}
 {\hspace*{-0.6cm}}234305
\\[-0.57cm] 
\item[\hspace{-0.3cm}]
Durig, J. R., \& Hannum, S. E. 1971, J. Cryst. Mol. Str., 1, 131
\\[-0.57cm] 
\item[\hspace{-0.3cm}]
Enzian, A., Cabot, H., \& Klinger, J. 1998, Plan.\ Space Sci., 46, 851
\\[-0.57cm]
%
%
\item[\hspace{-0.3cm}]
Fray, N., Bardyn, A., Cottin, H., et al. 2016, Nature, 538, 72
\\[-0.57cm]
\item[\hspace{-0.3cm}]
Friedman, A. S., \& Haar, L. 1954, J.\ Chem.\ Phys., 22, 2051
\\[-0.57cm]
\item[\hspace{-0.3cm}]
Froeschl\'e, C., \& Rickman, H. 1986, A\&A, 170, 145
\\[-0.57cm]
\item[\hspace{-0.3cm}]
Fulle, M., Cremonese, G., \& B\"{o}hm, C.\ 1998, AJ, 116, 1470
\\[-0.57cm]
\item[\hspace{-0.3cm}]
Fulle, M., Della Corte, V., Rotundi, A., et al.\ 2016, MNRAS, 462,{\linebreak}
 {\hspace*{-0.6cm}}S132
\\[-0.57cm]
\item[\hspace{-0.3cm}]
Galle, J. G. 1894, Cometenbahnen. (Leipzig: Engelmann Verlag)
\\[-0.57cm]
\item[\hspace{-0.3cm}]
George, D. 1995, IAU Circ.\ 6205
\\[-0.57cm]
\item[\hspace{-0.3cm}]
Green, D. W. E. (ed.) 1995, Int.\ Comet Q., 17, 193
\\[-0.57cm]
\item[\hspace{-0.3cm}]
Green, D. W. E. (ed.) 1996, Int.\ Comet Q., 18, 24, 58, 130
\\[-0.57cm]
\item[\hspace{-0.3cm}]
Green, D. W. E. (ed.) 1997, Int.\ Comet Q., 19, 126, 185, 255
\\[-0.57cm]
\item[\hspace{-0.3cm}]
Green, D. W. E. (ed.) 1998, Int.\ Comet Q., 20, 20, 185
\\[-0.57cm]
\item[\hspace{-0.3cm}]
Green, D. W. E. (ed.) 1999, Int.\ Comet Q., 21, 122
\\[-0.57cm]
\item[\hspace{-0.3cm}]
Gr\"{u}n, E., Hanner, M. S., Peschke, S. B., et al.\ 2001, A\&A, 377,{\linebreak}
 {\hspace*{-0.6cm}}1098
\\[-0.57cm] 
\item[\hspace{-0.3cm}]
Gunnarsson, M., Bockel\'ee-Morvan, D., Winnberg, A., et al. 2003,{\linebreak}
 {\hspace*{-0.6cm}}A\&A, 402, 383
\\[-0.57cm]
\item[\hspace{-0.3cm}]
Hale, A., \& Bopp, T. 1995, IAU Circ., 6187
\\[-0.57cm]
\item[\hspace{-0.3cm}]
Hanner, M. S. 1983, in Cometary Exploration, ed.\ T.\ I.\ Gombosi{\linebreak}
 {\hspace*{-0.6cm}}(Budapest:\ Hungarian Acad.\ Sci.), vol.\ 2, 1
\\[-0.57cm]
\item[\hspace{-0.3cm}]
Hanner, M. S. 1984, Adv. Space Res., 4(9), 189
\\[-0.57cm]
\item[\hspace{-0.3cm}]
Hanner, M. S., Gehrz, R. D., Harker, D. E., et al.\ 1999, Earth{\linebreak}
 {\hspace*{-0.6cm}}Moon Plan., 79, 247
\\[-0.57cm]
\item[\hspace{-0.3cm}]
Harker, D. E., Woodward, C. E., McMurtry, C. W., et al. 1999,{\linebreak}
 {\hspace*{-0.6cm}}Earth Moon Plan., 78, 259
\\[-0.57cm]
\item[\hspace{-0.3cm}]
Harker, D. E., Wooden, D. H., Woodward, C. E., \& Lisse, C. M.{\linebreak}
 {\hspace*{-0.6cm}}2002, ApJ, 580, 579 (Erratum: 2004, ApJ, 615, 1081)
\\[-0.57cm]
\item[\hspace{-0.3cm}]
Harris, W. M., Scherb, F., Mierkiewicz, E., et al.\ 2002. ApJ, 578,{\linebreak}
 {\hspace*{-0.6cm}}996
\\[-0.57cm]
\item[\hspace{-0.3cm}]
Hasegawa, I. 1980, Vistas Astron., 24, 59
\\[-0.57cm]
\item[\hspace{-0.3cm}]
Hayward, T. L., Hanner, M. S., \& Sekanina, Z. 2000, ApJ, 538, 428
\\[-0.57cm]
\item[\hspace{-0.3cm}]
Ho, P. Y. 1962, Vistas Astron., 5, 127
\\[-0.57cm]
\item[\hspace{-0.3cm}]
H\"{o}rz, F., Bastien, R., Borg, J., et al.\ 2006, Science, 314, 1716
\\[-0.57cm]
\item[\hspace{-0.3cm}]
Howard, R. A., Moses, J. D., Vourlidas, A., et al.\ 2008, SSRv,
 136,{\linebreak}
 {\hspace*{-0.6cm}}67
\\[-0.57cm]
\item[\hspace{-0.3cm}]
Jewitt, D., \& Matthews, H. 1999, AJ, 117, 1056
\\[-0.57cm] 
\item[\hspace{-0.3cm}]
Jewitt, D., Senay, M., \& Matthews, H. 1996, Science, 271, 1110
\\[-0.57cm] 
\item[\hspace{-0.3cm}]
Kidger, M. R. 1999, Earth Moon Plan., 78, 169
\\[-0.57cm]
%
%
\item[\hspace{-0.3cm}]
Kramer, E. A., Fern\'andez, Y. R., Lisse C. M., et al. 2014, Icarus,{\linebreak}
 {\hspace*{-0.6cm}}236, 136
\\[-0.57cm]
\item[\hspace{-0.3cm}]
Kr\'olikowska, M. 2004, A\&A, 427, 1117
\\[-0.57cm]
\item[\hspace{-0.3cm}]
Kronk, G. W. 1999, Cometography, Vol.\ 1:\ Ancient--1799. (Cam-{\linebreak}
 {\hspace*{-0.6cm}}bridge, UK:\ Cambridge University Press)
\\[-0.57cm]
\item[\hspace{-0.3cm}]
Landgraf, W. 1986, A\&A, 163, 246
\\[-0.57cm]
\item[\hspace{-0.3cm}]
Lasue, J., Levasseur-Regourd, A. C., Hadamcik, E., \& Alcouffe,~G.{\linebreak}
 {\hspace*{-0.6cm}}2009, Icarus, 199, 129
\\[-0.57cm]
%
%
%
\item[\hspace{-0.3cm}]
Liller, W. 2001, Int.\ Comet Q., 23, 93
\\[-0.57cm]
\item[\hspace{-0.3cm}]
Lisse, C. M., Fern\'andez, Y. R., A'Hearn, M. F., et al. 1999, Earth{\linebreak}
 {\hspace*{-0.6cm}}Moon Plan., 78, 251
\\[-0.57cm]
\item[\hspace{-0.3cm}]
Liu, C., Liu, X., \& Ma, L. 2003, J.\ Astron.\ Hist.\ Herit., 6, 53
\\[-0.57cm]
\item[\hspace{-0.3cm}]
Marcus, J. N. 2007, Int.\ Comet Q., 29, 39
\\[-0.57cm]
\item[\hspace{-0.3cm}]
Marsden, B. G. 1969, AJ, 74, 720
\\[-0.57cm]
\item[\hspace{-0.3cm}]
Marsden, B. G. 1970, AJ, 75, 75
\\[-0.57cm]
\item[\hspace{-0.3cm}]
Marsden, B. G. 1995, IAU Circ., 6194
\\[-0.57cm]
\item[\hspace{-0.3cm}]
Marsden, B. G. 1999, Earth Moon Plan., 79, 3
\\[-0.57cm]
\item[\hspace{-0.3cm}]
Marsden, B. G. 2007, Minor Plan.\ Circ., 61436
\\[-0.57cm]
\item[\hspace{-0.3cm}]
Marsden, B. G., \& Williams, G. V. 2008, Catalogue of Cometary
 {\hspace*{-0.6cm}}Orbits 2008, p.\ 108 (17th ed.; Cambridge, MA:
 Smithsonian {\hspace*{-0.6cm}}Astrophysical Observatory, 195pp)
\\[-0.57cm]
\item[\hspace{-0.3cm}]
Marsden, B.\,G., Sekanina, Z., \& Yeomans, D.\,K.\ 1973, AJ,\,78,\,211
\\[-0.57cm]
\item[\hspace{-0.3cm}]
Marsden, B. G., Sekanina, Z., \& Everhart, E. 1978, AJ, 83, 64
\\[-0.57cm]
\item[\hspace{-0.3cm}]
McDonnell, J. A. M., Lamy, P. L., \& Pankiewicz, G. S. 1991, in{\linebreak}
 {\hspace*{-0.6cm}}Comets in the Post-Halley Era, ed.\ R.\ L.\ Newburn, Jr.,
 M.\ Neu-{\linebreak}
 {\hspace*{-0.6cm}}gebauer, \& J.\ Rahe (Dordrecht, Netherlands:\ Kluwer),
 1043
\\[-0.57cm]
\item[\hspace{-0.3cm}]
McNaught, R. H. 1995, IAU Circ., 6198
\\[-0.57cm]
\item[\hspace{-0.3cm}]
Min, M., Hovenier, J. W., de Koter, A., et al.\ 2005, Icarus, 179, 158
\\[-0.57cm]
\item[\hspace{-0.3cm}]
Moreno, F., Mu\~{n}oz, O., Vilaplana, R., \& Molina, A. 2003, ApJ,{\linebreak}
 {\hspace*{-0.6cm}}595, 522
\\[-0.57cm]
\item[\hspace{-0.3cm}]
Murphy, D. M., \& Koop, T. 2005, QJRMS, 131, 1539
\\[-0.57cm]
\item[\hspace{-0.3cm}]
Pang, K. D., \& Yau, K. K. 1996, in Dynamics, Ephemerides,~and{\linebreak}
 {\hspace*{-0.6cm}}Astrometry of the Solar System, IAU Symp.\ 172, ed.\ S.\
 Ferraz-{\linebreak}
 {\hspace*{-0.6cm}}Mello, B.\ Morando, \& J.-E.\ Arlot (Dordrecht,
 Netherlands:{\linebreak}
 {\hspace*{-0.6cm}}Kluwer), 113
\\[-0.57cm]
\item[\hspace{-0.3cm}]
Pingr\'e, A. G. 1783, Com\`etographie; ou, Trait\'e historique et{\linebreak}
 {\hspace*{-0.6cm}}th\'eorique des com\`etes.  (Paris:\ L'Imprimerie Royale)
\\[-0.57cm]
\item[\hspace{-0.3cm}]
Porter, J. G. 1961, Mem. Brit. Astron. Assoc., 39, No. 3
\\[-0.57cm]
\item[\hspace{-0.3cm}]
Prialnik, D. 1997, ApJ, 478, L107
\\[-0.57cm]
\item[\hspace{-0.3cm}]
Probstein, R. F. 1969, in Problems of Hydrodynamics~and~Con-{\linebreak}
 {\hspace*{-0.6cm}}tinuum Mechanics, ed.\ F. Bisshopp \& L. I. Sedov
 (Philadelphia:{\linebreak}
 {\hspace*{-0.6cm}}Soc.\ Ind.\ Appl.\ Math.), 568
\\[-0.57cm]
\item[\hspace{-0.3cm}]
Rauer, H., Arpigny, C., Boehnhardt, H., et al. 1997, Science, 275,{\linebreak}
 {\hspace*{-0.6cm}}1909
\\[-0.57cm]
\item[\hspace{-0.3cm}]
Rickman, H., \& Froeschl\'e, C. 1983, in Cometary Exploration, ed.{\linebreak}
 {\hspace*{-0.6cm}}T.\ I.\ Gombosi (Budapest:\ Hungarian Acad.\ Sci.), vol.\ 3,
 109
\\[-0.57cm]
\item[\hspace{-0.3cm}]
Rickman, H., \& Froeschl\'e, C. 1986, A\&A, 170, 161
\\[-0.57cm]
\item[\hspace{-0.3cm}]
S\'arneczky, K., Szabo, G., \& Kiss, L. 2011, Minor Plan. Circ., 74775
\\[-0.57cm]
\item[\hspace{-0.3cm}]
Schleicher, D. G., Lederer, S. M., Millis, R. L., \& Farnham, T. L.{\linebreak}
 {\hspace*{-0.6cm}}1997, Science, 275, 1913
\\[-0.57cm] 
%
%
\item[\hspace{-0.3cm}]
Sekanina, Z. 1988a, AJ, 95, 911
\\[-0.57cm]
\item[\hspace{-0.3cm}]
Sekanina, Z. 1988b, AJ, 96, 1455
\\[-0.57cm]
\item[\hspace{-0.3cm}]
Sekanina, Z. 1996, A\&A, 314, 957
\\[-0.57cm]
\item[\hspace{-0.3cm}]
Sekanina, Z., \& Farrell, J. A. 1982, AJ, 87, 1836
\\[-0.57cm]
\item[\hspace{-0.3cm}]
Sekanina, Z., \& Kracht, R. 2014, eprint arXiv:1404.5968
\\[-0.57cm]
\item[\hspace{-0.3cm}]
Sekanina, Z., \& Kracht, R. 2015, ApJ, 801, 135
\\[-0.57cm]
\item[\hspace{-0.3cm}]
Sekanina, Z., \& Kracht, R. 2016, eprint arXiv:1607.3440
\\[-0.57cm]
\item[\hspace{-0.3cm}]
Senay, M., Rownd, B., Lovell, A., et al. 1997, BAAS, 29, 1034
\\[-0.57cm]
\item[\hspace{-0.3cm}]
Shaw, I.\,(ed.) 2000, The Oxford History of Ancient
 Egypt.~(Oxford:{\linebreak}
 {\hspace*{-0.6cm}}Oxford University Press)
\\[-0.57cm]
\item[\hspace{-0.3cm}]
Sitarski, G. 1990, Acta Astron., 40, 405
\\[-0.57cm]
\pagebreak
\item[\hspace{-0.3cm}]
Sitarski, G. 1994a, Acta Astron., 44, 91
\\[-0.57cm]
\item[\hspace{-0.3cm}]
Sitarski, G. 1994b, Acta Astron., 44, 417
\\[-0.57cm]
\item[\hspace{-0.3cm}]
Sosa, A., \& Fern\'andez, J. A. 2011, MNRAS, 416, 767
\\[-0.57cm]
\item[\hspace{-0.3cm}]
Stern, S. A., Colwell, W. B., Festou, M. C., et al.\ 1999, AJ, 118,{\linebreak}
 {\hspace*{-0.6cm}}1120
\\[-0.57cm] 
\item[\hspace{-0.3cm}]
Stratton, F. J. M. 1928, in Handbuch der Astrophysik (vol.\ 6),{\linebreak}
 {\hspace*{-0.6cm}}ed.\ G. Eberhard, A. Kohlsch\"{u}tter, \& H. Ludendorff
 (Berlin:{\linebreak}
 {\hspace*{-0.6cm}}Springer), 251
\\[-0.57cm]
\item[\hspace{-0.3cm}]
Szab\'o, Gy. M., Kiss, L. L., \& S\'arneczky, K. 2008, ApJ, 677, 121
\\[-0.57cm]
\item[\hspace{-0.3cm}]
Szab\'o, Gy. M., S\'arneczky, K., \& Kiss, L. L. 2011, A\&A, 531, A11
\\[-0.57cm]
\item[\hspace{-0.3cm}]
Szab\'o, Gy. M., Kiss, L. L., P\'al, A., et al. 2012, ApJ, 761, 8
\\[-0.57cm]
\item[\hspace{-0.3cm}]
Szutowicz, S., Kr\'olikowska, M., \& Sitarski, G. 2002, Earth Moon{\linebreak}
 {\hspace*{-0.6cm}}Plan., 90, 119
\\[-0.57cm]
\item[\hspace{-0.3cm}]
Touloukian,\,Y.\,S.\,(ed.)\,1970a, \mbox{Thermophysical\,Properties\,of\,Matter.}
 {\hspace*{-0.6cm}}Vol.\,4:\ \mbox{Specific\,Heat\,--\,Metallic\,Elements\,and\,Alloys.}\ Thermophys.{\linebreak}
 {\hspace*{-0.6cm}}Prop.\ Res.\ Center Data Series, Purdue Univ. (New York:\
 Plenum{\linebreak}
 {\hspace*{-0.6cm}}Press)
\\[-0.57cm]
\item[\hspace{-0.3cm}]
Touloukian,\,Y.\,S.\,(ed.)\,1970b,\,Thermophysical\,Properties\,of\,Matter.
 {\hspace*{-0.6cm}}Vol.\,5:\ Specific Heat -- Nonmetallic Solids. Thermophys.\
 Prop.{\linebreak}
 {\hspace*{-0.6cm}}Res.\,Center\,Data\,Series,\,Purdue\,Univ. (New\,York:\ Plenum Press)
\\[-0.57cm]
%
%
\item[\hspace{-0.3cm}]
Vasundhara, R., \& Chakraborty, P. 1999, Icarus, 140, 221
\\[-0.57cm]
\item[\hspace{-0.3cm}]
Vsekhsvyatsky, S. K. 1958, Fizicheskie kharakteristiki komet.{\linebreak}
 {\hspace*{-0.6cm}}(Moscow:\ Gosud.\ izd-vo fiz.-mat.\ lit.); translated: 1964,
 Physical{\linebreak}
 {\hspace*{-0.6cm}}Characteristics of Comets, NASA TT-F-80 (Jerusalem:\
 Israel{\linebreak}
 {\hspace*{-0.6cm}}Program for Scientific Translations)
\\[-0.57cm]
\item[\hspace{-0.3cm}]
Wagner, W., \& Pruss, A. 2002, J.\ Phys.\ Chem.\ Ref.\ Data, 31, 387
\\[-0.57cm]
%
%
\item[\hspace{-0.3cm}]
Weaver, H.\,A., Feldman, P.\,D., A'Hearn, M.\,F., et al.\,1997, Science,{\linebreak}
 {\hspace*{-0.6cm}}275, 1900
\\[0.77cm] 
\item[\hspace{-0.3cm}]
Weaver, H. A., Brooke, T. Y., Chin, G., et al.\ 1999a, Earth Moon{\linebreak}
 {\hspace*{-0.6cm}}Plan., 78, 71
\\[-0.7cm]
\item[\hspace{-0.3cm}]
Weaver, H.\,A., Feldman, P.\,D., A'Hearn, M.\,F., et al.\ 1999b,
 Icarus,{\linebreak}
 {\hspace*{-0.6cm}}141, 1
\\[-0.7cm]
\item[\hspace{-0.3cm}]
Weiler, M., Rauer, H., Knollenberg, J., et al.\ 2003, A\&A, 403, 313
\\[-0.7cm]
\item[\hspace{-0.3cm}]
Williams, D. M., Mason, C. G., Gehrz, R. D., et al. 1997, ApJ,~489,{\linebreak}
 {\hspace*{-0.6cm}}91
\\[-0.7cm]
\item[\hspace{-0.3cm}]
Williams, G. V. 2011, Minor Plan. Circ., 75007
\\[-0.7cm]
%
%
\item[\hspace{-0.3cm}]
Williams, J. 1871, Observations of Comets from B.C.\ 611 to~A.D.{\linebreak}
 {\hspace*{-0.6cm}}1640, Extracted from the Chinese Annals.
 (London:~Strange-{\linebreak}
 {\hspace*{-0.6cm}}ways \& Walden)
\\[-0.7cm]
\item[\hspace{-0.3cm}]
Womack, M., Festou, M. C., \& Stern, S. A. 1997, AJ, 114, 2789
\\[-0.7cm] 
\item[\hspace{-0.3cm}]
Wooden, D. H., Harker, D. E., Woodward, C. E., et al. 1999,~ApJ,{\linebreak}
 {\hspace*{-0.6cm}}517, 1034
\\[-0.7cm]
\item[\hspace{-0.3cm}]
Wooden, D. H., Butner, H. M., Harker, D. E., \& Woodward, C.~E.{\linebreak}
 {\hspace*{-0.6cm}}2000, Icarus, 143, 126
\\[-0.7cm]
\item[\hspace{-0.3cm}]
Wright, I. P., Sheridan, S., Barber, S. J., et al. 2015, Science,
 349,{\linebreak}
 {\hspace*{-0.6cm}}aab0673
\\[-0.7cm]
\item[\hspace{-0.3cm}]
Wu, Q., Zhao, Z., Liu, L., et al. 2016, Science, 353, 579
\\[-0.7cm]
\item[\hspace{-0.3cm}]
Wylie, L. M. 1958, The Vapor Pressure of Solid Argon, Carbon{\linebreak}
 {\hspace*{-0.6cm}}Monoxide, Methane, Nitrogen, and Oxygen from Their
 Triple{\linebreak}
 {\hspace*{-0.6cm}}Points to the Boiling Point of Hydrogen.  Thesis. (Atlanta,
 GA:{\linebreak}
 {\hspace*{-0.6cm}}Georgia Institute of Technology)
\\[-0.7cm]
\item[\hspace{-0.3cm}]
Yeomans, D. K. 1984, in Cometary Astrometry, ed.\ D.\ K.\ Yeomans,{\linebreak}
 {\hspace*{-0.6cm}}R.\ M.\ West, R.\ S.\ Harrington, \& B.\ G.\ Marsden
 (Pasadena:\ Jet{\linebreak}
 {\hspace*{-0.6cm}}Propulsion Laboratory), 167
\\[-0.73cm]
\item[\hspace{-0.3cm}]
Yeomans, D. K., \& Chodas, P. W. 1989, AJ, 98, 1083
\\[-0.81cm]
\item[\hspace{-0.3cm}]
Yomogida, K., \& Matsui, T. 1983, J.\ Geophys.\ Res., 88, 9513}
%
\vspace*{0.55cm}
\end{description}
\end{document}